\documentclass[twocolumn, notitlepage, aps, pra, floatfix, superscriptaddress]{revtex4-1}

\usepackage{algorithm}
\usepackage{algpseudocode}
\usepackage{algorithmicx}

\usepackage{csvsimple}
\usepackage{datatool}
\usepackage{graphicx}
\usepackage{multirow}
\usepackage{amsmath}
\usepackage{amssymb}
\usepackage{dsfont}
\usepackage{amsthm}
\usepackage{eucal}
\usepackage{qcircuit}
\usepackage{bm}
\usepackage{upgreek}
\usepackage{mathrsfs}
\usepackage{color}
\usepackage{latexsym}
\usepackage[dvipsnames]{xcolor}
\usepackage{float}
\usepackage{enumerate}
\usepackage{stackengine}
\usepackage{braket}
\usepackage[titletoc]{appendix}
\usepackage{natbib}
\usepackage{xcolor}
\usepackage{array}
\usepackage{cancel}
\usepackage{xpatch}
\xpatchbibdriver{misc}
  {\printtext[parens]{\usebibmacro{date}}}
  {\iffieldundef{year}
    {}
    {\printtext[parens]{\usebibmacro{date}}}}
  {}
  {\typeout{There was an error patching biblatex-ieee (specifically, ieee.bbx's @online driver)}}

\usepackage [english]{babel}
\usepackage [autostyle, english = american]{csquotes}
\MakeOuterQuote{"}

\theoremstyle{remark}

\usepackage{hyperref}
\hypersetup{
colorlinks,
linkcolor={blue},
citecolor={red},
urlcolor={blue},
linktocpage
}

\usepackage[caption=false]{subfig}
\algrenewcommand{\algorithmiccomment}[1]{\# #1}

\newcommand{\bb}{\begin{equation}}
\newcommand{\ee}{\end{equation}}
\newcommand{\bbb}{\begin{equation*}}
\newcommand{\eee}{\end{equation*}}

\newcommand{\id}{\mathds{1}}

\stackMath

\DeclareMathOperator{\Tr}{Tr}
\DeclareMathOperator{\sech}{sech}
\begin{document}

\title{Progress towards practical qubit computation using approximate Gottesman-Kitaev-Preskill codes}
\author{Ilan Tzitrin}
\thanks{These two authors contributed equally.}
\affiliation{Department of Physics, University of Toronto, Toronto, Canada}
\affiliation{Xanadu, 777 Bay Street, Toronto ON,  M5G 2C8, Canada}
\author{J. Eli Bourassa}
\thanks{These two authors contributed equally.}
\affiliation{Department of Physics, University of Toronto, Toronto, Canada}
\affiliation{Xanadu, 777 Bay Street, Toronto ON,  M5G 2C8, Canada}
\author{Nicolas C. Menicucci}
\affiliation{Centre for Quantum Computation and Communication Technology, School of Science, RMIT University, Melbourne, Victoria, Australia}
\author{Krishna Kumar Sabapathy}
\email{krishna@xanadu.ai}
\affiliation{Xanadu, 777 Bay Street, Toronto ON,  M5G 2C8, Canada}

\begin{abstract}
Encoding a qubit in the continuous degrees of freedom of an oscillator is a promising path to error-corrected quantum computation. One advantageous way to achieve this is through Gottesman-Kitaev-Preskill (GKP) grid states, whose symmetries allow for the correction of any small continuous error on the oscillator. Unfortunately, ideal grid states have infinite energy,  so it is important to find finite-energy approximations 
that are realistic, practical, and useful for applications.  
In the first half of this work we investigate the impact of imperfect GKP states on computational circuits independently of the physical architecture. To this end, we analyze the behaviour of the physical and logical content of normalizable GKP states through several figures of merit, employing a recently-developed modular subsystem decomposition. By tracking the errors that enter into the computational circuit due to imperfections in the GKP states, we are able to gauge the utility of these states for NISQ (Noisy Intermediate-Scale Quantum) devices. In the second half, we focus on a state preparation approach in the photonic domain wherein photon-number-resolving measurements on some modes of Gaussian states produce non-Gaussian states in others. We produce detailed numerical results for the preparation of GKP states alongside estimating the resource requirements in practical settings and probing the quality of the resulting states with the tools we develop. Our numerical experiments indicate that we can generate any state in the GKP Bloch sphere with nearly equal resources, which has implications for magic state preparation overheads.
\end{abstract}

\maketitle 

\tableofcontents

\section{Introduction}
\label{sec:intro}
Bosonic codes are quantum codes where logical qubits are encoded in states of continuous-variable systems. The most well-known among bosonic codes are those introduced by Gottesman, Kitaev and Preskill (GKP)~\cite{GKP}, also known as grid states. These codes were constructed to correct displacement errors in phase space with the help of their inherent translation symmetry. An error analysis of finite energy GKP states was presented in~\cite{glancy}. The GKP code has garnered much interest in recent years for various applications. 

Fault-tolerant quantum computation (FTQC) using GKP qubits was shown to be possible in the context of continuous-variable measurement-based quantum computation, along with a conservative upper bound of $20.5$ dB on the squeezing required~\cite{clusterFT}.  This result has spurred research in this direction. It has been demonstrated that the only non-Gaussian resource required for universal FTQC using GKP codes is the ability to faithfully prepare
the logical zero state~\cite{all-gauss}. It has also been shown that noise in the anti-squeezing quadrature of the squeezed states is immaterial in such an implementation~\cite{walshe2019}.  The authors in~\cite{walshe2019} further pointed out that practical squeezing thresholds were closer to 15 -- 17~dB depending on the choice of qubit-level error-correcting code, a point previously made in~\cite{clusterFT} but rarely appreciated.  The squeezing requirement was further reduced to $\sim$10~dB~\cite{fukui2018high} alongside methods to track the error correction~\cite{fukui2018tracking} and perform a hybrid error correction procedure using digital and analog information~\cite{fukui2017analog}. Further refinements to the techniques and the incorporation of practical imperfections has led to a squeezing requirement that is less than $10$ dB for a certain loss parameter range~\cite{fukui-real}. 

There are several other applications of the GKP code. The performance of GKP codes, including the hexagonal GKP code, was compared to other codes with respect to transmission through a bosonic loss channel in Refs.~\cite{albert2018performance} and~\cite{noh2018quantum}. Comparison of the structure of the GKP code and the rotation-symmetric code was presented in~\cite{grimsmo2019quantum}. The use of GKP qubits for quadratic noise suppression in continuous variables was considered in~\cite{noh2019}. The surface code using GKP qubits has been considered in~\cite{fukui2018high,vuillot2019quantum,mastersthesis,nohgkpsurface}. The use of GKP qubits as a displacement sensor was explored in~\cite{duivenvoorden2017single}. A probabilistic preparation of GKP qubits and its role in FTQC and sampling problems was presented in~\cite{douce2019probabilistic}. The sign-altered GKP states were studied as a means for decoherence suppression in~\cite{1901.05358}. 
Further, the fundamental mathematical structure of the GKP code states was presented  in~\cite{englert2006periodic,ketterer2016quantum,Pantaleoni2019}, connections among the different approximation schemes were analytically explored in \cite{matsuura2019}, and aspects of the syndrome extraction was considered in \cite{wan2019}. 

Considering the extensive utility and optimality~\cite{noh2018quantum} of the GKP code, preparing high quality GKP states is broadly desirable. There are several proposals and demonstrations for generating GKP qubits across various platforms, including ion traps~\cite{travaglione2002preparing,pirandola2006continuous,fluhmann2019encoding,fluhmann2018}, superconducting cavities~\cite{1907.12487}, light-matter interactions~\cite{pirandola2006generating,motes2017encoding}, cross-Kerr interactions~\cite{pirandola2004constructing}, using phase estimation~\cite{terhal2016encoding} (including its fault-tolerant version~\cite{ibm2019, blog}), and an analysis of practical aspects of modular quadrature measurements \cite{weigand2019} all in circuit-QED, using cat states~\cite{weigand2018generating,puri19}, optical methods~\cite{vasconcelos2010all,Etesse_2014}, photon catalysis~\cite{eaton2019gottesman}, 
a machine learning method that uses self-Kerr interactions~\cite{arrazola2018machine}, realization in time-frequency encodings~\cite{fabre2019}, as eigenstates of tailored Hamiltonians in superconducting circuits~\cite{arne}, a plausible attempt using quantum walks~\cite{su2018encoding}, and a measurement-free proposal~\cite{hastrup2019}.

We focus on the photonic method using the minimal resources of squeezed displaced vacuum states, interferometers, and photon number-resolving detectors first introduced in~\cite{saba1} and subsequently developed in \cite{su1,su2,1905.07011}. Our scheme is probabilistic in nature due to the postselection on specific photon detection outcomes. Photonic platforms have a proven track-record for implementing various forms of error correction circuits such as two-qubit encoding in polarization modes~\cite{pittman2005}, topological error correction with an 8-photon cluster state~\cite{yao2012experimental}, graph state codes~\cite{bell2014experimental,1811.03023}, Shor's 9-qubit code~\cite{aoki2009quantum}, 5 wave-packet error correcting code~\cite{hao2015five}, qubit codes preventing loss errors~\cite{lu2008experimental}, and certain continuous-variable error correction~\cite{lassen2013gaussian,lassen2010quantum} with applications to quantum sensing \cite{Zhang2019}. 

In this work, we focus on the role played by physical approximations to ideal grid states. The approximations reveal themselves in various figures of merit and circuit characterizations whose understanding is central for NISQ devices. We use the recently-introduced modular subsystem decomposition~\cite{Pantaleoni2019} to monitor and mitigate errors on the logical qubit. We then provide an extensive analysis of the resource requirements for preparing these approximate states and useful benchmarks for employing them in computational circuits.

The paper is organized as follows. In Sec.~\ref{sec:formalism}, we review the formalism of the GKP code independently of the platform for its implementation. This includes a synthesis of results for the ideal (non-normalizable) GKP states in~\ref{subsec:ideal_GKP} and the finite-energy GKP states in~\ref{subsec:norm_gkp}; an overview of the modular decomposition in~\ref{subsec:mod_decomp} and its relevance to analyzing GKP states; a set of prescriptions for tracking, quantifying, and alleviating errors in computation induced by imperfect GKP states in~\ref{subsec:ops_errors}; and a discussion of error correction and recovery with the normalizable states in \ref{subsec:error_correction}.

Sec.~\ref{sec:stateprep} is dedicated to the preparation of approximate GKP states in the optical domain. In~\ref{subsec:nongauss} we discuss an architecture analogous to \textsf{Gaussian} \textsf{BosonSampling} devices for the generation of general non-Gaussian states; in \ref{subsec:core_states} we discuss a framework for approximating GKP states; in~\ref{subsec:charac_app} we characterize the approximate states using the formalism described in Sec.~\ref{sec:formalism}; in~\ref{subsec:circuits} we present our strategy for the preparation of these approximate GKP states with optical circuits and provide our numerical results for the optimization of these circuits; and finally, in~\ref{subsec:exp_imperf}, we give comments on incorporating experimental imperfections into our scheme.

\section{Formalism of bosonic grid codes} \label{sec:formalism}

Our physical landscape consists of continuous-variable systems whose Hilbert space is $L^2(\mathbb{R}^n)$, the square-integrable functions defined over the space of $n$ real variables corresponding to $n$-mode systems. Examples of continuous-variable systems include modes of an electromagnetic field, harmonic oscillator chains, phonon modes in materials, continuous modes of ion traps and superconducting circuits. In what follows we work in units where $\hat{q} = \tfrac 1 {\sqrt{2}} (\hat{a} + \hat{a}^\dag)$ and $\hat{p} = \tfrac {-i} {\sqrt{2}} (\hat{a} - \hat{a}^\dag)$, so that $[\hat{q},\hat{p}]=i$, and $\hbar=1$. This means the measured variance of a vacuum state is $\langle \hat{q}^2 \rangle = \langle \hat{p}^2 \rangle = \tfrac 1 2$. For more details on our conventions, see App.~\ref{sec:convention}.

\subsection{Ideal GKP states}\label{subsec:ideal_GKP}

The ideal GKP states~\cite{GKP} are defined as the simultaneous eigenstates of
the continuous-variable stabilizer elements
\begin{subequations}
\label{eq:stabgeneral}
\begin{align}
S_{q} & =e^{i\left(2\sqrt{\pi}\right)(S_{11}\hat{q}+S_{21}\hat{p})}\\
S_{p} & =e^{i\left(2\sqrt{\pi}\right)(S_{12}\hat{q}+S_{22}\hat{p})}
\end{align}
\end{subequations}
such that $\boldsymbol{S}=\begin{bmatrix}S_{11} & S_{12}\\
S_{21} & S_{22} 
\end{bmatrix} \in Sp(2,\mathbb{R})$, the real symplectic group in two dimensions. The Wigner function (see App.~\ref{subsec:non-class} for the definition) of the ideal GKP states
is a sum of delta functions located at the points of a lattice in
phase space; the shape and spacing of this lattice is determined by
$\boldsymbol{S}$. For example, rectangular GKP states are associated
with the matrix
\begin{equation}
\boldsymbol{S}_{\text{rect}}=\begin{bmatrix}\sqrt{\pi}/\alpha & 0\\
0 & \alpha/\sqrt{\pi}
\end{bmatrix},
\end{equation}
for $\alpha \in \mathbb{R}_{\neq{0}}$, so that the unit cell of the lattice has dimensions $\frac{\sqrt{\pi}}{\alpha}\times\frac{\alpha}{\sqrt{\pi}}$
and an area of $1$, while hexagonal GKP states~\cite{GKP, noh2018quantum} have
\begin{equation}
\boldsymbol{S}_{\text{hex}}=\left(\frac{2}{\sqrt{3}}\right)^{1/2}\begin{bmatrix}1 & 1/2\\
0 & \sqrt{3}/2
\end{bmatrix}.
\end{equation}

We be focusing on square-lattice states -- that is,
rectangular GKP states with $\alpha=\sqrt{\pi}$. For these states,
the stabilizer elements become
\begin{subequations}
\label{eq:stabsquare}
\begin{align}
S_{q}^{\text{square}} & =D\left(i\sqrt{2\pi}\right) \equiv Z\left(2\sqrt\pi\right)\\
S_{p}^{\text{square}} & =D\left(\sqrt{2\pi}\right) \equiv X\left(2\sqrt\pi\right),
\end{align}
\end{subequations}
where $D\left(\beta\right)\equiv e^{\beta \hat{a}^{\dagger}-\beta^{*}\hat{a}}$
is the displacement operator, $X(q) \equiv D(q / \sqrt{2})$ is a displacement in position by $q$ and $Z(p) \equiv D(ip/\sqrt{2})$ is a displacement in momentum by $p$. After identifying the logical $Z$
and $X$ operations for this code (logical operators are denoted by an overline),
\begin{equation}
\bar{Z}=Z\left(\sqrt{\pi}\right)\text{ and }\bar{X}=X\left(\sqrt{\pi}\right),
\end{equation}
one can infer that the ideal encoded logical square GKP states, which we label
with the subscript $I$, are infinite superpositions of infinitely
squeezed states -- delta spikes arranged like a comb -- spaced by $2\sqrt{\pi}$ in position:
\begin{equation}
\Ket{0_{I}}\equiv\sum_{n=-\infty}^{\infty}\Ket{2n\sqrt{\pi}}_{q}. \label{eq:0I}
\end{equation}

The encoded logical 1 state is then just a $q$-displacement by $\sqrt{\pi}$ of
the $0$ state: $\Ket{1_{I}}\equiv X\left(\sqrt{\pi}\right)\Ket{0_{I}}$.
The periodicity of the delta spikes implies that the ideal GKP code
can correct displacements of position and momentum of up to $\sqrt{\pi}/2$,
i.e., those displacements that do not confuse a logical 0 for a logical
1~\cite{GKP}. The $\bar{X}$ eigenstates are then just
\begin{equation}
\Ket{+_{I}}\equiv\sum_{n=-\infty}^{\infty}\Ket{n\sqrt{\pi}}_{q},
\end{equation}
and $\Ket{-_{I}}\equiv Z\left(\sqrt{\pi}\right)\Ket{+_{I}}.$

\subsection{Normalizable GKP states}\label{subsec:norm_gkp}
\begin{figure}
    \centering
\includegraphics[width=\linewidth]{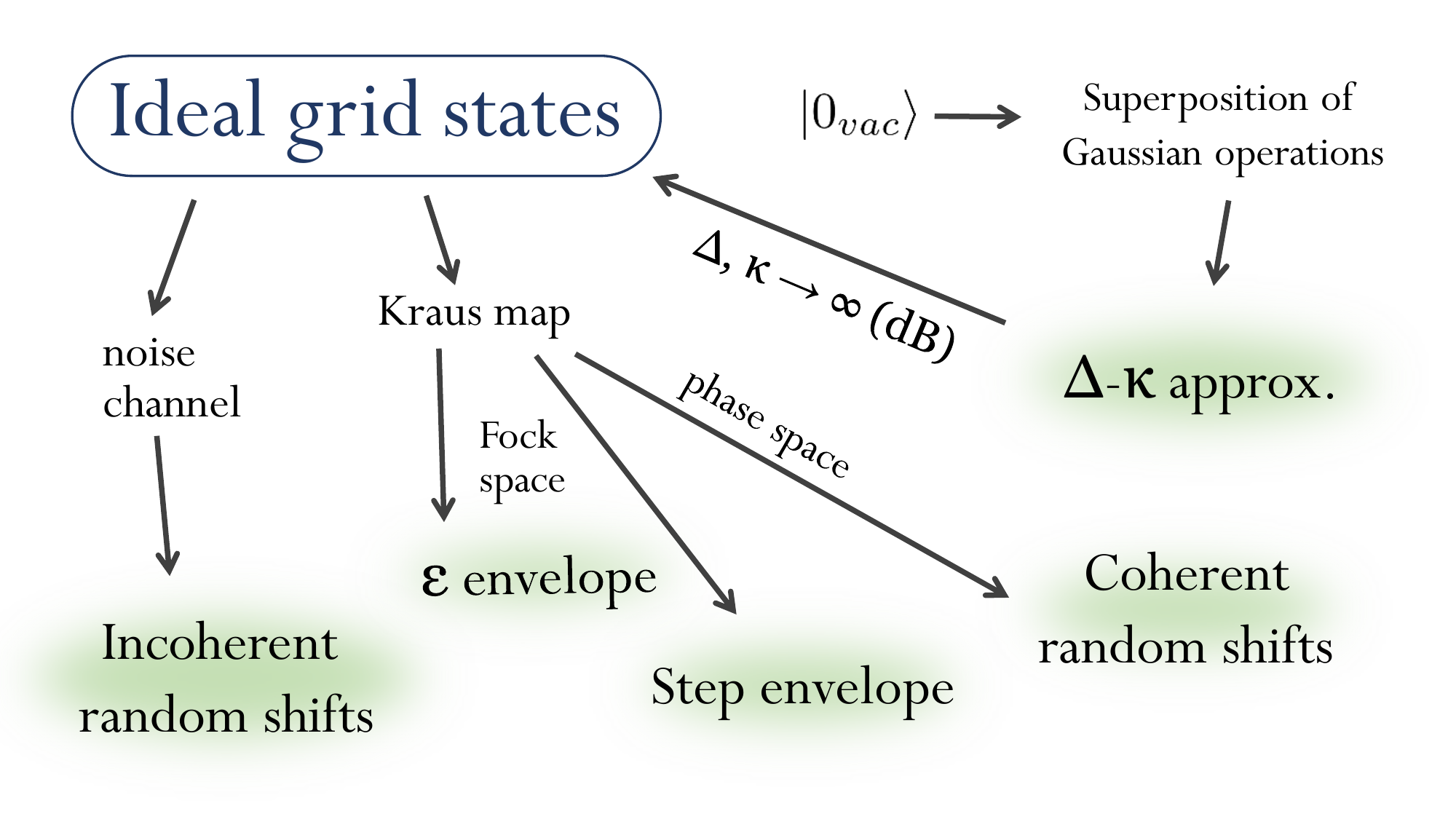}
    \caption{Various ways to obtain physical states starting from ideal grid states. From left to right, the approximations in green clouds correspond to discussions around Eq.~\eqref{eq:class_noise}, \eqref{eq:0Nen}, \eqref{eq:step_del}, \eqref{eq:coh_shifts}, and \eqref{eq:0Ndeltakappa}.}
    \label{fig:idealtoreal}
\end{figure}

\begin{figure}
    $$
    \Qcircuit @C=1.0em @R=.7em { 
    \lstick{\Ket{\psi_I}} & \qw & \multigate{1}{B\left(t = -\ln{\epsilon}\right)} & \qw & \push{\Ket{\psi_{\epsilon}}} \\
    \lstick{\Ket{0_{vac}}} & \qw & \ghost{B\left(t = -\ln{\epsilon}\right)} & \qw & \measureD{n=0}
    }
    $$
    \caption{A schematic of how to obtain the normalizable GKP state $\Ket{\psi_\epsilon} \equiv e^{-\epsilon \hat{n}}\Ket{\psi_I}$ through an optical circuit, as noted in~\cite{noh2019} and shown in App.~\ref{subsec:envelope}. An ideal GKP state and the vacuum state pass through the first and second mode of a beamsplitter $B(\theta, \phi)$ with transmissivity $-\ln{\epsilon}$ (see Eq.~\eqref{eq:beamsplitter}); the second mode is measured and post-selected on a vacuum state.}  
    \label{fig:env_as_bs}
\end{figure}
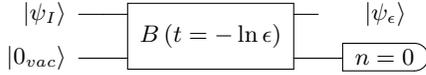

In an experimental setting one will be dealing with finitely squeezed states. The
canonical way to generate normalizable GKP states \footnote{What we call normalizable is generally referred to as approximate
in the literature. We prefer this nomenclature because there are several
different approximations to ideal GKP states we will be considering.}, which we label with a subscript corresponding to the normalization scheme, is to replace the delta
functions with Gaussians of width $\Delta$
and then introduce an overall Gaussian envelope of width~$\kappa^{-1}$:
\begin{align}
\Ket{0_{\Delta, \kappa}} & \propto\sum_{n=-\infty}^{\infty}e^{-\frac{1}{2}\kappa^{2}\left(2n\sqrt{\pi}\right)^{2}}X\left(2n\sqrt{\pi}\right)\Ket{\Delta}_{q}\label{eq:0Ndeltakappa}\\
\Ket{1_{\Delta, \kappa}} & \propto\sum_{n=-\infty}^{\infty}e^{-\frac{1}{2}\kappa^{2}\left[\left(2n+1\right)\sqrt{\pi}\right]^{2}}X\left[\left(2n+1\right)\sqrt{\pi}\right]\Ket{\Delta}_{q}, \label{eq:0Ndeltakappa1}
\end{align}
where $\Ket{\Delta}$ is defined so that
\begin{equation}
    \Braket{q|\Delta} = \left(\frac{1}{\pi\Delta ^ 2}\right)^{\frac{1}{4}} e^{-\frac{q^2}{2\Delta^2}}.
\end{equation}
This is just one prescription to transition from infinite to finite energy states, as shown in Fig.~\ref{fig:idealtoreal}. Recently, three conventional approximations of the GKP codes were analytically shown to be equivalent in \cite{matsuura2019}.

It is common to work in the regime where $\Delta=\kappa$, so that
in the $\Delta\to0$ limit the number of spikes increases while the
width of each spike decreases, and the normalizable states approach the ideal ones. In this case we omit the $\kappa$ from the subscript and write simply 
\begin{equation}
\label{eq:delta=kappa}
    \Ket{\psi_{\Delta}} \equiv \Ket{\psi_{\Delta, \kappa = \Delta}}.
\end{equation}
A less cumbersome way than Eq.~\eqref{eq:0Ndeltakappa} of expressing $\Ket{\psi_{\Delta}}$, as pointed out in~\cite{clusterFT,noh2018quantum},  is to apply a non-unitary envelope operator $E\left(\epsilon\right)\equiv e^{-\epsilon\hat{n}}$
to the ideal state, where  $\hat{n}\equiv \hat{a}^{\dagger}\hat{a}=\frac{1}{2}(\hat{q}^{2}+\hat{p}^{2})$
is the number operator:
\begin{equation}
\Ket{\psi_{\epsilon}} \equiv E\left(\epsilon\right)\Ket{\psi_{I}}.
\label{eq:0Nen}
\end{equation}
One can think of this operator as being the result of interfering an ideal GKP state and a vacuum state at a beamsplitter, measuring one of the modes, and postselecting on the vacuum~\cite{noh2019}. We provide an illustration of this in Fig.~\ref{fig:env_as_bs} and a derivation in App.~\ref{subsec:envelope}.  Note that $\Ket{\psi_{\epsilon}}\approx\Ket{\psi_{\Delta}}$ whenever $\Delta=\kappa$ and $\epsilon$ are small. From~\cite{nohgkpsurface}, the more precise regime is whenever
\begin{equation}
\tanh \frac \Delta 2  \approx\frac{\Delta}{2}.
\label{eq:tanh}
\end{equation}
Since $\tanh x = x - \tfrac 1 3 x^3 + \ldots$,  this will occur whenever $\frac{1}{24}\Delta^{3}$ is negligible.

\begin{figure}
\centering
\includegraphics[width=\linewidth]{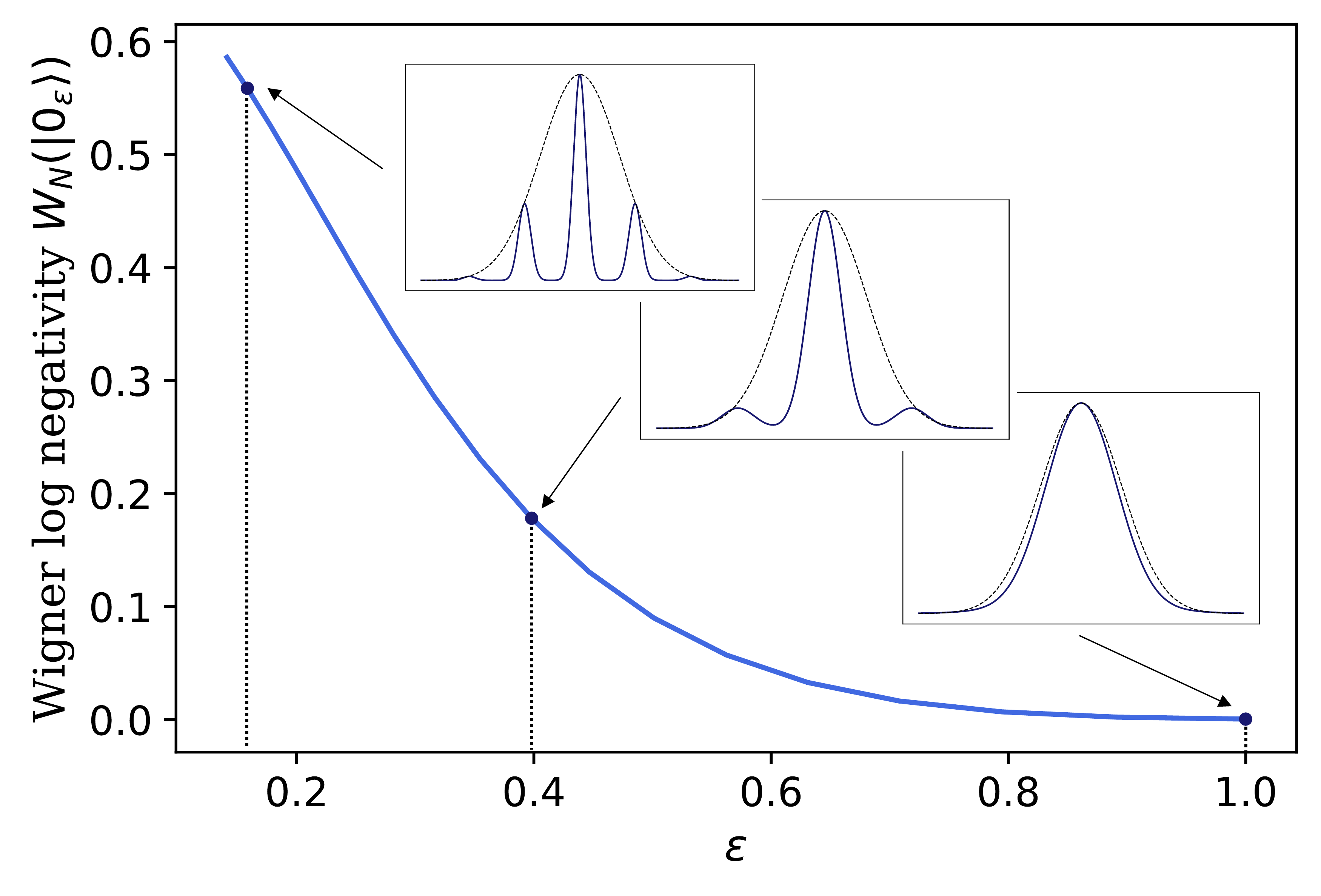}
    \caption{The Wigner logarithmic negativity $W_N$ (defined in App.~\ref{subsec:non-class}) of the normalizable GKP state $\Ket{0_{\epsilon}}$ from Eq.~\eqref{eq:0Nen} as a function of
    $\epsilon$ shown as solid (blue) line. {\bf Insets\,:} Highlighted from left to right are wavefunctions corresponding to $\epsilon$ values of 0.158, 0.398 and 1 ($\Delta$ values of 8, 4, and 0 dB). The overall Gaussian envelope of width $\kappa = \Delta^{-1}$ is also drawn. The Gaussian in the third inset noticeably does not perfectly envelop the $\epsilon=1$ state because here there is an error of order $0.04$ in the approximation between $\Delta$ and $\epsilon$ (as shown in Eq.~\eqref{eq:tanh} and Table {\ref{tab:e_db}). $W_N$ decreases with increasing $\epsilon$, i.e. as the number of peaks in our GKP state decreases, the width of each one increases, and the state becomes more Gaussian.}
\label{fig:wig_log_neg_0e}}
\end{figure}

Since we require non-Gaussian resource states for universal quantum computation, it is important to quantify the non-Gaussianity of the normalizable GKP states. For this, we plot the Wigner logarithmic negativity $W_N$ (see App.~ \ref{subsec:non-class} for the definition) of the $\Ket{0_\epsilon}$ state as a function of $\epsilon$ in Fig.~\ref{fig:wig_log_neg_0e}. We see that $W_N$ decreases for increasing $\epsilon$, as expected.

We can, in principle, come up with other ways of approximately normalizing the ideal
states. In general, we have that
\begin{equation}
\Ket{\psi_{G}} \equiv G\Ket{\psi_{I}},\label{eq:0NG}
\end{equation}
for some operation $G$, which we can regard as an error or a single
Kraus operator. Note that $G$ will not be trace-preserving in general; even though we have written the left-hand-side of \eqref{eq:0NG} as a ket, it is understood that the state will need to be normalized.
For example, $G$ can be defined as the operator
\begin{equation}
\label{eq:coh_shifts}
    G = \sqrt{\frac{2}{\pi \Delta^2}} \int d^2 \alpha \, e^{-|\alpha|^2/\Delta^2} D(\alpha),
\end{equation}
in other words a Gaussian distribution of displacements, as in \cite{GKP, albert2018performance, matsuura2019}.

$G$ can also be chosen so that the normalizable state
is a weighted superposition of Fock states, as in
\begin{equation}
\Ket{\mu_{G}}=\sum_{n=0}^{\infty}g\left(n\right)\Braket{ n|\mu_{I}}\Ket{n},\label{eq:step_del}
\end{equation}
for some envelope function $g\left(n\right)$. For example, we can
set $g$ to be a step function; or, to obtain (\ref{eq:0Nen}), we can make $g(n) = e^{-\epsilon n}$ \footnote{Note that we avoid ambiguity in notation because the subscript of an ideal state \unexpanded{$\Ket{\psi_{I}}$} can be interpreted as a normalizable state with an identity envelope}.  We will explore in Sec.~\ref{sec:stateprep} which envelope functions are best to target in the state preparation scheme we consider.

There are several advantages to using Eq.~\eqref{eq:0Nen} to denote
the normalizable states. Beyond its compactness, it provides us with
a convenient way to explore the effects of deviating from ideal GKP
states on Gaussian operations, as required for implementing Clifford gates, and on the logical content of the states, as we will see. It is also important to point out that the displacement
and envelope operators will, in general, not commute (see App.~\ref{subsec:comm_rel_E} for explicit commutation and conjugation relations with $E(\epsilon)$). This means that
the logical state obtained by displacing $\Ket{0_{G}}$ by $\sqrt{\pi}$ in position will not be the same as the state obtained by applying an envelope
to $\Ket{1_{I}}$. For example,
\begin{align}
X\left(\sqrt{\pi}\right)\Ket{0_{\epsilon}} & =X\left(\sqrt{\pi}\right)E\left(\epsilon\right)\Ket{0_{I}}\\
 & =E\left(\epsilon\right)e^{\sqrt{\pi}\left(e^{\epsilon}\hat{a}^{\dagger}-e^{-\epsilon}\hat{a}\right)}\Ket{0_{I}}\\
 & \neq E\left(\epsilon\right)X\left(\sqrt{\pi}\right)\Ket{0_{I}} = E\left(\epsilon\right)\Ket{1_{I}}.
\end{align}

To avoid ambiguity, we will use the prescription implied by Eq.~(\ref{eq:0NG})
for our normalizable states and explicitly write, e.g., $X\left(\sqrt{\pi}\right)\Ket{0_{G}}$
where necessary. We will investigate in the following sections what
impact this difference will have on practical considerations and
on the logical information encoded in the states.

Note that a more general way of writing down normalizable states is through some noise channel, $\mathcal K$, acting on the ideal states:
\begin{equation}
\label{eq:class_noise}
\rho_{\mathcal{\kappa}} = \mathcal{K}\left(\Ket{\mu_I}\Bra{\mu_I}\right)
\end{equation}
This approach is taken in~\cite{noh2019} with $\mathcal{K}$ corresponding to random Gaussian displacement errors, also known as the Gaussian classical noise channel. 

\subsection{Modular subsystem decomposition} \label{subsec:mod_decomp}

As one deviates from the ideal GKP states, it becomes less obvious what the logical state is, where the information resides, and how to address and access it. Answers to these questions are facilitated by an important tool for analyzing states wherein a qubit is encoded periodically in a infinite-dimensional Hilbert space, the
\emph{modular subsystem decomposition}, investigated by Pantaleoni et al.~\cite{Pantaleoni2019}. Here
we briefly review its formalism and discuss its application to approximate GKP states.

\begin{figure}

\subfloat[\label{fig:bloch-traj-basis}]{\includegraphics[width=\linewidth]{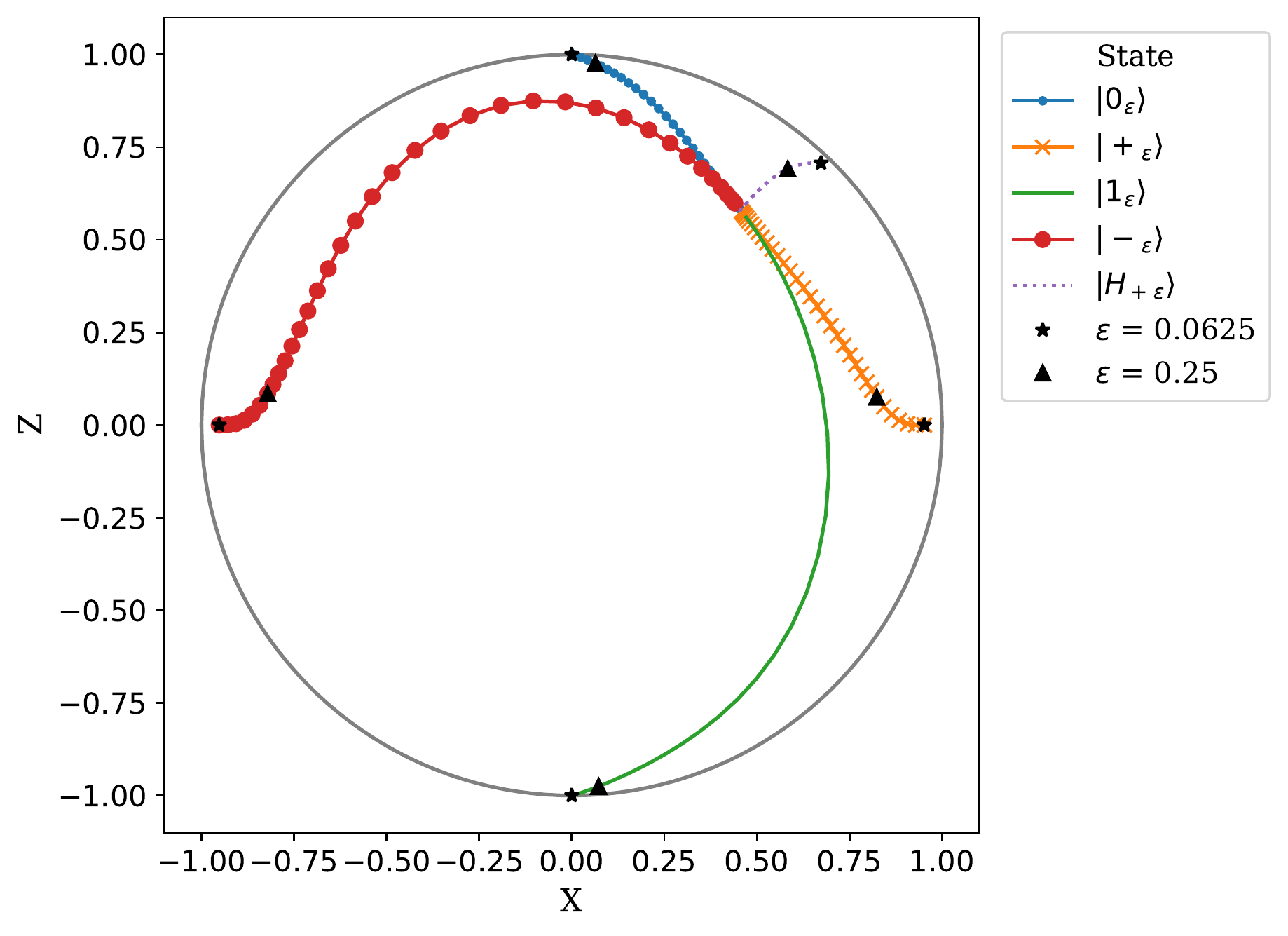}
}
\newline
\subfloat[\label{fig:E-effect-bloch}]{\includegraphics[width=\linewidth]{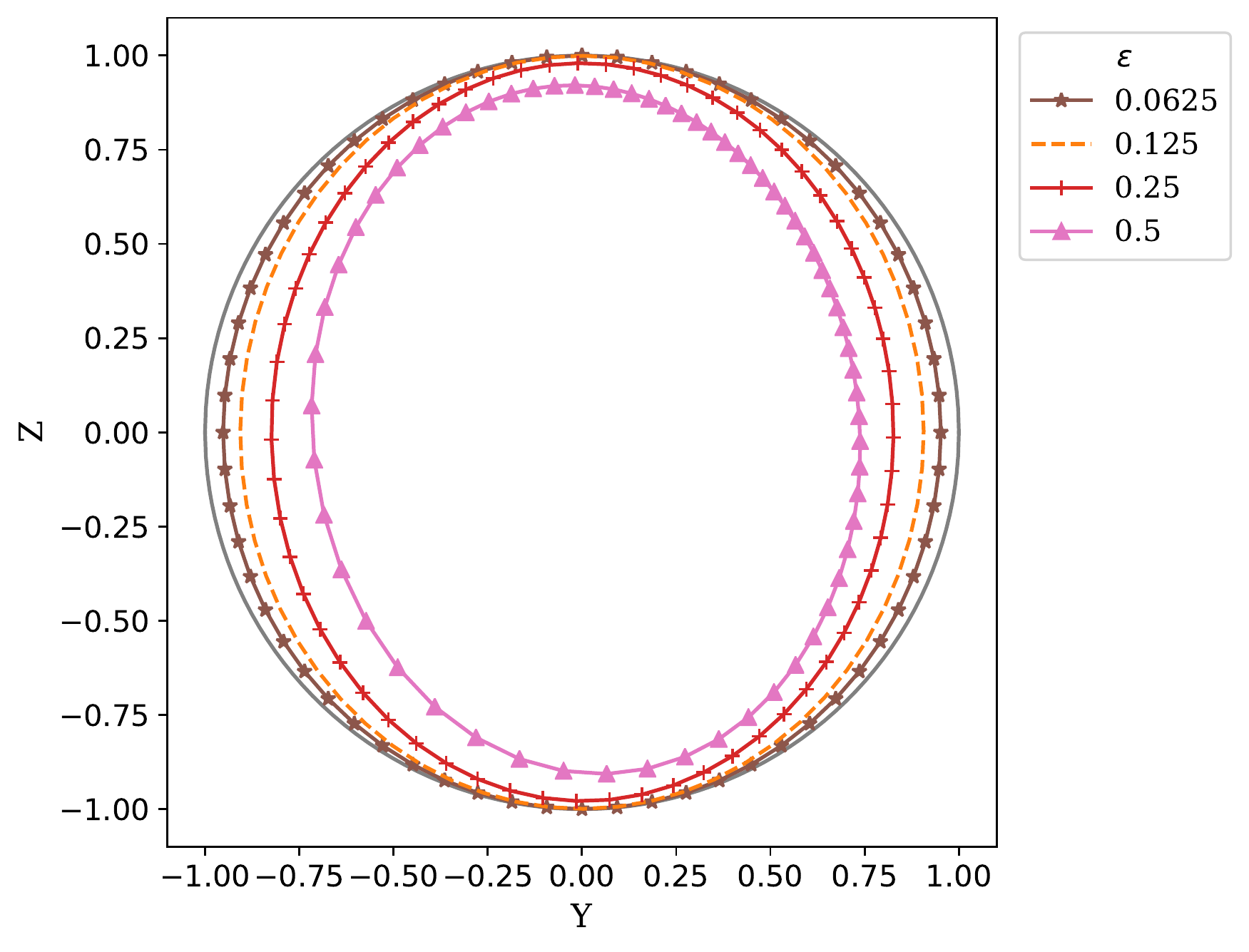}

}\caption{Effects of the normalization envelope $E\left(\epsilon\right)=e^{-\epsilon\hat{n}}$
on the logical content of GKP states. In (a), trajectories of $\Ket{0_{\epsilon}}, \Ket{1_{\epsilon}}, \Ket{\pm_{\epsilon}}$, and $\Ket{{H_+}_{\epsilon}}$
as a function of $\epsilon$ confined to the $X$-$Z$ plane of the Bloch sphere (here ${H}_+$ is the +1 eigenstate of the Hadamard operator; see \eqref{eq:magic}). For large $\epsilon$, all Fock states but $\Ket{0}$ are exponentially suppressed, causing all the trajectories to converge at the logical subsystem point corresponding to the physical vacuum state. We additionally show locations for reasonable values of $\epsilon$: $\epsilon = 0.0625$  $\approx \Delta = 12$~dB, and $\epsilon = 0.25$  $\approx \Delta = 6$ dB are denoted by a star and triangle, respectively (see Eq.~\ref{eq:tanh} and App.~\ref{sec:convention} for conversions). We see that the required $\epsilon$ (dB) values to achieve higher purity states are lower (higher) towards the equator of the Bloch sphere. In (b), the modification of the $X$-$Z$ plane of the ideal Bloch sphere
as a function of $\epsilon$. We see that points towards the equator are more distorted.}

\end{figure}

Given some real number $\alpha$ corresponding to the spacing between
the logical basis states in position, established in Sec.~\ref{subsec:ideal_GKP}, we can decompose any position
eigenket $\Ket{s}_{q}$ in an infinite-dimensional Hilbert space $\mathcal{H}$
as
\begin{equation}
\Ket{s}_{q}=\Ket{\alpha m+u}\equiv\Ket{m,u},
\end{equation}
where $m\in\mathbb{Z}$ and $u\in\left[-\alpha/2,\alpha/2\right)$.
We call $\alpha m$ the \emph{integer part }of $s$ (mod $\alpha$)
and $u$ the \emph{fractional part }of $s$ (mod $\alpha$). We can
subsequently decompose the physical space $\mathcal{H}$ into $\mathcal{H}=\mathcal{L}\otimes\mathcal{\mathcal{G}}$:
a two-dimensional \emph{logical }space $\mathcal{L}$ corresponding
to our qubit and another (virtual) infinite-dimensional \emph{gauge }space $\mathcal{G}$
corresponding to everything else. Position eigenkets break down as
\begin{equation}
\Ket{s}_{q}=\Ket{\mu}_{\mathcal{L}}\otimes\Ket{\tilde{m},\tilde{u}}_{\mathcal{G}},
\end{equation}
where $\mu=\text{parity}\left(m\right)$, $\tilde{m}=\frac{1}{2}\left(m-\mu\right)$,
and $\tilde{u}=u$. Effectively, this decomposition amounts to stitching
together the wavefunction sitting within position bins corresponding
to the logical $\mu$. In this subsystem picture, the ideal GKP states can
be written
\begin{equation}\label{eq:log_decomp}
\Ket{\psi_{I}}=\Ket{\bar{\psi}}_{\mathcal{L}}\otimes\Ket{+_{I}}_{\mathcal{G}},
\end{equation}
so that the logical mode is completely separable from gauge mode and
we can recover the logical information perfectly through a trace over
$\mathcal{G}$. Physically, we can access the logical information with a binned homodyne measurement, as shown in App.~\ref{subsec:readout}. For the normalizable states, however, we see that
\begin{equation}
\Ket{\psi_{\epsilon}}=E\left(\epsilon\right)\left(\Ket{\bar{\psi}}_{\mathcal{L}}\otimes\Ket{+_{I}}_{\mathcal{G}}\right),
\end{equation}
for example. Since $E\left(\epsilon\right)$ acts on the full space $\mathcal{H}$,
it will generally entangle the logical and gauge modes, leaving our
logical qubit in a mixed state
\begin{equation}\label{eq:logical state}
\rho^{\mathcal{L}}\left(\epsilon\right)=\Tr_{\mathcal{G}}\left[E\left(\epsilon\right)\Ket{\psi_{I}}\Bra{\psi_{I}}E\left(\epsilon\right)\right].
\end{equation}
As $\epsilon$ grows, $\rho^{\mathcal{L}}$ will find itself somewhere inside
the Bloch sphere; we plot its ``trajectory'' as a function of $\epsilon$ for
the $\bar{Z}$ and $\bar{X}$ basis states in Fig.~\ref{fig:bloch-traj-basis}. Notice
that, for $\epsilon\gg0$, the states converge at the same point:
this is the vacuum state, the only state picked out by the envelope
in this regime \footnote{Note that in Pantaleoni et al., Fig.~2, the Bloch sphere trajectory
for the \unexpanded{$\Ket{+_{G}}$} state differs form ours. This is because a
different approximation is used there, effectively corresponding to
a different envelope operator.}. We can also see how the Bloch sphere itself changes as a function of $\epsilon$, as depicted in Fig.~\ref{fig:E-effect-bloch}. \\

\noindent {\bf Remark.} We calculate this efficiently by first noting that every unnormalized qubit operator can be decomposed as $\rho^{\mathcal{L}}\left(\epsilon\right) = \frac{1}{2}\sum_{i=0}^3 s_i \sigma_i$, where $\sigma_i$ correspond to the identity and the Pauli matrices, and $\boldsymbol{s}$ is the Stokes vector for the state (see for e.g. \cite{kks_thesis}). If $s_0 = 1$ then $(s_1,s_2,s_3)$ corresponds to a Bloch vector. Thus, we can find the matrix that transforms the Stokes vectors corresponding to the ideal GKP qubits under application of the completely positive (but not trace preserving) map $E\left(\epsilon\right)$. Given the Stokes vectors corresponding to the states before renormalizing, we can now individually renormalize them by simply dividing by the first coefficient of each Stokes vector, yielding Bloch vectors in the remaining three components.

\subsection{Operations on normalizable GKP states and error-tracking}\label{subsec:ops_errors}

The implementation of Clifford gates is a critical step for universal quantum computation. The Clifford group on $n$ qubits, $\mathcal{C}_n$, is defined through its action on the Pauli group, $\mathcal{P}_n$, which consists of $n$-fold tensor products of Pauli gates. Any $U \in \mathcal{C}_n$ maps the Pauli group to itself under conjugation:
\begin{equation}
    A \in \mathcal{P}_n \implies U A U^{\dagger} \in \mathcal{P}_n.
\end{equation}
 In principle, it  is enough to consider a set of generators of the Clifford
group, for example the Hadamard gate, \begin{equation}
\bar{H}=\frac{1}{\sqrt{2}}\begin{bmatrix}1 & 1\\
1 & -1
\end{bmatrix},
\end{equation}
and the phase gate,
\begin{equation}
\bar{P}=\sqrt{\bar{Z}}=\begin{bmatrix}1 & 0\\
0 & i
\end{bmatrix},
\end{equation}
along with the two-qubit CNOT gate (also known as a \text{CX} or controlled-$X$ gate),
\begin{equation}
\overline{\text{CNOT}}=\Ket{\bar{0}}\Bra{\bar{0}}\otimes I+\Ket{\bar{1}}\Bra{\bar{1}}\otimes\bar{X}.
\end{equation}
Any Clifford element can then be obtained, in principle,  from applications of the above gates in the generator set. In practice, and certainly with GKP states, one should also explicitly define other fundamental Clifford gates like $X$ and $Z$ rather than relying on compositions of the minimal set of generators.

\renewcommand{\arraystretch}{1.5}
\begin{table}
\begin{centering}
\begin{tabular}{c|c|c}
\hline 
$\bar{U}$ & $U$ physical (symbol) & $U$ physical (name) \\
\hline 
\hline 
$\bar{X}$ & $D(\sqrt{\pi/2})=X\left(\sqrt{\pi}\right)$ & $q$ displacement \\
\hline 
$\bar{Z}$ & $D(i\sqrt{\pi/2})=Z\left(\sqrt{\pi}\right)$ & $p$ displacement \\
\hline 
$\bar{H}$ & $F=R\left(\pi/2\right)=e^{i\frac{\pi}{2}\hat{n}}$ & Fourier gate; $\frac{\pi}{2}$ rotation \\
\hline 
$\bar{P}$ & $P=e^{i\frac{1}{2}\hat{q}^{2}}$ & Phase gate \\
\hline 
$\overline{\text{CNOT}}$ & $\text{SUM}=e^{-i\hat{q}_{1}\otimes \hat{p}_{2}}$ & SUM gate \\
\hline 
\end{tabular}
\par\end{centering}
\caption{Conventional association between logical qubit operations and physical
Gaussian transformations for ideal GKP encoding.  Note that the physical gates are not unique due to fact that the stabilizers~\eqref{eq:stabsquare} and a $\pi$ phase shift~$e^{i \pi \hat n}$ act trivially on the code space. This means that any displacement by a Gaussian-integer multiple of $\sqrt{\pi/2}$ acts as a Pauli operator; $F^\dag$ also represents $\bar H$; and a SUM gate of any odd-integer weight is also a CNOT. (A SUM gate of weight~$g$ is $e^{-i g \hat{q}_1 \otimes \hat{p}_2}$. The one shown in the table is weight~1.) \label{tab:logic-Gauss}}
\end{table}

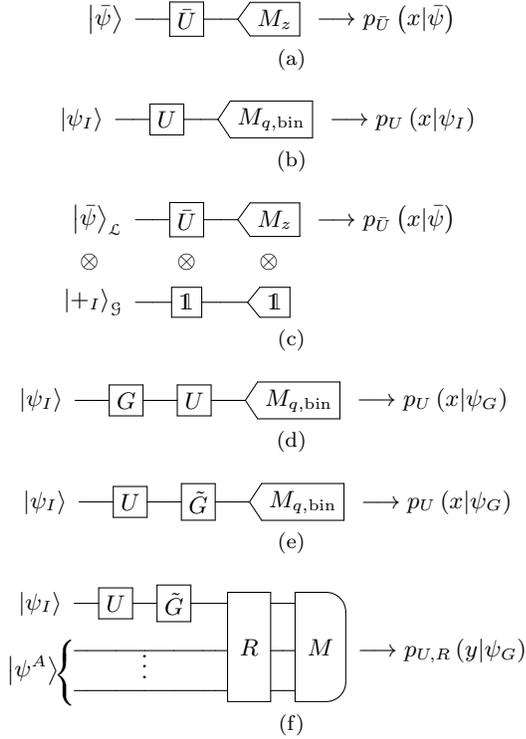
\begin{figure}
\subfloat[\label{subfig:ideal_l}]{
$$
\Qcircuit @C=.7em @R=.7em { 
\lstick{\Ket{\bar{\psi}}} & \qw & \gate{\bar{U}} & \qw & \measuretab{M_z} & \push{\longrightarrow p_{\bar{U}}\left(x \vert \bar{\psi}\right)} 
}
$$
}

\subfloat[\label{subfig:ideal_p}]{
$$
\Qcircuit @C=.7em  @R=.7em { 
\lstick{\Ket{\psi_I}} & \qw & \gate{U} & \qw & \measuretab{M_{q, \text{bin}}} & \push{\longrightarrow p_{U}\left(x \vert \psi_I \right)} 
}
$$
}

\subfloat[\label{subfig:ideal_p_sub}]{$$
\Qcircuit @C=.7em  @R=.7em { 
\lstick{\Ket{\bar{\psi}}_\mathcal{L}} & \qw & \gate{\bar{U}} & \qw & \measuretab{M_z} & \push{\longrightarrow p_{\bar{U}}\left(x \vert \bar{\psi}\right)}   \\
\lstick{\otimes \quad} & \push{} & \push{\otimes} & \push{} & \push{\otimes} \\
\lstick{\Ket{+_I}_\mathcal{G}} & \qw & \gate{\id} & \qw & \measuretab{\id} 
}
$$}

\subfloat[\label{subfig:real}]{$$
\Qcircuit @C=.7em @R=.7em { 
\lstick{\Ket{\psi_I}} & \qw & \gate{G} & \qw & \gate{U} & \qw & \measuretab{M_{q, \text{bin}}} & \push{\longrightarrow p_{U}\left(x \vert \psi_G\right)}
}
$$

}

\subfloat[\label{subfig:real_env}]{$$
\Qcircuit @C=.7em  @R=.7em { 
\lstick{\Ket{\psi_I}} & \qw & \gate{U} & \qw & \gate{\tilde{G}} & \qw & \measuretab{M_{q, \text{bin}}} & \push{\longrightarrow p_{U}\left(x \vert \psi_G\right)}
}
$$}

\subfloat[\label{subfig:recov_anc}]{$$
\Qcircuit @C=.5em  @R=.7em { 
\lstick{\Ket{\psi_I}} & \qw & \gate{U} & \qw & \gate{\tilde{G}} & \qw &  \qw & \multigate{2}{R} & \qw & \multimeasureD{2}{M} \\
\lstick{} & \qw & \qw & \qw & \qw & \qw &  \qw & \ghost{R} & \qw & \ghost{M} & \push{\longrightarrow p_{U, R}\left(y \vert \psi_G\right)} \\
\lstick{} & \qw & \qw & \ustick{\vdots} \qw & \qw & \qw &  \qw & \ghost{R} & \qw & \ghost{M} \inputgroupv{2}{3}{.8em}{.8em}{\Ket{\psi^A}}\\
}
$$}

\caption{
\protect\subref{subfig:ideal_l} The ideal logical circuit: a Clifford unitary, $\bar{U}$, is applied to a qubit, $\Ket{\bar{\psi}}$, followed by Pauli $Z$ measurements resulting in a bit, $x$, with probability $p_{\bar{U}}\left(x \vert \bar{\psi}\right)$. 
\protect\subref{subfig:ideal_p} The ideal encoded physical circuit: a Gaussian unitary, $U$, which implements $\bar{U}$ is applied to an ideal GKP state, followed by (binned) homodyne measurements to produce a bit with probability $p_{U}(x \vert {\psi_I})$; \protect\subref{subfig:ideal_p_sub} the circuit in \protect\subref{subfig:ideal_p} redrawn with a modular subsystem decomposition (note that we are only able to decompose the $U$ in this way for an input ideal GKP state). The probability distributions in \protect\subref{subfig:ideal_l},  \protect\subref{subfig:ideal_p_sub}, and \protect\subref{subfig:ideal_p} are identical:  $p_{\bar{U}}(x \vert \bar{\psi})= p_{U}(x \vert {\psi_I})$;  \protect\subref{subfig:real} The realistic GKP circuit, where the initial states are subject to a finite energy envelope, $G$. The unitary remains the same and a homodyne measurement produces a bit with probability $p_{U}\left(x \vert \psi_G\right)$; \protect\subref{subfig:real_env} the circuit in \protect\subref{subfig:real} but we highlight that the envelope can be conjugated with the Gaussian with appropriate modifications. \protect\subref{subfig:recov_anc} A general GKP circuit: the same as \protect\subref{subfig:real_env}, but now we include a unitary recovery operation on the data states and ancillae $\Ket{\psi^A}$. The measurement operator -- which could be homodyne or PNR -- is generalized to be on all modes, now producing a bit string of some composite variable, $y$. See the remark in the main text for a generalization of this schematic to a multimode circuit.}
\label{fig:circuit_models}
\end{figure}

In addition to these, one also requires an operation beyond the
Clifford group (referred to as non-Clifford element), such as the $\pi/8$ gate or the T gate,
\begin{equation}
\bar{T}=\sqrt{\bar{P}}=\frac{1}{\sqrt{2}}e^{i\frac{\pi}{8}}\begin{bmatrix}e^{-i\frac{\pi}{8}} & 0\\
0 & e^{i\frac{\pi}{8}}
\end{bmatrix},
\end{equation}
for fault-tolerant universal quantum computation.

Alternatively, non-Clifford gates can be effected by preparing special
resource states called \emph{magic states} that can be used in a type of gate teleportation circuit~\cite{Bravyi2005}. We can implement the $\bar{T}$
gate, for example, with a supply of \emph{H-type} magic states, which
are equivalent through Clifford unitaries to the Hadamard
eigenstates. Refer to Fig.~5 of \cite{GKP} for an example gate teleportation circuit. The eigenstate corresponding to the +1 eigenvalue is
\begin{equation}\label{eq:magic}
\Ket{\bar{H}_+}\equiv\cos\frac{\pi}{8}\Ket{\bar{0}}+\sin\frac{\pi}{8}\Ket{\bar{1}}.
\end{equation}
Here we will only consider Gaussian gates, because non-Clifford gates that correspond to non-Gaussian gates can be realized via gate teleportation with magic states and Gaussian gates. This means all the gate resources remain Gaussian, and the non-Gaussian resources are imposed on the state preparation (of logical and magic GKP states), which we will discuss in Sec.~\ref{sec:stateprep}.

Suppose we would like to apply a gate $\bar{U}\in\mathcal{C_1}$
to a logical qubit $\Ket{\bar{\psi}}$. If we encode $\Ket{\bar{\psi}}$
in an ideal GKP state $\Ket{\psi_{I}}$, $\bar{U}$ will correspond
to a Gaussian operation $U$ on the physical space -- that is, a combination of a
symplectic transformation on the quadrature operators and a displacement. We recall that not all Gaussian operations correspond to Clifford gates on the ideal GKP codes, just particular non-unique ones. Writing $\left|\psi_{I}\right\rangle $ in its modular
subsystem decomposition, we have that
\begin{equation}
U\Ket{\psi_{I}}=\left(\bar{U}\Ket{\bar{\psi}}_\mathcal{L}\right)\otimes\Ket{+_{I}}_\mathcal{G}.
\end{equation}
We can see that this is true by applying $U$ to both sides of Eq.~\eqref{eq:log_decomp}: on the right-hand side we must obtain an ideal-GKP-encoded state (since $U$ implements a Clifford unitary $\bar{U}$). Thus the  resulting state can also be written in the form (\ref{eq:log_decomp}), and the action on the logical subsystem must be that of $\bar{U}$. Importantly, however, note that this does \emph{not} mean that the physical operation can be modelled as $\bar{U}_{\mathcal{L}}\otimes I_{\mathcal{G}}$ since the decomposed operator is entangling in general~\cite{Pantaleoni2019}. The reason the gauge mode is unchanged in this particular case is that the entangling pieces of the decomposed gate do nothing when the gauge mode is exactly $\Ket {+_{I}}$. In general, there are many physical operations $U$ corresponding to a given logical gate $\bar{U}$; the standard mapping between $\bar{U}$ and $U$ is given in Table \ref{tab:logic-Gauss}.\\

\noindent {\bf Remark.} The formalism above can be generalized to multimode states, which is necessary for generating entanglement and implementing gate teleportation. In the ideal case, a logical gate $\bar{U} \in \mathcal{C}_n$ acting on $n$ qubits can be implemented as a Gaussian operation $U$ on $n$ oscillator modes. Otherwise, $U$ will cause the logical and gauge subsystems of the different modes to interact \cite{Pantaleoni2019}. Note further that we ought to allow for non-terminal measurements and classical feedforward within the computation. Although the unitaries will then generally depend on the result of these measurements, they will remain Gaussian. \\

For normalizable states, we have
\begin{align}
U\Ket{\psi_{\epsilon}}=UE\left(\epsilon\right)\Ket{\psi_{I}} & =\tilde{E}\left(\epsilon\right)U\Ket{\psi_{I}},\label{eq:U-modifiedE}
\end{align}
where $\tilde{E}\left(\epsilon\right)\equiv UE\left(\epsilon\right)U^{\dagger}$.
Therefore, applying a Gaussian operation to a normalizable state can
be viewed as applying this operation to an ideal state followed by
a modified envelope. If we write the operation $U$ as some function $u$
of the creation and annihilation operators, $U=u\left(\hat{a}^{\dagger}, \hat{a}\right)$,
we can also see, as shown in App.~\ref{subsec:comm_rel_E}, that
\begin{equation}
u\left(\hat{a}^{\dagger},\hat{a}\right)E\left(\epsilon\right)=E\left(\epsilon\right)u\left(e^{\epsilon}\hat{a}^{\dagger},e^{-\epsilon}\hat{a}\right).\label{eq:E-modifiedU}
\end{equation}
Expressions (\ref{eq:U-modifiedE})
and (\ref{eq:E-modifiedU}) show that the interplay between perfect
Gaussian operations and imperfect GKP states causes injury to both.
The damage increases with increasing $\epsilon$ -- corresponding
to fewer and broader peaks in the GKP state -- and with increasing
powers of quadrature operators that feature in $U$. 

In this section
we explore ways to quantify the severity of this damage and approaches
to mitigating it. For this, we consider models for computational circuits
with increasing complexity, as in Fig.~\ref{fig:circuit_models}.
First, we will overview figures of merit for normalizable GKP states
before any kind of recovery operation. This will be important for
ensuring that the state input to the recovery is the best possible,
thereby easing the requirements on the error correction that follows.
Armed with these figures of merit, we will analyze the effects of an
important class of Gaussian operations on the normalizable GKP states.
In Sec.~\ref{subsec:error_correction}, we will discuss the notion of error correction with approximate
GKP states and revisit the metrics.

\subsubsection{Prerecovery figures of merit}\label{subsec:pre-recov-metrics} 
\begin{figure}
\begin{centering}
\includegraphics[width=\linewidth]{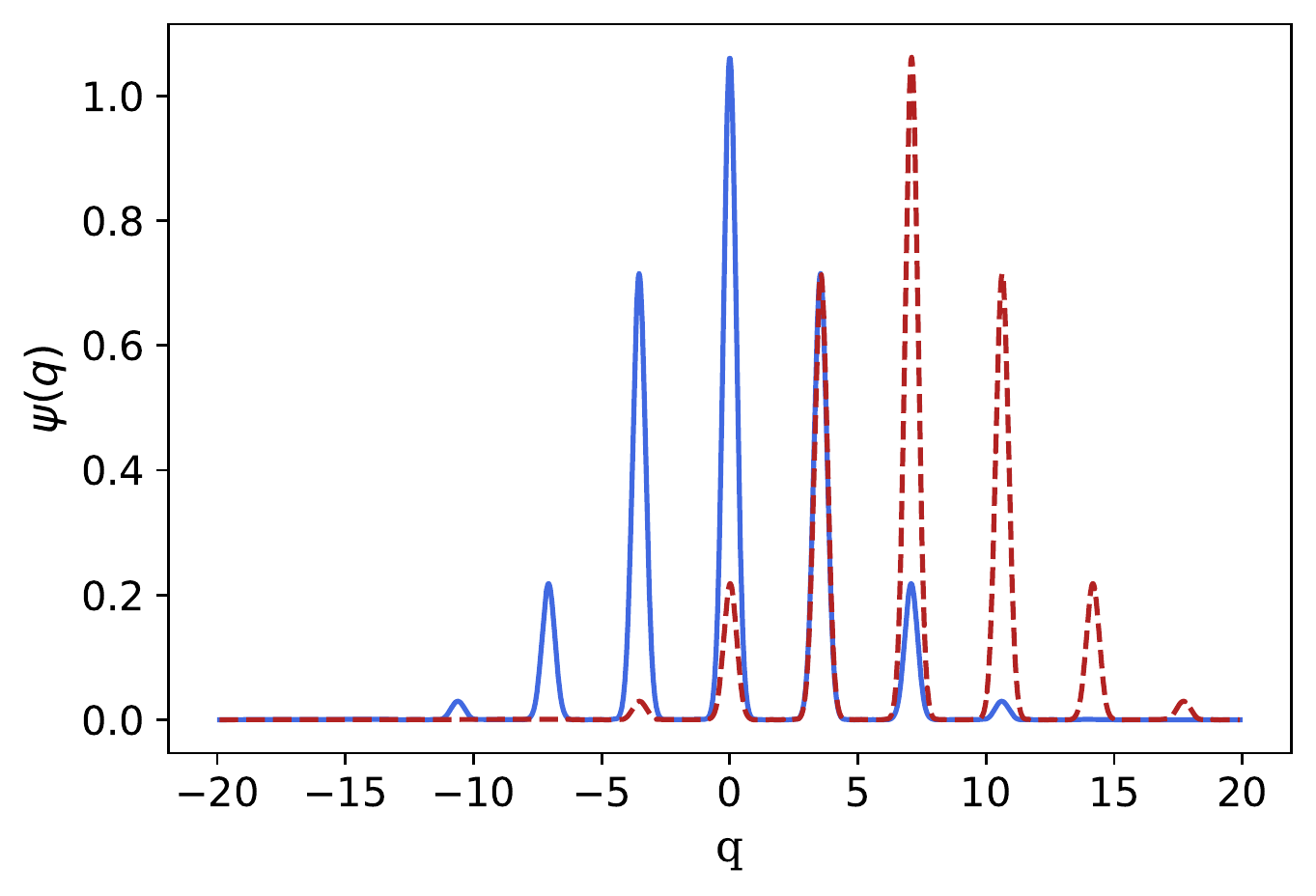}
\par\end{centering}
\caption{The normalizable GKP state $\left|0_\epsilon\right>$ (blue, solid) and the same state after two applications of the $X(2\sqrt{\pi})$ gate (red, dashed) for $\epsilon = 0.063$ $(\Delta \approx 12 \text{ dB})$. Because finite-energy GKP states have finite support in position space and hence lack a translational symmetry, the physical fidelity between the displaced state and the original state decreases with each application of $X(2\sqrt{\pi})$ even though this gate preserves the logical content.}
\label{fig:disp0}
\end{figure}

Here we define several metrics to probe the quality and usefulness
of the normalizable GKP states. For each metric, we will use as a
sanity check a toy circuit consisting of an even number, $2k$, of
logical $\bar{X}$ operations:
$$
\Qcircuit @C=0.8em @R=.7em { 
\lstick{\Ket{\bar{0}}} & \qw & \gate{\bar{X}} & \qw & \cdots && \gate{\bar{X}} & \qw  & \push{\equiv} && \gate{\bar{I}}
}
$$
This circuit is equivalent to a logical identity. If we would like
to implement it on an oscillator, we can encode $\Ket{\bar{0}}\to\Ket{0_{I}}$
and use the standard map $\bar{X}\to X(\sqrt{\pi})$ from Table~\ref{tab:logic-Gauss}.
The ideal oscillator circuit is thus
$$
\Qcircuit @C=0.8em @R=.7em { 
\lstick{\Ket{0_I}} & \qw & \gate{X(\sqrt{{\pi}})} & \qw & \cdots && \gate{X(\sqrt{{\pi}})} & \qw  & \push{\equiv} && \gate{I}
}
$$
Thanks to the complete translational symmetry of the ideal GKP state,
this circuit is also equivalent to an identity on the oscillator space
(and hence the logical space), despite it effecting a net translation
of $n\alpha$ in position. In a more practical setting we have
$$
\Qcircuit @C=0.8em @R=.7em { 
\lstick{\Ket{0_G}} & \qw & \gate{X(\sqrt{{\pi}})} & \qw & \cdots && \gate{X(\sqrt{{\pi}})} & \qw & \push{\equiv} && \gate{X(n\sqrt{{\pi}})}
}
$$
Because the normalized GKP states break the translational symmetry
of the ideal states, the latest circuit is no longer an identity on
the state space (see Fig.~\ref{fig:disp0}). The extent of the problems that this causes will
be made clearer using the following figures of merit. Note that we always
treat $U$ as a unitary operation on the physical space that effects
$\bar{U}$ on the logical mode of an ideal GKP state.

\paragraph{Physical Fidelity.}

A straightforward figure of merit to consider is to compare physical
states, for which we use the \emph{physical fidelity }
\begin{equation} 
F^{\mathcal{P}}\left(\Ket{\psi},\Ket{\phi}\right)\equiv\left|\Braket{\phi|\psi}\right|^2.
\end{equation}
We will want to compare the state before a transformation to the state after, and so we can use the shorthand notation
\begin{equation}
F_{U}^{\mathcal{P}}\left(\Ket{\phi}\right)\equiv F^{\mathcal{P}}\left(U\Ket{\phi},\Ket{\phi}\right).
\end{equation}

If $U_{\id}$ is a unitary for which $\bar{U}=\bar{\id}$, then
$F_{U}^{\mathcal{P}}\left(\Ket{\psi_{I}}\right)=1$ for an ideal GKP
state $\Ket{\psi_{I}}$. We might therefore demand for a normalizable state that
\begin{equation}
F_{U_{\id}}^{\mathcal{P}}\left(\Ket{\psi_{G}}\right)\approx1.
\end{equation}
 However, this requirement is too stringent: to see this, we can refer
back to our toy circuit, where $U_{\id}=X\left(2k\sqrt\pi\right)$.
In this case
\begin{align}
F_{X\left(2k\sqrt{\pi}\right)}^{\mathcal{P}}\left(\Ket{\psi_{\epsilon}}\right) & =\Bra{\psi_{\epsilon}}X\left(2k\sqrt{\pi}\right)\Ket{\psi_{\epsilon}}\\
 & =\Bra{\psi_{I}}E\left(\epsilon\right)\tilde{E}\left(\epsilon\right)\Ket{\psi_{I}},
\end{align}
where $\tilde{E}\left(\epsilon\right)=e^{-\epsilon\left[\left(\hat{q}-2k\sqrt{\pi}\right)^{2}+\hat{p}^{2}\right]}$ (see Table \ref{tab:e_conj} for more conjugation relations).
For $k$ high enough, our normalizable state can be displaced so much
that the physical fidelity vanishes; however, the functional form
of the wavefunction has not changed save for a rigid translation in position.
This indicates that, while potentially useful for some applications,
the physical fidelity does not adequately reveal the presence of information
encoded in our normalizable state.

\paragraph{Logical Fidelity.}

Since we are interested in the logical content of the state rather
than the content of the gauge mode, we might instead consider \emph{logical
fidelity, }i.e., the fidelity between the reduced logical states:
\begin{equation}
F^{\mathcal{L}}\left(\Ket{\psi},\Ket{\phi}\right)\equiv F\left[\Tr_{\mathcal{G}}\left(\Ket{\phi}\Bra{\phi}\right),\Tr_{\mathcal{G}}\left(\Ket{\psi}\Bra{\psi}\right)\right],
\end{equation}
where $F$ on the right-hand-side is the fidelity for mixed states, given by
\begin{equation}
F\left(\rho, \sigma \right) \equiv \Tr\left|\sqrt{\rho}\sqrt{\sigma}\right|^2.
\end{equation}
Again, we will make use of the notation
\begin{equation}
F_{U}^{\mathcal{L}}\left(\Ket{\phi}\right)\equiv F^{\mathcal{L}}\left(U\Ket{\phi},\Ket{\phi}\right),
\end{equation}
and if $U_{\id}$ is such that $\bar{U}_{\mathds{}}=I$,
then $F_{U}^{\mathcal{\mathcal{L}}}\left(\Ket{\psi_{I}}\right)=1$,
and we might require
\begin{equation}
F_{U_{\id}}^{\mathcal{\mathcal{L}}}\left(\Ket{\psi_{G}}\right)\approx1.
\end{equation}

Returning to our toy circuit, if $U_{\id}=X\left(2k\sqrt{\pi}\right)$,
this amounts to a displacement of the gauge mode only:
\begin{align}
    X(2k\sqrt\pi)
&
=
    \id_{\mathcal{L}}
\otimes
    X_{\mathcal{G}} (2k \sqrt \pi)
    ,
\end{align}
so  the logical fidelity is expected to be close to unity. This shows the utility of the modular subsystem decomposition in an analysis of imperfect GKP states. 

\paragraph{Distribution Distance.}
Since we are ultimately interested in the result of the computation being accurate, we can also define a post-readout metric. Returning to Fig.~\ref{fig:circuit_models}, we can compare the output of the ideal circuit \ref{subfig:ideal_p} with input state $\Ket{\psi_I}$, which will be a bit string following some probability distribution $p_{U}(x|\psi_I)$, with the output $p_{U}(x|\psi_{G})$ of the circuit \ref{subfig:real}, initialized with a normalizable GKP state $\Ket{\psi_{G}}$. For this we define the \emph{distribution
distance} through
\begin{equation}
\mathscr{D}^{p}_{U}\left(\Ket{\phi_G}\right)=d[p_{U}(x|\phi_{G}),p_{U}(x|\phi_I)],
\end{equation}
where $d$ is a statistical distance, i.e., a generalized metric on the space of probability distributions. For a listing and comparison of probability metrics, see, for example, \cite{Gibbs2002}.

In general, we want that $\mathscr{D}^{p}_{U}(\Ket{\psi_G}) \approx 0$ for any $U$. By focusing on the probability distribution, which we find after the readout, we do not require the state before the measurement in circuit \ref{subfig:recov_anc} to be the same as the state before
the measurement in circuit \ref{subfig:ideal_p}. For example, in circuits \ref{subfig:real}, any operation
that modifies the ideal GKP states but keeps the probability distribution within the bin structure unchanged,
e.g. a displacement less than $\sqrt{\pi}/2$, will still yield the
same measurement statistics at readout.

With our toy circuit, if $U=X\left(2k\sqrt{\pi}\right)$,
we see that a binned homodyne measurement (see \ref{subsec:readout}) will extract the logical
information from $\left|\psi_{G}\right\rangle $, implying that $\mathscr{D}^{p}_{U}$ will be invariant with $k$.

\paragraph{Mean and variant energy cost.}
Given a state $\Ket{\phi}$, we can characterize its energy through
the quantities
\begin{align}
\bar{n}\left(\Ket{\phi}\right)&\equiv\Braket{\phi|\hat{n}|\phi},
\\
\sigma^{2}_{\hat{n}}\left(\Ket{\phi}\right)&\equiv\Braket{\phi|\hat{n}^{2}|\phi}-\bar{n}^{2}\left(\Ket{\phi}\right),
\end{align}
which correspond to the occupation number mean and variance \footnote{In \cite{albert2018performance}, the authors show that the GKP code (i.e. the codespace spanned by \unexpanded{$\Ket{0_\Delta}$ and $\Ket{1_\Delta}$}) is geometrically distributed in Fock space and has a mean occupation number of \unexpanded{$\bar{n}\left(\Ket{0_\Delta} + \Ket{1_\Delta}\right)  \approx \frac{1}{\Delta^2} -1$}}. Then we define the
the mean $\mathscr{D}_{U}^{\hat{n}}$ and variant $\mathscr{D}_{U}^{\sigma^2}$ energy cost of a gate $U$ on a state $\Ket{\phi}$ to be
\begin{align}
    \mathscr{D}_{U}^{\bar{n}}\left(\Ket{\phi}\right) & \equiv \bar{n}\left(U\Ket{\phi}\right)-\bar{n}\left(\Ket{\phi}\right),\\
    \mathscr{D}_{U}^{\sigma^2}\left(\Ket{\phi}\right) & \equiv\sigma^2_{\hat{n}}\left(U\Ket{\phi}\right)-\sigma^2_{\hat{n}}\left(\Ket{\phi}\right),
\end{align}
and we write $\mathscr{D}_{U}^{\sigma}$ when we use the standard deviation. We would like that $\left|\mathscr{D}_{U}^{\bar{n}}\right|$ and $|\mathscr{D}_{U}^{\sigma^2}|$ are as small as possible. Note that the energy costs are physical figures of merit that translate to resource requirements. It is therefore no surprise that $\mathscr{D}_U^{\bar{n}}$ from our toy circuit with $U=X(2k\sqrt{\pi})$ increases monotonically with $k$ as our state finds itself further and further from the origin, while the physical fidelity decreases.  

\paragraph{Extremized and averaged measures.}
\begin{figure}
\centering
\includegraphics[width=\linewidth]{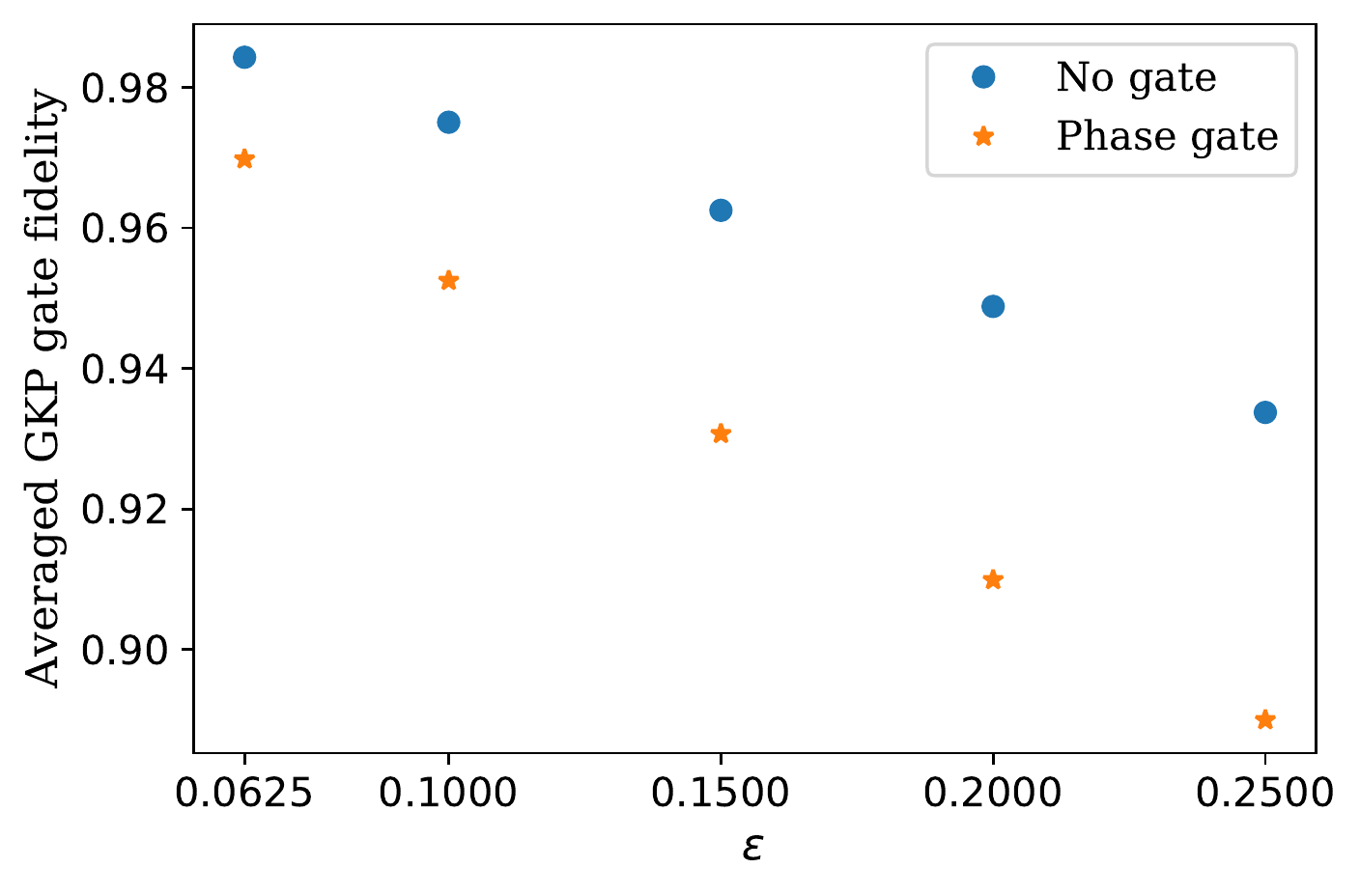}
\caption{Averaged GKP gate fidelity  $\bar{F}^{\mathcal{L}}_{U, G}\left(\bar{U}\right)$ from \eqref{eq:av_log_fid} evaluated for the logical and physical identity \{$\bar{U} = \bar{\id}$, $U=\id$\} (blue), and logical and physical phase gate \{$\bar{U} = \bar{P}$, $U=P$\} (orange) assuming a normalizable GKP state with $G=E(\epsilon)$ as a function of $\epsilon$. This plot shows that even without the application of any gate,
the error envelope $E$ creates a deviation, on average, from unity fidelity with the ideal data qubit state. The deviation is even greater in the case of a phase gate, the reasons for which are given in Sec.~\ref{subsubsec:shearing}. In both cases the fidelity becomes worse as $\epsilon$ becomes bigger, as expected.}
\label{fig:av_gate_fid}
\end{figure}

The figures of merit just described are single-shot measures; for a more complete picture, one can minimize, maximize, or average these measures over a set of a states. Let us attempt to do so for the logical fidelity:
\begin{align}
    \bar{F}^{\mathcal{L}}_{U, G}\left(\bar{U}\right)
    &\equiv \int d\bar{\phi} F^{\mathcal{L}}(U\Ket{\phi_I}, U\Ket{\phi_G}) \label{eq:av_log_fid} \\
    & = \int d\bar{\phi} \Braket{\bar{\phi}|\bar{U}^\dagger \mathcal{E}_{U,G}\left(\bar{\phi}\right)\bar{U}|\bar{\phi}},
\end{align}
where
\begin{align}
    \Ket{\phi_I} &= \Ket{\bar\phi}_\mathcal{L} \otimes \Ket{+_I}_\mathcal{G} \\
    \Ket{\phi_G} &= G\left(\Ket{\bar\phi}_\mathcal{L}  \otimes \Ket{+_I}_\mathcal{G}\right),
\end{align}
we define the channel
\begin{equation}
    \mathcal{E}_{U,G}\left(\bar{\phi}\right) \equiv \Tr_{\mathcal{G}}{\left[U \frac{G\left(\Ket{\bar{\phi}}_{\mathcal{L}}\Bra{\bar{\phi}} \otimes \Ket{+}_{\mathcal{G}}\Bra{+}\right)G^{\dagger}}{N\left(G, \bar{\phi}\right)}U^{\dagger}\right]}
\end{equation}
with normalization factor $N(G,\bar{\phi})$, and the integral is taken over the normalized uniform Haar measure on the two-dimensional state space. We call the quantity (\ref{eq:av_log_fid}) the \emph{averaged GKP gate fidelity} (of a physical gate $U$); it measures the average logical fidelity of the imperfect implementation $U\Ket{\phi_G}$ to the perfect implementation $U\Ket{\phi_I}$ 
\footnote{In \cite{Nielsen2002} there is a simplified, easy-to-compute expression for averaged gate fidelity. However, it assumes that the error channel is trace-preserving, so we can only use it in our setting with appropriate modifications.}. Setting $U=\id$ gives us an application-independent metric of the quality of our GKP states. In Fig.~\ref{fig:av_gate_fid} we plot $\bar{F}^{\mathcal{L}}_{U, G}\left(\bar{U}\right)$ for the identity gate and phase gate with the choice of envelope $G=E(\epsilon)$. Varying $\epsilon$ produces the expected behaviour. 

Similarly, for a fixed $\bar{U}$, $U$, and $G$ we can compute an averaged or maximized distribution distance or energy cost over the normalizable states. 
\subsubsection{Displacements in position} \label{subsubsec:q_disps}

\begin{figure}
\begin{centering}
\includegraphics[width=\linewidth]{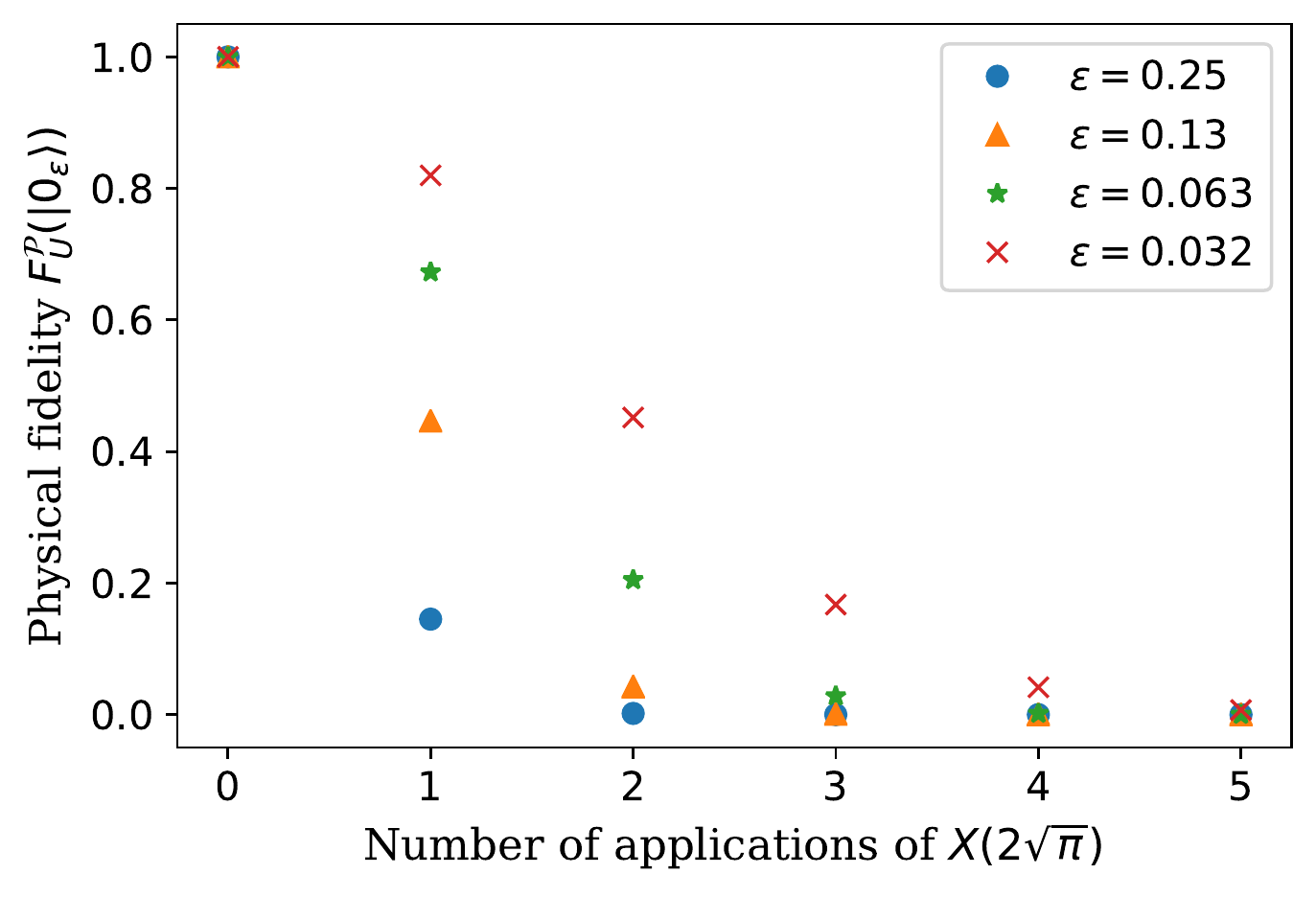}
\par\end{centering}
\caption{Physical fidelity of the normalizable GKP state $\Ket{0_\epsilon}$ after sequential applications
of the $X\left(2\sqrt{\pi}\right)$ gate, which in the ideal case implements a logical identity. The lower the $\epsilon$, the better the approximation to the ideal GKP state, the better the translational symmetry of the state, and the slower the physical fidelity falls as a function of the number of gate applications. \label{fig:x-unit-test}}
\end{figure}

\begin{figure}
\begin{centering}
\includegraphics[width  = \linewidth]{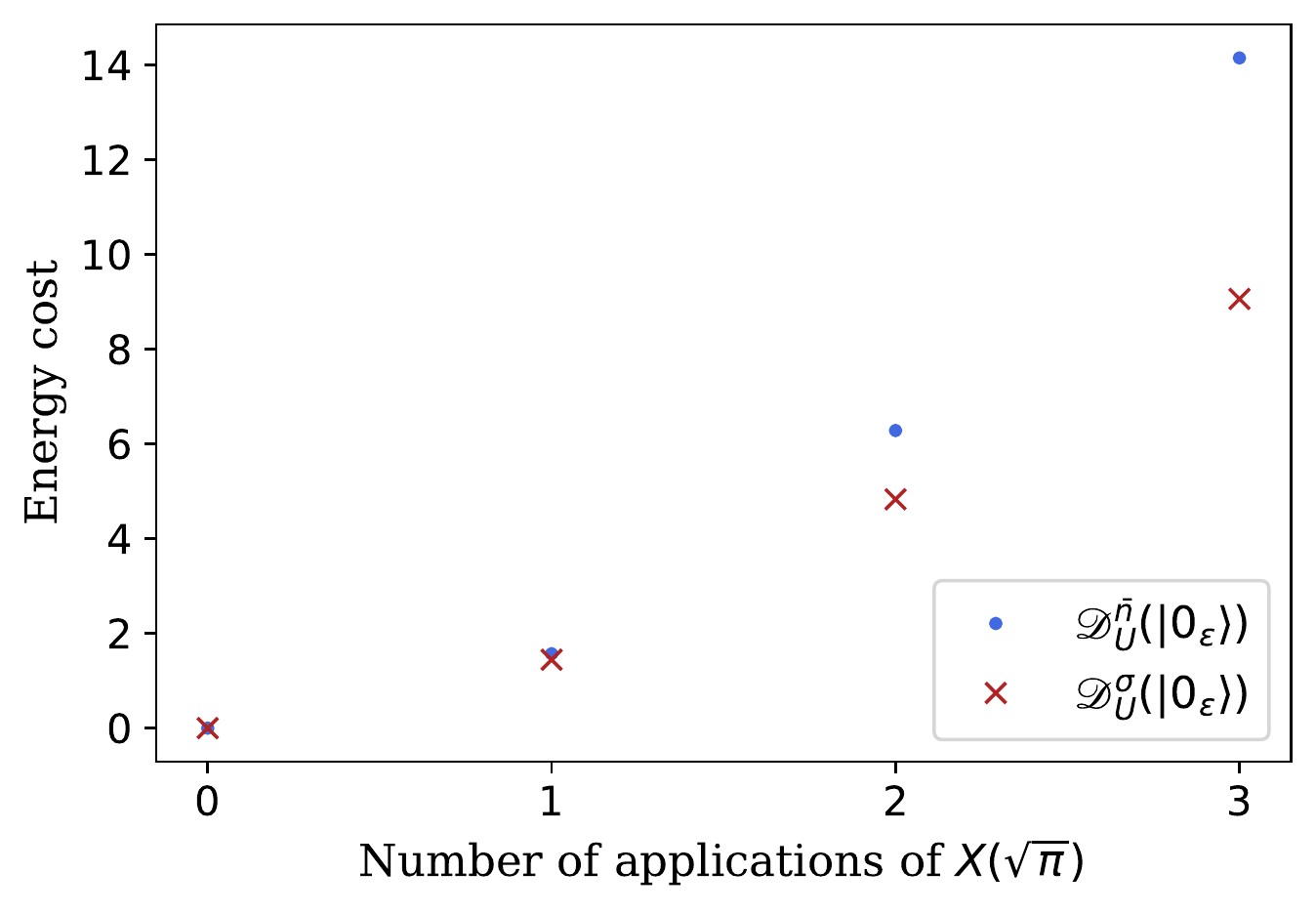}
\par\end{centering}
\caption{The energy cost mean $\mathscr{D}_{U}^{\bar{n}}$
and standard deviation $\mathscr{D}_{U}^{\sigma}$
of multiple applications of the $X(\sqrt{\pi})$
gate to the state $\Ket{0_{\epsilon}}$ for $\epsilon = 0.063$ (corresponding to a $\Delta \approx$ 12 dB). Each subsequent displacement comes at an increased energy cost, and the spread increases more slowly than the mean. Note that an analytic expression for $\mathscr{D}_{U}^{\bar{n}}$ for an arbitrary state $\Ket{\psi}$ yields $\frac{\pi}{2}k^2 - \Braket{\psi|\hat{q}|\psi}k$, where $k$ is the number of gate applications. Therefore applying $X$ gates to zero-mean states will result in the same behaviour.}
\label{fig:en_test_X}
\end{figure}

A displacement in position by $\sqrt{\pi}$ -- half the period of $\left|0_I\right>$ -- is the standard way to implement a logical $\bar{X}$ operation on GKP-encoded qubits. However, we saw above that a physical $X$ gate acting on a normalizable state quickly lowered the
physical fidelity of the initial state while preserving the logical
fidelity and the distribution distance. See Fig.~\ref{fig:x-unit-test}
for the decay in physical fidelity with the number of gate applications for various envelope strengths. But there
are reasons wanting to preserve the physical fidelity. For one, large
displacements to a state from the origin require large amounts of energy. This can be seen through the increase in mean and standard deviation energy cost, as shown in Fig.~\ref{fig:en_test_X}. Notably, higher energy can make the states more susceptible to loss.
We propose several approaches for dealing with this.

First, we might wish to monitor the changes to the mean of our wavefunction
caused by the gates in the circuit. Every Gaussian operation effects
the map $\bar{\boldsymbol{r}}\to\boldsymbol{S}\bar{\boldsymbol{r}}+\boldsymbol{d}$
on the vector of means, $\bar{\boldsymbol{r}}$, where $\boldsymbol{S}$
is a symplectic matrix and $\boldsymbol{d}$ is some displacement.
Finding the updated mean following an application of $k$ Gaussian unitaries on $n$ modes is therefore equivalent to multiplying $k$ matrices of size $2n \times 2n$. After the circuit has been specified, we can thus classically compute the changes in the mean of our state.  With this information in hand, we can then reoptimize the circuit to minimize the maximal displacement. For purely Gaussian circuits we suspect this will be an inexpensive compilation that can be done prior to the computation. The details are left to future research.

Second, we can modify the standard logical-to-physical mapping $\bar{U}\to U$; a naive approach is through
\begin{equation}
\bar{X}\to FPPF,
\end{equation}
where $F$ is the Fourier gate featured in Table~\ref{tab:logic-Gauss}. While this new prescription has no explicit displacements, it requires
a larger number of physical operations, including a shearing of position
and momentum that will damage the imperfect GKP state in a worse way,
as we will see shortly. A better mapping is
\begin{equation}
\bar{X} \to
\begin{cases}
X\left(\sqrt{\pi}\right)\\
X\left(-\sqrt{\pi}\right)
\end{cases}
\end{equation}
This mapping could be \emph{probabilistic}, for example, alternating between the physical gates with probability $\frac{1}{2}$. In this case one randomizes forward and backward displacement
by $\sqrt{\pi}$. This is an example of a one-dimensional simple random
walk: at the end of the circuit, our state is expected to remain undisplaced
with a standard deviation of $\sqrt{n\sqrt{\pi}}$. Roughly, this
means that if there is a threshold displacement $k$ after which the
computation becomes practically untenable, on average a circuit of
depth at most $k^{2}/\sqrt{\pi}$ can accommodate the computation. A better mapping is \emph{deterministic}, that is, ensuring that a forward displacement is always followed by a backward displacement and vice versa.

As an alternative,
we can recalibrate the state after the mean reaches a certain threshold,
$\bar{r}=k$, that is, apply a displacement $X\left(-k\right)$ at
pre-specified points in the circuit. We will discuss combining this
step with error correction in Sec.~\ref{subsec:error_correction}.

We see that the new prescriptions for the physical implementation of the $\bar{X}$ gate maintain both the logical and physical fidelities while minimizing the energy cost. In fact, there is evidence here that demanding that energy costs be minimized generally results in a prescription for the physical implementation of a gate that also improves the other figures of merit.

\subsubsection{Displacements in momentum}\label{subsubsec:p_disps}

Normally, the logical $\bar{Z}$ is physically realized on GKP states by a $\sqrt{\pi}$ displacement in momentum. The impact of momentum displacements $Z\left(\sqrt{\pi}\right)$ on the normalizable GKP states will be equivalent to that of the $X$ gate. This can be seen in two different ways: First, the $Z$ gate is the Fourier transform of the $X$ gate, $Z=FXF^{\dagger}$, and $F$ commutes with the envelope operator $E\left(\epsilon\right)$ as they are both exponentials of $\hat{n}$. Second, the $X$ and $Z$ gate are both linear in the quadrature operators, and so the damage inflicted by the envelope will be on the order of $e^{\epsilon}$ in both cases (both points are verified in App.~\ref{subsec:comm_rel_E}). Thus the physical fidelity seen in Fig.~\ref{fig:x-unit-test} will be the same in both cases.

As with position, we can monitor the changes to the average momentum and recompile the circuit. We ought also to apply the mapping
\begin{equation}
\bar{Z} \to
\begin{cases}
 Z\left(\sqrt{\pi}\right)\\  Z\left(-\sqrt{\pi}\right)
\end{cases}
\end{equation}
deterministically. We do not consider $\bar{Z}\to PP$, as this prescription requires two applications of the quadratic phase gate and is thus a worse solution.

\subsubsection{Shearing operations}
\label{subsubsec:shearing}
\begin{figure}
\centering
\includegraphics[width=\linewidth]{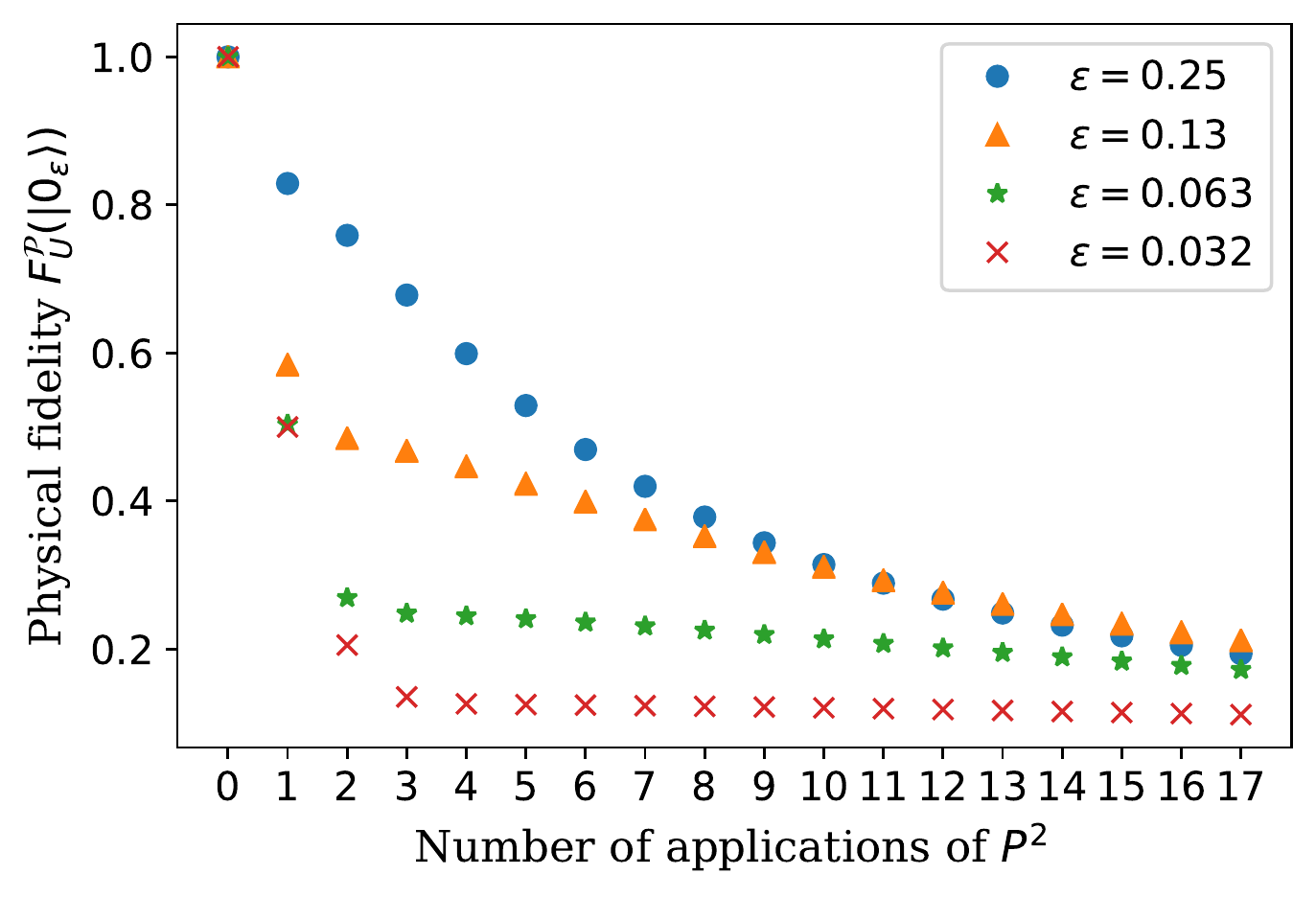}
\caption{
    Physical fidelity of normalizable GKP state $\Ket{0_{\epsilon}}$ after sequential applications
    of $P^2$ gates for various values of $\epsilon$. In the ideal case this gate implements the logical $\bar{Z}$, i.e., an identity on the $\Ket{0_{I}}$ state. The physical fidelity falls more slowly when $\epsilon$ is higher, that is, when $\Ket{0_{\epsilon}}$ is further from the ideal state. In this regime the state has a lower squeezing and hence a wider peak at $q=0$, where the phase gate introduces a close-to-trivial phase $e^{i\frac{1}{2}}q^2$.
}
\label{fig:p-unit-test}
\end{figure}

\begin{figure}
\includegraphics[width = \linewidth]{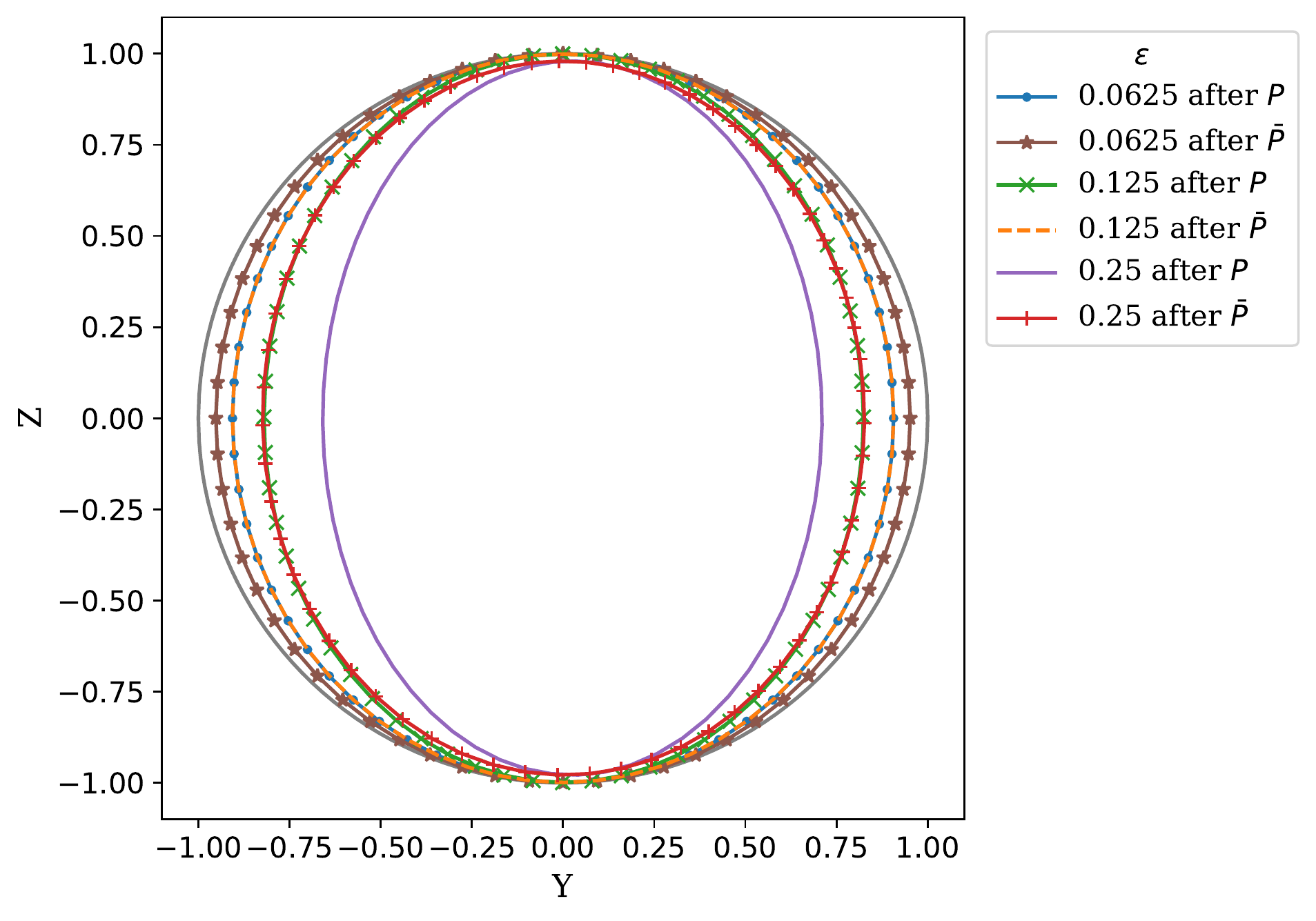}
\caption{ 
    Warping of the logical Bloch sphere after application of the physical phase gate $P$ to the normalizable GKP states $E\left(\epsilon\right)\Ket{0_{I}}$, as compared to the action of a true logical phase gate $\bar{P}$. We show the result for various values of $\epsilon$ (for conversion to dB see App.~\ref{sec:convention}).  We apply the gate to states in the $X-Z$ plane, so the $Y-Z$ plane is depicted since the ideal $\bar{P}$ gate rotates the sphere about the $Z$ axis. Whereas $\bar{P}$ simply applies the rotation to an already distorted Bloch sphere, $P$ additionally warps the Bloch sphere, notably near the equator where states lose more purity.
}
\label{fig:bloch_sphere_ph}
\end{figure}

\begin{figure}
\centering
\includegraphics[width=\columnwidth]{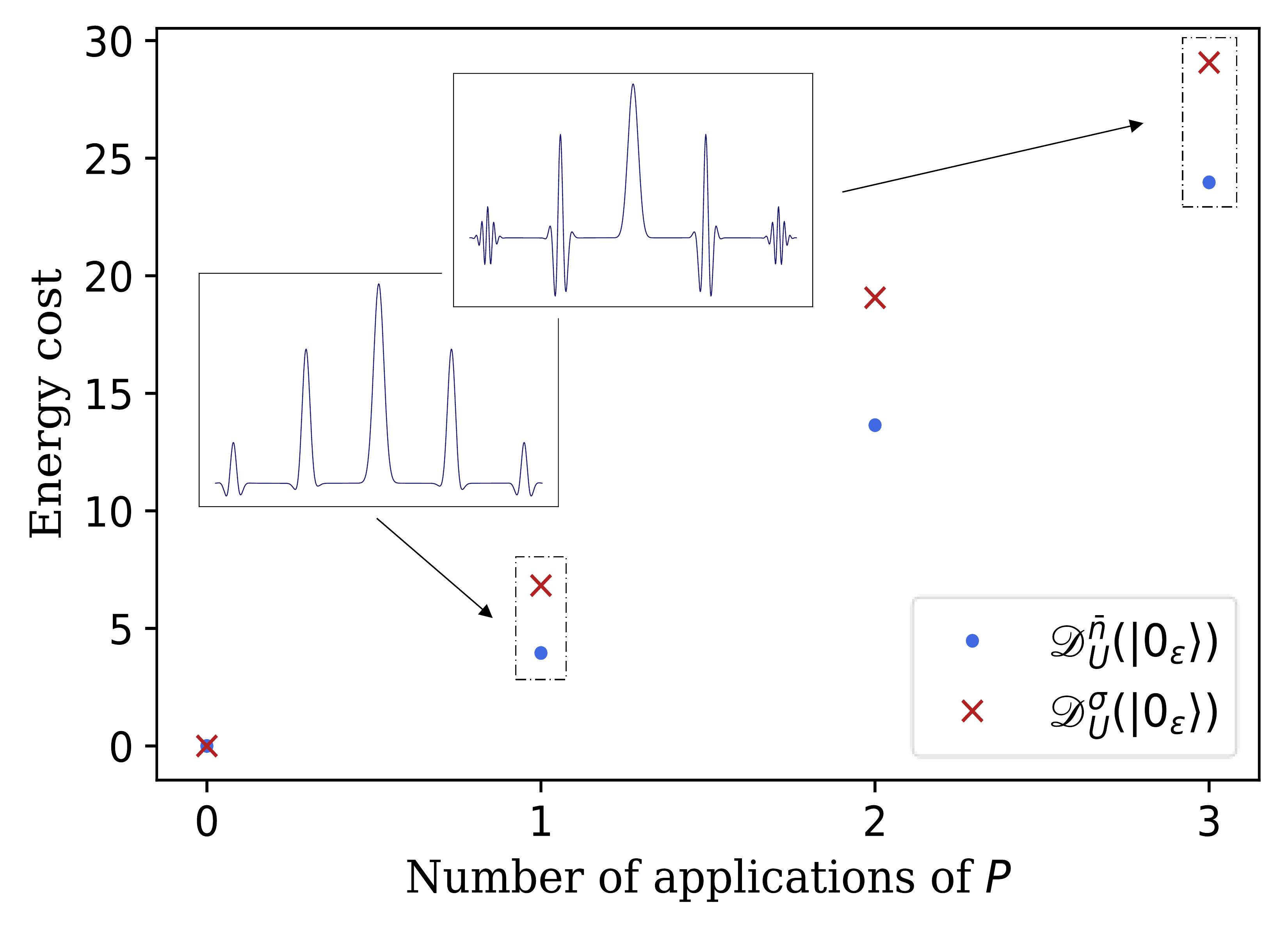}
\caption{
    The energy cost mean and standard deviation of successive applications of the $P$ gate to the state $\Ket{0_{\epsilon}}$ for $\epsilon = 0.063$ ($\Delta \approx 12$~dB). In contrast with the $X$ gate in Fig.~\ref{fig:en_test_X}, the spread grows faster than the mean. The analytic expression for $\mathscr{D}_{U}^{\bar{n}}$ after $k$ applications of the phase gate on the state $\Ket{0_\epsilon}$ is $\frac{1}{2}\left(k^2\Braket{0_\epsilon|\hat{q}^2|0_\epsilon}+k\Braket{0_\epsilon|\{\hat{q},\hat{p}\}|0_\epsilon}\right)$. {\bf Insets\,:} Highlighted are the wavefunctions after one (left) and three (right) applications of the gate to visualize the damage the gate inflicts on the state in the form of shearing, that is, increasing oscillatory behaviour near the peaks. }
\label{fig:en_test_p}
\end{figure}

The logical phase gate $\bar{P}$ is effected on GKP states by the physical phase gate $P$, which shears the wavefunction by introducing a position-dependent phase. But the state $\Ket{0_{I}}$ only has
support on $q=2k\sqrt{\pi}$ for $k\in\mathbb{Z}$, where the phase
gate introduces a trivial factor. Since $P^{2}=Z\left(\sqrt{\pi}\right)$,
this means that an application of $P^{n}$ to $\left|0_{I}\right\rangle $
is an identity on both the logical and physical space for any $n$. On the other
hand, the gate $P^{n}$ applied to $\left|0_{G}\right\rangle $ will
introduce a phase for every position value, and in particular a complex
phase away from $q=k\sqrt{\pi}$. The damage inflicted on the state
$E\left(\epsilon\right)\left|0_{G}\right\rangle $ will be on the
order of $e^{2\epsilon}$ (App.~\ref{subsec:comm_rel_E}).

From Fig.~\ref{fig:p-unit-test}, it looks like better states -- those with higher squeezing -- actually do worse under the phase gate when considering physical fidelity. This is because states with lower squeezing have a wider central peak, which is located in a region ($q \approx 0$) where the phase gate has little effect. The warping of the Bloch sphere as a result of the phase gate is shown
in Fig.~\ref{fig:bloch_sphere_ph}, where we see that states with higher squeezing better preserve logical information. The mean and standard deviation energy cost are plotted in Fig.~\ref{fig:en_test_p}.
As with displacements, we can deal with this problem by noting the flexibility in the logical-to-physical mapping:
\begin{align}
\bar{P}^{k}&=\begin{cases}
\bar{I} & k\equiv0\left(\text{mod 4}\right)\\
\bar{P} & k\equiv1\left(\text{mod 4}\right)\\
\bar{Z} & k\equiv2\left(\text{mod 4}\right)\\
\bar{P}^{-1} & k\equiv3\left(\text{mod 4}\right)
\end{cases} \\ &\to\begin{cases}
I & k\equiv0\left(\text{mod 4}\right)\\
P & k\equiv1\left(\text{mod 4}\right)\\
Z\left(\sqrt{\pi}\right) & k\equiv2\left(\text{mod 4}\right)\\
P^{-1} & k\equiv3\left(\text{mod 4}\right).
\end{cases}
\end{align}
Thus, in any circuit recompilation, we ought to treat even applications of $\bar{P}$ as the identity
or a $\bar{Z}$ gate, which are easier to implement. In the latter case,
we can rely on the prescription for the $Z$ gate we have provided.
Similarly, 3 (mod 4) successive applications of the $\bar{P}$ gate should be replaced with a single application of $\bar{P}^{-1}$, which corresponds to the physical gate $P^{-1}\equiv e^{-i\frac{1}{2}\hat{q}^{2}}$.

In analogy with monitoring the position mean, we ought to keep track of how many shearing operations (positive exponents of $P$) and anti-shearing (negative exponents of $P$) we have used. For example, if the computation calls for four non-successive applications of $\bar{P}$, one might consider the physical pattern $(P,P,P,P^{-3})$ so that the net shear is 0. Another possibility is $(P, Z(\sqrt\pi) P^{-1}, P, Z(-\sqrt\pi) P^{-1})$, so that no more than one unit of shear is ever applied to the state. The best optimization strategy for any particular circuit is left to future research.

\subsubsection{Rotations}
The phase gate (a counter-clockwise rotation), 
\begin{equation}
R(\phi) = e^{i \frac \phi 2 \left( \hat{q}^2 + \hat{p}^2 \right)},
\end{equation}
which implements the logical Hadamard operation whenever $\phi = \pi/2$ (and is called in this case the Fourier gate) commutes with the envelope operator $E(\epsilon)$. Therefore repeated applications of a perfect rotation will not damage the GKP state.

\subsubsection{Squeezing and beamsplitters}

Although they do not effect Clifford operations per se, quadrature squeezers
\begin{equation}
S\left(z\right) \equiv e^{\frac{1}{2}(z^{*}\hat{a}^2-z\hat{a}^{\dagger2})}
\end{equation}
and beamsplitters
\begin{equation}
\label{eq:beamsplitter}
B\left(\theta,\phi\right)\equiv e^{\theta(e^{i\phi}\hat{a}_{1}\hat{a}_{2}^{\dagger}-e^{-i\phi}\hat{a}_{1}^{\dagger}\hat{a}_{2})},
\end{equation}
are used in the implementation of gates, such as the SUM gate below, in the continuous-variable optical domain. For later use, note that $t = \cos{\theta}$ is the beamsplitter transmissivity. Since the squeezing gate has terms $\hat{a}^{2}$ and $\hat{a}^{\dagger 2}$, its impact on normalizable GKP states will be comparable to that of the phase gate. On the other hand, beamsplitter gates have only products of $\hat{a}$ and $\hat{a}^{\dagger}$, which means they commute with the envelope operator $E(\epsilon)$ (see App.~\ref{subsec:comm_rel_E}) and will not be harmful to GKP states, like the rotation gates described above.

\subsubsection{SUM gates}
For universal quantum computation we require at least one kind of two-qubit entangling operation, for example the $\overline{\text{CNOT}}$ gate. For ideal GKP states, this can translate to the continuous-variable SUM gate,
\begin{equation}
\text{SUM}(g) \equiv e^{-i g \hat{q}_1 \otimes \hat{p}_2},
\end{equation}
with $g=1$ being the standard weight. However, one is free to choose any odd-integer weight to effect a $\overline{\text{CNOT}}$:
\begin{equation}
\overline{\text{CNOT}} \to \text{SUM}(2k + 1) \text{ for } k \in \mathbb{Z}. 
\end{equation}
This can be seen by noting that the SUM gate effects the following quadrature transformations:
\begin{align}
    \hat{q}_1 &\to \hat{q}_1 \\
    \hat{p}_1 &\to \hat{p}_1 - g\hat{p}_2 \\
    \hat{q}_2 &\to \hat{q}_1 +  g\hat{q}_2 \\
    \hat{p}_2 &\to \hat{p}_2.
\end{align}
When the control mode is $\Ket{0_I}$, then $\hat{p}_1$  and $\hat{q}_2$ are both shifted by an even multiple of $\sqrt{\pi}$ for any integer weight, meaning neither the control nor target mode are changed, as desired. On the other hand, when the control is $\Ket{1_I}$, then $\hat{p}_1$ and $\hat{q}_2$ are shifted by an odd multiple of $\sqrt{\pi}$ for odd-integer weight; since $\Ket{0_I}$ and $\Ket{1_I}$ are $\sqrt{\pi}$-periodic in momentum, this implements a bit flip on the target. Therefore, like for the previous gates, one should update the weight of the SUM gate in a computational circuit depending on the weight of the previously applied SUM gate. To see how this translates to physical requirements, consider the decomposition of the SUM gate into squeezers and beamsplitters:
\begin{equation}
    \text{SUM(g)} = B(\pi/2 + \theta, 0) \left(S(r, 0) \otimes S(-r, 0)\right) B(\theta, 0),
\end{equation}
where $\sin(2\theta) = -\sech(r)$, $\cos(2\theta) = \tanh(r)$, and $\sinh(r) = -g/2$. From the discussion above, the harmful element is not the beamsplitter but the squeezer. As expected, $r$ is a monotonic function of $g$; and positive values of $g$ correspond to negative values of $r$.

\begin{figure}
\includegraphics[width = \linewidth]{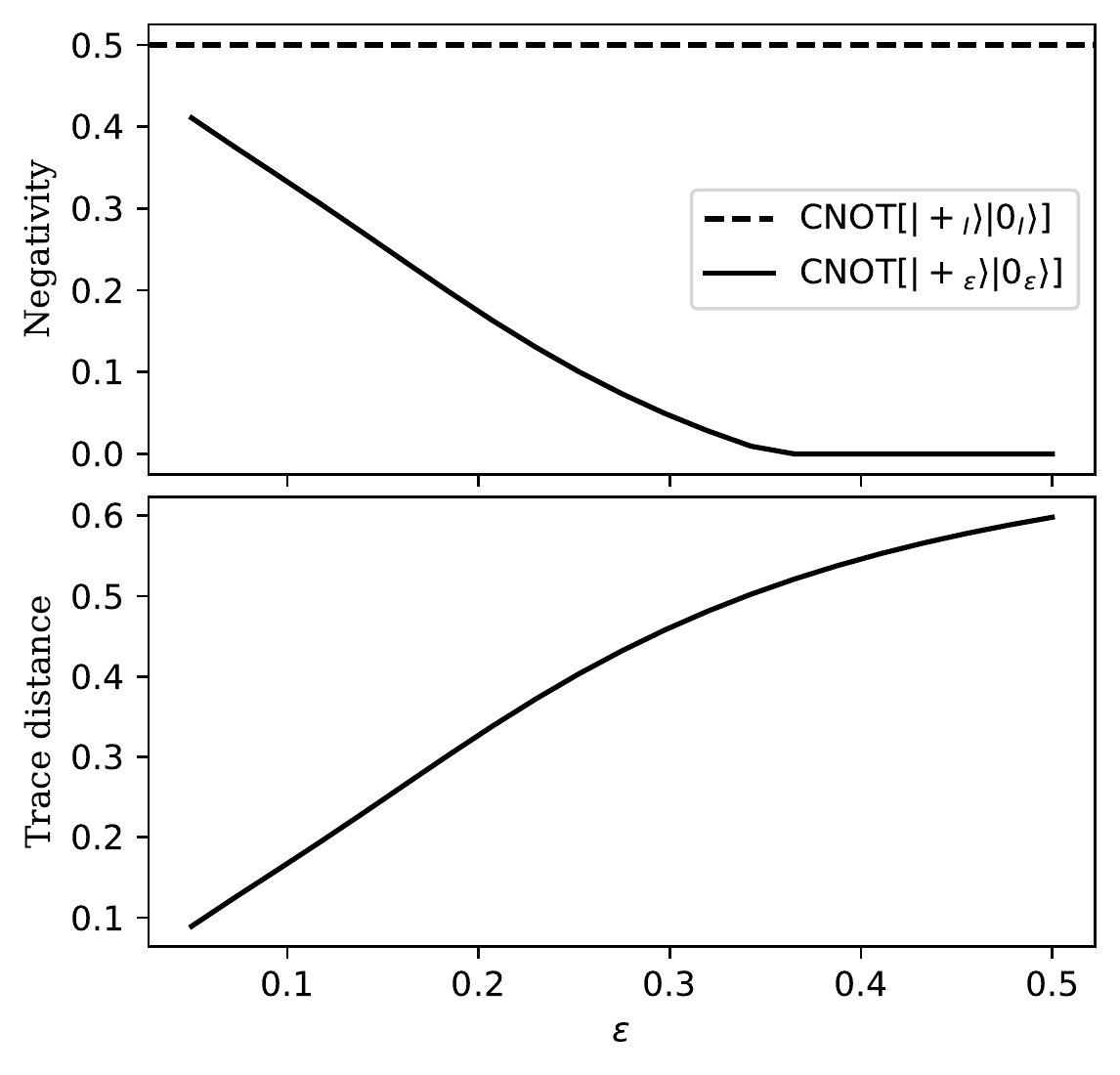}\caption{ (a) Entanglement negativity (defined in main text) of a two-qubit system initialized in $\Ket{+_\epsilon}\Ket{0_\epsilon}$ and then subjected to the application of a CNOT [SUM(1)] gate, as compared to the ideal (maximal) negativity for perfect qubits. We see that to generate entanglement in the logical subsystem, we require $\epsilon \approx 0.36$, which corresponds to roughly $\Delta \approx 4.5$ dB of squeezing; however, to generate maximal entanglement, significantly more stringent thresholds are required, with $\epsilon \approx $ 0.05 only generating 80\% of the maximal entanglement negativity. (b) The logical subsystem trace distance between  CNOT$[\Ket{+_\epsilon}\Ket{0_\epsilon}]$ and CNOT$[\Ket{+_I}\Ket{0_I}]$. Again, we see that to achieve the ideal distance of 0, smaller values of $\epsilon$ are required.}
\label{fig:CNOT_neg}
\end{figure}

One benchmark for how the normalizable states perform under the SUM gate is how entangled the modes become in the logical subsystem. Ideally, we have that $\overline{\text{CNOT}} \Ket{\bar{+}}\Ket{\bar{0}}$ is a maximally-entangled state. In  Fig.~\ref{fig:CNOT_neg}, we plot the entanglement negativity of a two-qubit system in the logical subspace, a well-known measure for determining if two systems are entangled \cite{negativity}. Entanglement negativity is defined to be the sum of the absolute values of the negative eigenvalues of the partial transpose of a bipartite density matrix with respect to one of the subsystems. For two-qubit systems, negativity ranges from 0 (no entanglement) to 1/2 (maximal entanglement). 
We initialize the two modes in the $\Ket{+_\epsilon}\Ket{0_\epsilon}$, and then apply the CNOT [SUM(1)] gate. Interestingly, we find a threshold corresponding to $\sim$ 4.5~dB of squeezing is required to produce entanglement in the logical subsystem.

\subsection{Error correction with normalizable GKP states} \label{subsec:error_correction}

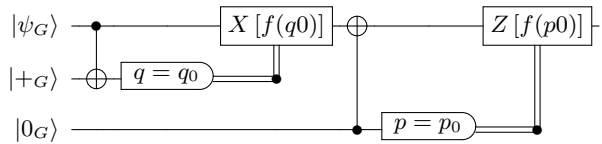
\begin{figure}
$$
\Qcircuit @C=0.3em @R=.7em { 
\lstick{\Ket{\psi_G}} & \qw & \ctrl{1} & \qw & \qw & \gate{X\left[f(q0)\right]} & \qw & \targ & \qw & \qw & \gate{Z\left[f(p0)\right]} & \qw \\
\lstick{\Ket{+_G}} & \qw  & \targ & \qw & \measureD{q=q_0} & \cctrl{-1}  \\
\lstick{\Ket{0_G}} & \qw & \qw & \qw & \qw & \qw & \qw & \ctrl{-2} & \qw & \measureD{p=p_0} & \cctrl{-2}
}
$$
\caption{Error syndrome measurement with normalizable GKP states following the Steane approach. First, shifts
in $q$ are corrected: an encoded data qubit $\Ket{\psi_{G}}$ and
an ancilla $\Ket{+_{G}}$ are sent through a SUM gate, and $\Ket{\psi_{G}}$
is displaced according to the result of a homodyne $q$ measurement
on the ancilla. A similar procedure follows for shifts in $p$.  \label{fig:syndrome_meas}}
\end{figure}

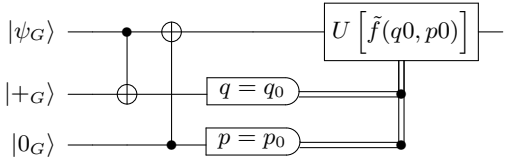
\begin{figure}
$$
\Qcircuit @C=1.0em @R=.7em { 
\lstick{\Ket{\psi_G}} & \qw & \ctrl{1}  & \targ & \qw & \gate{U\left[\tilde{f}(q0, p0)\right]} & \qw \\
\lstick{\Ket{+_G}} & \qw  & \targ & \qw  & \measureD{q=q_0} & \cctrl{-1}  \\
\lstick{\Ket{0_G}} & \qw & \qw & \ctrl{-2} & \measureD{p=p_0} & 
\cctrl{-2}
}
$$
\caption{The circuit in Fig.~\ref{fig:syndrome_meas} but with the measurements pushed to the end. Now a unitary  is applied that encompasses both the correcting position and momentum shifts determined by a new function, $\tilde{f}$, of the homodyne measurement results $q_0$ and $p_0$. This unitary shifts the projected state back onto the GKP grid. When the error is too big, these shifts are mistakenly performed in the wrong direction, and a logical error results.
}
\label{fig:steane_meas_end}
\end{figure}

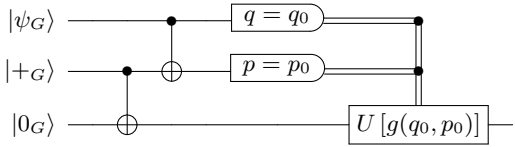
\begin{figure}
$$
\Qcircuit @C=1.0em @R=.7em { 
    \lstick{\Ket{\psi_G}} & \qw & \qw & \ctrl{1} & \qw & \measureD{q=q_0} & \cctrl{1} \\
    \lstick{\Ket{+_G}} & \qw & \ctrl{1} & \targ & \qw & \measureD{p=p_0} &
    \cctrl{1} 
    \\
    \lstick{\Ket{0_G}} & \qw & \targ & \qw & \qw & \qw & \gate{U\left[g(q_0, p_0)\right]} & \qw
}
$$
\caption{GKP error correction using the Knill approach. Here, the ancillary states $\Ket{+_G}$ and $\Ket{0_G}$ are entangled and then the data state $\Ket{\psi_G}$ is sent through a SUM gate with $\Ket{+_G}$ as the target. Homodyne position and momentum measurements are conducted on $\Ket{\psi_G}$ and $\Ket{+_G}$, respectively; this teleports the logical data to $\Ket{0_G}$ up to a unitary, applied to $\Ket{0_G}$ depending on the measurement results. Notice that, unlike in the Steane approach (Fig.~\ref{fig:steane_meas_end}), the output state after the projection is already on the GKP grid. The purpose of this unitary, therefore, is to correct the logical-Pauli byproduct operators that result from the teleportation. When the error is too big, these byproduct operators are misidentified, and a logical error results. \label{fig:knill_ec}}
\end{figure}
As a continuous limit of shift-resistant qudit codes, GKP encoding
allows one to correct for small displacement errors in the encoded
data state. In fact, the GKP codes accommodate arbitrary errors on
the oscillator, since displacements -- i.e., the   Weyl-Heisenberg  operators $X\left(\alpha\right)$
and $Z\left(\beta\right)$ -- form a complete operator basis. The error syndrome
measurement in the Steane approach \cite{steane96} is shown and described in Figs. \ref{fig:syndrome_meas} and \ref{fig:steane_meas_end} and in the Knill teleportation approach \cite{knill2005pra,knill2005quantum} in Fig.~\ref{fig:knill_ec}. The main difference between the two pictures is that, in the former, the data qubit interacts with two separable ancillae, whereas in the latter, the data qubit interacts with only one of two entangled ancillae. If one can guarantee a supply of high-quality ancillae, the Knill approach could be advantageous, as it precludes two applications of the SUM gate to a noisy data state. There may exist other scenarios involving GKP states in which one or the other circuit is preferable; for our purposes, we focus on the Steane approach.

In the Steane error correction circuit, the SUM gate preserves the position of the data qubit and transforms the ancilla $A$ through $q_{A}\to q_{\psi}+q_{A}$. This means that
a measurement of the ancilla will yield $n\sqrt{\pi}+\delta q_{\psi}+\delta q_{A}$ for $n \in \mathbb{N}$, where the $\delta q$'s denote the position shift errors. Therefore
the function $f$ that ought to be applied to the measurement outcomes
before the correcting shift is
\begin{equation}
f\left(r\right)=-\text{mod}_{\sqrt{\pi}}\left(r\right),
\end{equation}
where $\text{mod}_{t}(x)\in\left[-\frac{t}{2},\frac{t}{2}\right)$. Whenever
$\delta_{\psi}$ and $\delta_{A}<\frac{\sqrt{\pi}}{2}$, we have that $f\left(q_{\text{net}}\right)=-\delta_{\psi}-\delta_{A}$,
and we have corrected our $q$ displacement errors. However, we have
now introduced a new shift error in momentum, since the CNOT gate
also effects $p_{\psi}\to p_{\psi}+p_{A}$. Given a perfect ancilla,
we would be able to correct displacements of magnitude $\sqrt{\pi}/2$
in position and momentum. But the finite energy envelope introduces
errors on the ancilla that restrict the range of correctable errors.

In this setting, since we accumulate three errors after a complete
round of error correction, Glancy and Knill~\cite{glancy}
found that error correction will only be successful if the magnitude
of all shifts is less than $\sqrt{\pi}/6$. This ensures that the total error on the data qubit is within correctable region. For a position wavefunction
$\psi\left(q\right)$ of any of the noisy states, the probability
of successful error correction is thus:
\begin{align}
P_{\text{no error}}= & \frac{\pi}{3}\sum_{s,t}\text{sinc}\left(\frac{\pi t}{3}\right)\times\nonumber \\
 & \int_{\sqrt{\pi}(2s-\frac{1}{6})}^{\sqrt{\pi}(2s+\frac{1}{6})}du\psi^{*}(2t\sqrt{\pi}+u)\psi(u).
\label{eq:glancy}
\end{align}
where we provide a short derivation of this expression in App.~\ref{subsec:GKderiv} and explain its meaning there. Note that although we are assuming approximate GKP ancillae, the formalism just described is general enough to accommodate arbitrary ancillary states, with their usefulness quantified by Eq.~(\ref{eq:glancy}). For error correction to succeed with high probability, $P_{\text{no error}}$ must be high; this is satisfied by close-to-ideal GKP states. Conversely, a generic state with a high $P_{\text{no error}}$ must have little modular spread in both position and momentum, implying that it approaches the form of an ideal GKP state. In this sense, the Glancy-Knill condition is both necessary and sufficient for correcting displacement errors using the standard Steane and Knill schemes described above. We explore this point further in \ref{subsubsec:glancy}.

If during the $q$ ($p$) error correction the total magnitude of
position shifts is greater than $\sqrt{\pi}/2$, the procedure results
in a bit-flip (phase flip) error on the logical qubit. These additional
errors can be corrected by concatenating the GKP code with standard
qubit codes, for example.

Note that one does not need a physical SUM gate to implement GKP error correction; other gates that mix the physical modes can also work. For example, the gate corresponding to a beamsplitter in the optical domain
will suffice, in addition to a squeezing gate and an appropriate modification of the function $f$. This approach is taken in \cite{glancy}; its benefit is that it is less demanding to implement than the canonical optical CNOT gate, for example, which requires two beamsplitters and two squeezers.

\begin{figure}
\begin{centering}
\subfloat[\label{subfig:gk_abs_prob}]{\includegraphics[width=\linewidth]{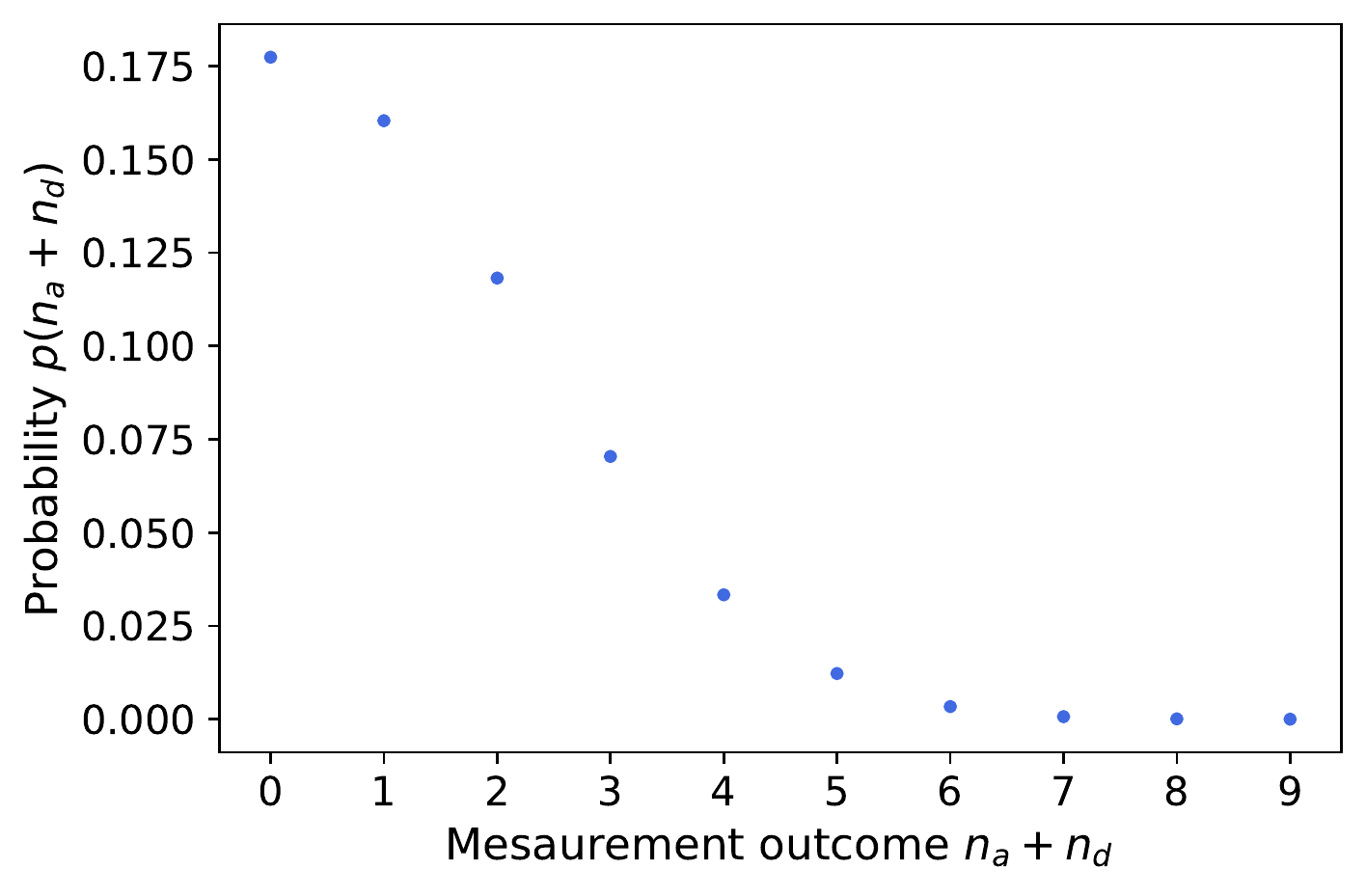}}
\\
\subfloat[\label{subfig:gk_cond_prob}]{\includegraphics[width=\linewidth]{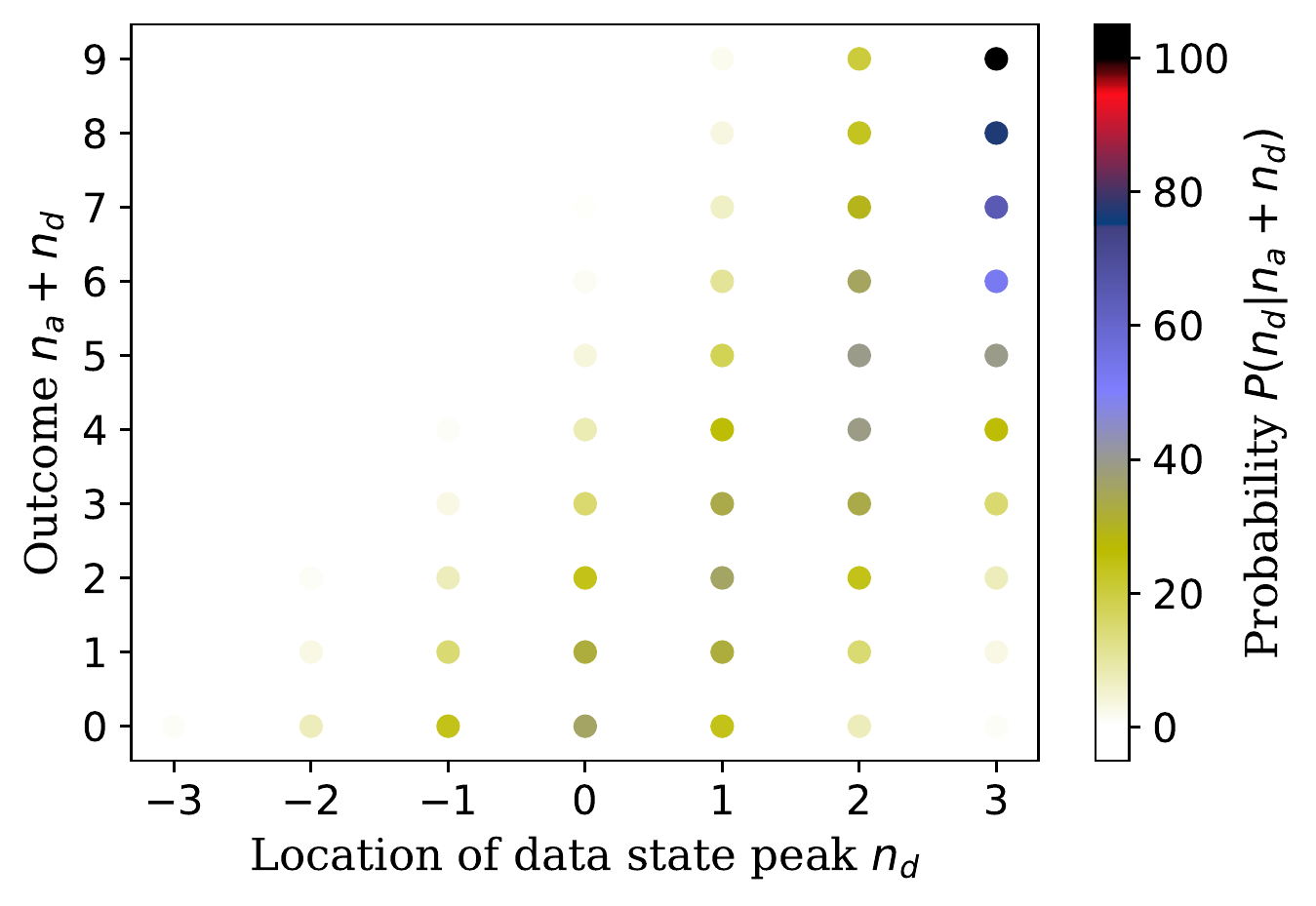}}
\end{centering}
\caption{Homodyne measurement probabilities in a successful position error correction step with data state
$\Ket{0_{\epsilon}}$ and ancilla $\Ket{+_{\epsilon}}$ with $\epsilon=0.063$ ($\Delta \approx 12~\text{dB}$);
\protect\subref{subfig:gk_abs_prob} the probability of the integer part of the measurement outcome
equalling $n_{a}+n_{d}$, where $n_a \sqrt{\pi}$ and $2 n_d \sqrt{\pi}$ are peak locations in the ancilla and data state, respectively, and \protect\subref{subfig:gk_cond_prob} the probability of the data state
having a peak at $q = 2n_{d}\sqrt{\pi}$ given that the integer part of
the measurement outcome was $n_{a}+n_{d}$. After several rounds of error correction, we may be able to use data like this to conclude, with some probability, how far the mean of our data state is from 0 so that we may recalibrate it.}
\label{fig:GK_prob}
\end{figure}

\paragraph{Incorporating $q$ and $p$ calibration within error correction.}

GKP error correction gives information not only about the fractional
part of the gauge mode -- the small errors that enter haphazardly
in the computation -- but also the integer part, which is related
to the center of the encoded state, the periodicity of the peaks,
and the net displacement from the origin of a finite-energy GKP state,
as discussed in \ref{subsubsec:q_disps} and \ref{subsubsec:p_disps}.
Therefore we might consider ``feeding two birds from one hand''
and using the outcomes of the homodyne mesaurements in the error correction
procedure to recenter our data state. This would entail modifying
the function $f$ above so that the $X$ and $Z$ gates both correct
and calibrate the state. 

However, we run into a problem: the SUM gate
mixes the quadratures of the ancillary and data qubit, limiting the
information we can extract about the data state. To see this, suppose that
$\delta q_{\psi}+\delta q_{A}<\sqrt{\pi}$; then the integer part
of the measurement outcome above is $\left(n_{a}+n_{d}\right)\sqrt{\pi}$,
corresponding to the $n_{d}^{\text{th}}$ position peak of the data
state and the $n_{a}^{\text{th}}$ position peak of the ancillary
state ($n_a, n_d \in \mathbb{Z}$). In the limiting case of an ideal GKP state, no useful information
is obtained, since $n_{a}$ and $n_{d}$ could have come from any
peaks in the two states with equal probability.

In the normalizable
case, we can say a little more, since the probability of finding oneself
within a particular peak varies. Consider Fig.~\ref{fig:GK_prob}, where
we assume a normalizable data state with effectively seven peaks (so that $-3 \leq n_d \leq 3$)
and an ancilla with thirteen peaks  ($-6 \leq n_d \leq 6$). We can see in \ref{subfig:gk_cond_prob}, for example, that if we find ourselves
in $n_{a}+n_{d}=9$, then $n_{d}=3$ with unity probability. However,
not only does this outcome happen with vanishingly small probability (\ref{subfig:gk_abs_prob}),
but it also does not necessarily mean that the center of our data
state is at $2\sqrt{\pi}$. Nevertheless, one may still use Bayesian reasoning to establish the maximum-likelihood mean of the data qubit after the error-correction step and then use $2n\sqrt\pi$ displacements to return that mean to as close to zero as possible.

We note that the above GKP error correction cannot be expected to correct all errors. However, one can aim to construct more sophisticated codes built on the GKP qubits, such as surface codes, which can protect more general errors as considered by a few references mentioned in Sec.~\ref{sec:intro} \cite{fukui2018high,vuillot2019quantum,mastersthesis,nohgkpsurface}. We believe that the modular decomposition picture could play an important role in developing this further.

\paragraph{Post-recovery metrics.}

Having outlined the GKP error correction procedure, we can revisit
the metrics in Sec.~\ref{subsec:pre-recov-metrics} with an additional
recovery step. For a unitary $U_{\id}$ that effects $\bar{U}=I$
on the logical qubit, followed by a recovery operation, $R$, on a
data state together with ancilla, we can compute
the updated physical fidelity, logical fidelity, and distribution
distance:
\begin{align}
F_{RU_{\id}}^{\mathcal{P}}\left(\rho^D\right) &=  F\left[\rho^{D}, \Tr_{A}\left\{RU_{\id}\rho^{DA} U_{\id}^{\dagger}R^{\dagger}\right\}\right]\\
F_{RU_{\id}}^{\mathcal{L}}\left(\rho^{D}\right) &=  F\left[\Tr_{\mathcal{G}}\rho^{D}, \Tr_{\mathcal{G},A}\left\{RU_{\id}\rho^{DA} U_{\id}^{\dagger}R^{\dagger}\right\}\right]\\
\mathscr{D}^{p}_{RU_{\id}}\left(\rho^{D}\right) &= \left \|p_{RU}(x|\psi_G)-p_{U}(x|\psi_I)\right\|,
\end{align}
where $\rho^D \equiv \Ket{\psi^D}\Bra{\psi^D}$, $\rho^{DA} \equiv \rho^D \otimes \Ket{\psi^A}\Bra{\psi^A}$, and we refer back to Fig.~\ref{fig:circuit_models} for the definitions
of the probability distributions $p_{RU}$ and $p_{U}$. The energy cost is a path-dependent measure: it should only increase with an additional recovery step, even if the net change in energy might be smaller. This prompts the post-recovery energy cost definitions
\begin{align}
    \mathscr{D}_{RU}^{\bar{n}}\left(\rho^{D}\right) \equiv \left|\bar{n}\left(U\rho^{D}U^{\dagger}\right)-\bar{n}\left(\rho^{D}\right)\right| + \\
    \left|\bar{n}\left(RU\rho^{DA}U^{\dagger}R^{\dagger}\right)-\bar{n}\left(U\rho^DU^{\dagger}\right)\right|
    \nonumber \\
    \mathscr{D}_{RU}^{\sigma^2}\left(\rho^D\right) \equiv \left|\sigma^2_{\bar{n}}\left(U\rho^DU^{\dagger}\right)-\sigma^2_{\bar{n}}\left(\rho^D\right)\right| + \\
    \left|\sigma^2_{\bar{n}}\left(RU\rho^{DA}U^{\dagger}R^{\dagger}\right)-\sigma^2_{\bar{n}}\left(U\rho^DU^{\dagger}\right)\right|,
    \nonumber
\end{align}
where now $\bar{n}(\rho) \equiv \Tr({\hat{n}\rho})$ and $\sigma_{\bar{n}}^2(\rho) \equiv \Tr({\hat{n}^2\rho}) - (\bar{n}(\rho))^2$.

Having introduced various figures of merit to track and understand the way error propagates due to state preparation errors, we now set up the tools to prepare explicit optical circuits to produce the realistic grid states. 

\section{Photonic state preparation and characterization}\label{sec:stateprep}
\subsection{Preparation of non-Gaussian states}\label{subsec:nongauss}
\begin{figure}
    \centering
    \includegraphics[width=\linewidth]{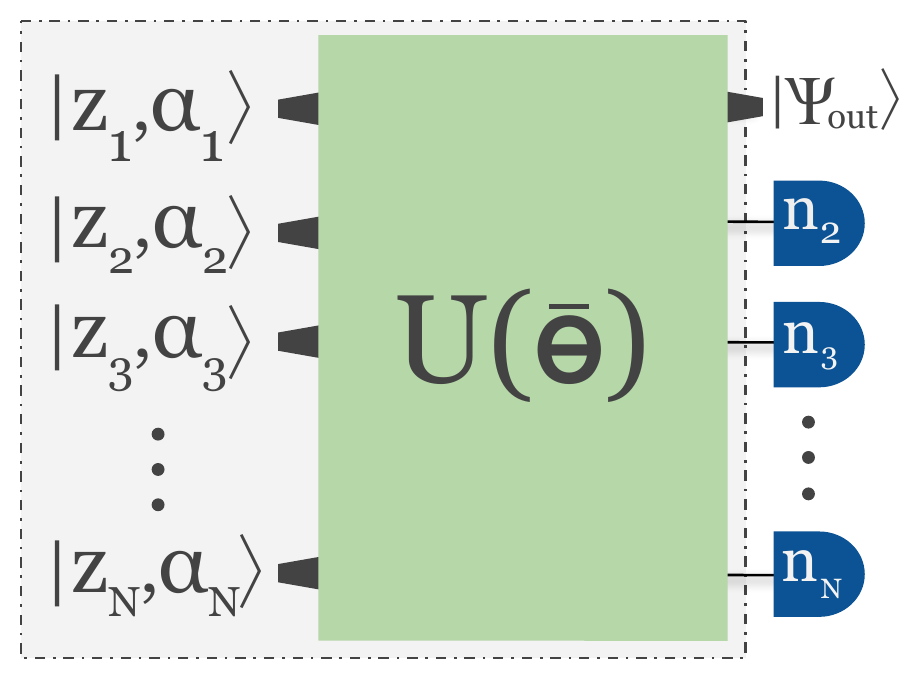}
    \caption{The model for the GBS-like device used for state preparation. Gaussian states consisting of squeezed, displaced vacuum states $|z_i,\alpha_i\rangle$ are sent through an $N$-mode interferometer parametrized by $U(\mathbf{\bar{\Theta}})$, followed by a PNR measurement on all but one of the modes. Given a PNR outcome of $\bar{\mathbf{n}}$, the desired output state, $\Ket{\psi_\text{out}}$, is produced in the unmeasured mode. Our task is to optimize the circuit components ($\boldsymbol{\alpha}$, $\boldsymbol{z}$, and $\mathbf{\bar{\Theta}}$) for a given $\bar{\mathbf{n}}$ to produce a desired approximate GKP state. We loop over $\bar{\mathbf{n}}$ (subject to constraints) to find the best $N$-mode circuit for the task.}
    \label{fig:GBS_circ}
\end{figure}

GKP states are highly non-Gaussian (see, for example, the trend in Fig.~\ref{fig:wig_log_neg_0e}), so their preparation requires non-Gaussian resources. For optical platforms, one such resource that is already experimentally accessible is the photon number-resolving (PNR) detector. A Gaussian multimode state can be prepared by applying a general interferometer, $U(\mathbf{\bar{\Theta}})$, with $N^2$ independent beam splitter and phase shift parameters $\mathbf{\bar{\Theta}}$, to a multimode input of displaced squeezed vacuum states $D(\boldsymbol{\alpha})S(\boldsymbol{z}\,)|\mathbf{0}\rangle$. Next, by making PNR measurements of $N-M$ of the $N$ modes, and obtaining results other than zero photon detections across the detectors, one  prepares an $M$-mode non-Gaussian state~\cite{saba1,su1,su2,1905.07011}. As the architecture is analogous to \textsf{Gaussian} \textsf{BosonSampling} (GBS), we will refer to the technique as state preparation with GBS devices. The circuit for generating a single-mode non-Gaussian state conditioned on measuring the remaining modes using PNR detectors is depicted in Fig.~\ref{fig:GBS_circ}. This general framework encompasses other state preparation schemes, such as photon subtraction and addition (see for example, Fig. 1 of \cite{su2}).

{}
When using GBS circuits for state preparation, the number of modes, the initial squeezing and displacement parameters, the interferometer beamsplitter angles, and the PNR measurement patterns can all become parameters for tailoring an output state according to a predefined cost function, such as closeness to a given target state and probability of successfully preparing the given state. Extensive analysis of this framework has been performed for the preparation of Fock, cat, NOON and weak-cubic-phase states~\cite{saba1,su1,su2,1905.07011}, and proof-of-concept calculations for the technique have been made for GKP states~\cite{su2}. Here, we provide a thorough analysis for preparing GKP states with GBS circuits. 

There are a few important results that guided our search for optimal state preparation using GBS circuits. First, if one wishes to prepare a Fock superposition of up to $N$ photons by measuring pure Gaussian states, then the number of photons detected in PNR measurements should sum to $N$~\cite{su2, Fiur2005}; this allows us to significantly restrict our search over post-selection patterns. Second, Su et. al. \cite{su1,su2} conjectured that measuring $(N-1)$ modes with PNR detectors in a GBS circuit outputs a Fock superposition with at most $(N+2)(N-1)/2$ independent coefficients. This implies that, by tuning $\boldsymbol{\alpha},\boldsymbol{z}$, and $\bar{\boldsymbol{\Theta}}$ and choosing a suitable photon number post-selection pattern $\bar{\mathbf{n}}$, one can always, in principle, prepare with perfect fidelity a single-mode target state of at most $(N+2)(N-1)/2$ photons using an $N$-mode circuit. For a given target state, this allows us to set an upper bound on the number of modes in the circuit for which we search for optimal $\boldsymbol{\alpha},\boldsymbol{z},\bar{\boldsymbol{\Theta}}$ and $\bar{\mathbf{n}}$.
 
\subsection{Core states for GKP} \label{subsec:core_states}
We now discuss a framework for approximating GKP states. In this section, we describe a formulation of GKP states that is platform-independent, so it can be applied to GKP state preparation in superconducting circuits or ion traps ~\cite{travaglione2002preparing,pirandola2006continuous,fluhmann2019encoding,fluhmann2018,1907.12487}, for example.

An arbitrary single-mode quantum state can be constructed from a core superposition of Fock states followed by Gaussian operations: squeezing, displacement and rotation~\cite{Menzies2009, Chabaud2019}. This is sometimes referred to as the stellar representation of the state~\cite{Chabaud2019}. An approximation to a given state can be found by truncating the core state with a suitable $n_{\max}$: 
\begin{equation}\label{eq:core}
    \Ket{\psi} \approx S(\zeta)D(\beta) \underbrace{\sum_{n=0}^{n_{\max}}\frac{c_n}{N(\boldsymbol{c},n_{\max})}  |n\rangle}_{\text{truncated core state}},
\end{equation}
where $N(\boldsymbol{c},n_{\max})$ is the normalization constant. The exact state can be recovered by taking $n_{\max} \rightarrow \infty$. 

The approximate representation (\ref{eq:core}) is particularly useful when using GBS circuits for state preparation. First, to prepare a target state with a given $n_{\max}$, we know how many modes and what set of PNR measurement patterns are required to guarantee production of a state with perfect fidelity to the target. Thus, we can find a circuit that optimally produces the truncated core state. Next, as the circuit consists of Gaussian operations on Gaussian states, the additional Gaussian operations, $S(\zeta)D(\beta)$, that we apply to the core state after it is produced can simply be absorbed into the Gaussian circuit, $U(\bar{\boldsymbol{\Theta}})$; moreover, the circuit can be re-decomposed, yielding new $\boldsymbol{\alpha'},\boldsymbol{z'},U(\bar{\boldsymbol{\Theta}}')$ \cite{euler}. Thus, operations, such as the squeezing on the core state, are implemented at the start of the circuit, eliminating the need for inline squeezing that is comparatively harder to implement~\cite{furu-sq}. Finally, if $n_{\max}$ is constrained by the available physical resources, such as the number of circuit modes, by targeting a core state from Eq.~\eqref{eq:core} rather than a truncation of the Fock expansion of $\Ket{\psi}$ directly, we are generally able to attain higher fidelities between the prepared state and $\Ket{\psi}$. Some of the extended support in Fock space is captured by the displacement and squeezing of the core state and may be unnecessarily discarded if one truncates the Fock expansion of $\Ket{\psi}$.\\

\noindent {\bf Remark.} The approximate representation \eqref{eq:core} is additionally valuable if one wants to prepare GKP states with a different lattice symmetry, such as the hexagonal GKP states. As noted in \ref{subsec:ideal_GKP}, the hexagonal GKP is related to the square GKP via a symplectic transformation, which can be decomposed into Gaussian operations. Thus, if one has a GBS circuit which can prepare an approximate square GKP state, one can use it prepare the hexagonal GKP state by appending the Gaussian operations to the end of circuit (or by redecomposing the circuit into a new one).\\

Before we examine how to prepare approximations to the GKP $X$, $Z$ and Hadamard eigenstates, let us clarify some nomenclature. \emph{Normalizable GKP states}, reviewed in Sec.~\ref{subsec:norm_gkp}, are ideal GKP states reduced to a finite energy state by the application of an envelope operator. Our first task is to find an approximation, in the form Eq.~\eqref{eq:core}, to a choice of normalizable GKP states, i.e., for a specific choice of envelope; approximations in the form (\ref{eq:core}) will be referred to as \emph{approximate GKP states}. Keeping with our notation, we can denote these states as $\Ket{\psi_A}$. Our second task is to find a GBS circuit that can optimally prepare the core state corresponding to approximate GKP states, and then to redecompose the circuit to include the Gaussian operations applied to the core state. We call the states output by the final circuit the \emph{circuit GKP states}. If the circuit GKP states, for which we can define a clear experimental prescription, have high enough fidelity to the approximate GKP states, and these states have high enough fidelity to the normalizable GKP states, which for a good enough choice of envelope capture the properties of the ideal GKP states, then the circuit GKP states will share the desired properties of the ideal states.

As they have been the most commonly studied form of normalizable GKP states, we choose the $\Ket{\mu_\Delta}$ states with $\mu \in \{0,1,+,H_+\}$ as the states for which we want to find approximate GKP states (such as $\Delta=\kappa$ case in Eqs. \eqref{eq:0Ndeltakappa}, \eqref{eq:0Ndeltakappa1}). We could have also chosen to target (\ref{eq:step_del}) with a step function $g$, that is, some finite cutoff of the ideal GKP states expressed directly in the Fock basis. However, we found that, for equal cutoffs, these states had greater support than the $\Ket{\mu_\Delta}$ in regions where the wavefunctions were supposed to vanish.

We note that since the wavefunctions of the above set of $\Ket{\mu_\Delta}$ are all real, we do not need to apply complex squeezing to the core state, meaning $\zeta = r\in\mathbb{R}$. As the wavefunctions are also symmetric, we do not need to displace the core state, so $\alpha = 0$, and only the even Fock coefficients will contribute, meaning $c_n = 0$ for $n$ odd. Thus, for a given $n_{\max}$, we want to find $r$ and $c_n$ ($n$ even) such that the fidelity between the squeezed, truncated core state is maximized with $\Ket{\mu_\Delta}$. We summarize our method for finding optimal parameters in Algorithm~\ref{algone},  with additional details provided in App.~\ref{subsec:alg1det}. We found the \textsf{scipy basinhopping} global optimization package \cite{scipy,basinhopping} to be particularly useful.

\begin{algorithm}[H]

\begin{algorithmic}
\Function{cost}{r, $\boldsymbol{c}$, $\Delta$, $\mu$} 

\textbf{initialize} $\Ket{\mu_\Delta}$

$\Ket{\psi} \gets S(r) \sum_n c_n |n\rangle$

\Return $|\langle \psi\Ket{\mu_\Delta}|^2$ 

\EndFunction

\Procedure{optimization}{$\Delta$, $\mu$, $n_{\max}$}

\textbf{initialize} r \#squeezing within desired range

\textbf{initialize} $\boldsymbol{c}$ \#normalized vector of dim $n_{\max}$

\# \textsf{basinhopping} is a global search algorithm

$r_\text{opt}, \boldsymbol{c}_\text{opt} \gets \textsf{basinhopping}[(r, \boldsymbol{c}),\texttt{cost},args = (\Delta, \mu)]$  
\EndProcedure
\end{algorithmic}
\caption{Optimal Approximate States}
\label{algone}
\end{algorithm}

In Fig.~\ref{fig:fid vs n,D}, we plot the fidelity between the normalizable and approximate GKP states for $\mu = 0$ as a function of $\Delta$ from 3 to 11 dB for even values of $n_{\max}$ from 2 to 12 photons.  We provide a comment on the near identical results for $n_{\max} = 4$ and 6 in App.~\ref{subsec:further0}.   Our results for the other $\mu = 1, +, H_+$ states are available in App.~\ref{subsec:1pm}. As can be expected, for a fixed $\Delta$, the fidelity improves monotonically with increasing $n_{\max}$, and for a fixed $n_{\max}$ the fidelity worsens monotonically with increasing $\Delta$.

\begin{figure}
    \centering
    \includegraphics[width=\linewidth]{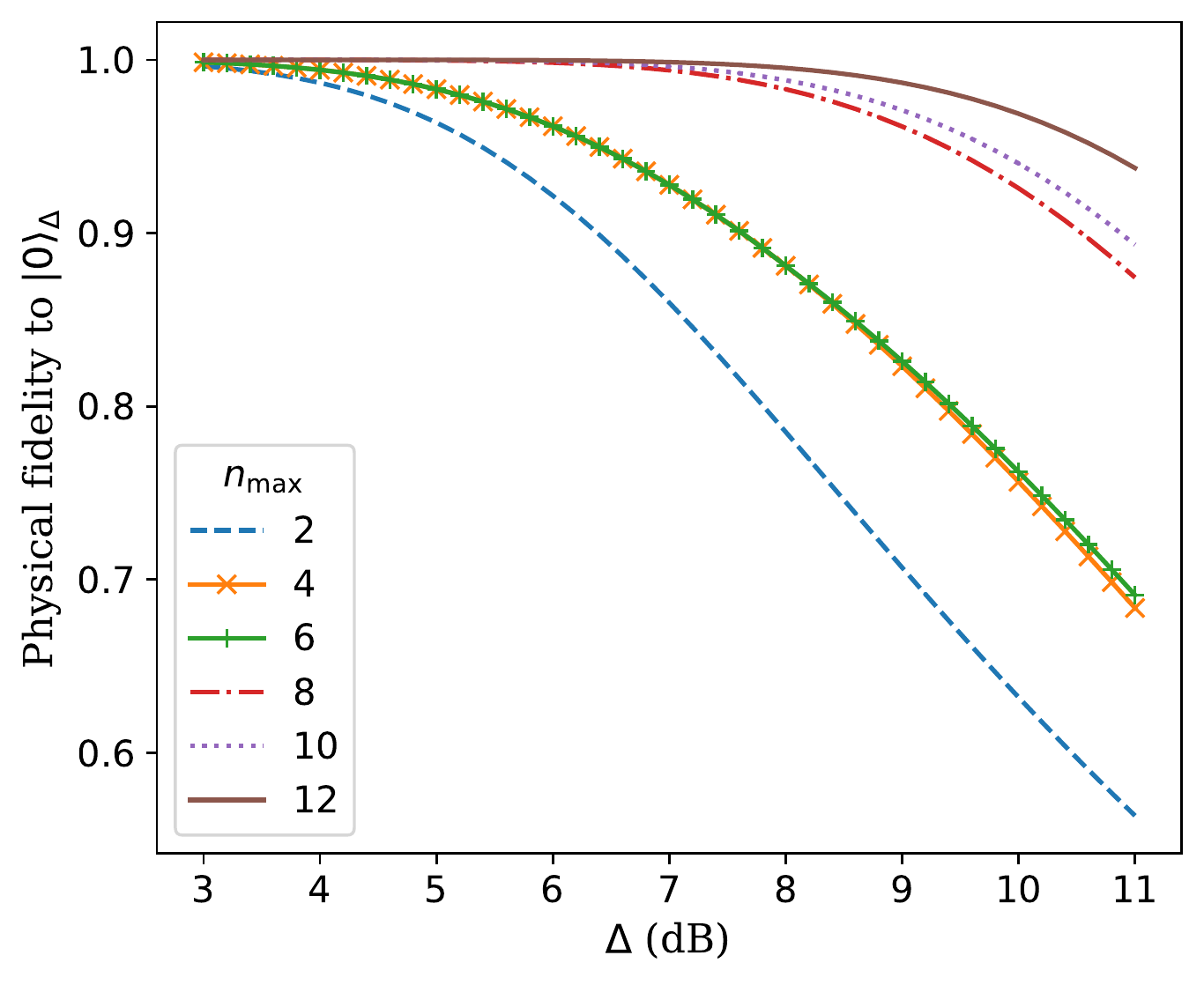}
    \caption{ Physical fidelity of our approximate states, $\Ket{0_A}$, to the target state, $\Ket{0_{\Delta}}$, as $\Delta$ (dB) is varied (see Eq.~\eqref{eq:0Ndeltakappa} for the definition of $\Ket{0_{\Delta}}$ and App.~\ref{sec:convention} for conversion of dB to other conventions). The $\Ket{0_A}$ states are constructed by applying squeezing to a core superposition of Fock states, so different line colours and styles correspond to different values of $n_{\max}$ for the core state. Note that, for different $n_{\max}$, the optimal squeezing parameter and Fock coefficients are generally different. The optimal $\Ket{0_A}$ states for different $n_{\max}$ are found by maximizing the fidelity to $\Ket{0_{\Delta}}$.}
    \label{fig:fid vs n,D}
\end{figure}

\subsection{Characterization of approximate states}\label{subsec:charac_app}
We can now look at various properties of the approximate states we have considered, namely the average photon number, orthogonality relations, the Glancy-Knill error correction condition, and behaviour in the modular decomposition. In App.~\ref{sec:add_feat}, we examine even more qualitative features of the approximate states, such as the projectors and quantum error correction matrices.

\subsubsection{Average energy}
The average energy of the approximate states will have repercussions for the resources required for making the state; for circuit GKP states, this translates to a demand on the initial squeezing applied to each mode. In Fig.~\ref{fig:navg vs n,D}, we plot the average photon numbers of $\Ket{0_A}$ for different $n_{\max}$ as a function of $\Delta$. We see that the average energy of the states is not too high; it increases with $\Delta$ and $n_{\max}$, but never exceeds five photons. We have already shown that increasing $n_{\max}$ is required for producing higher fidelities to the target GKP states, and we know additionally that states with higher $\Delta$ values (in dB) provide better error correction and encoding properties; thus, we learn a simple trend for the resources required, even in this method of preparing approximate GKP states: better states require more energy. It is important to remember this trend in addition to the energy costs already established for gate applications (recall Figs.~\ref{fig:en_test_X} and~\ref{fig:en_test_p}). Average photon number can also be used as a fundamental property to compare various bosonic codes; notably, when considering codes with average photon number less than five, GKP codes were shown to outperform other bosonic codes, including cat and binomial codes, for protecting against loss errors \cite{albert2018performance}. \\ 

\noindent{\bf Remark.} In App.~\ref{subsec:1pm}, we see that the average photon numbers for the $\Ket{1_A}$, $\Ket{+_A}$ and $\Ket{H_{+A}}$ states are all on the same order and follow similar trends. Therefore, given our approach for preparing states with optical circuits, we expect that this will mean preparing different approximate GKP states on the Bloch sphere will require comparable resources, i.e., the same required order of magnitude for squeezing, and the same number of interferometer elements, and PNR detectors (see also Ref. \cite{yamasaki2019}). This means one has the options when designing a computation of only preparing $\Ket{0_A}$ states and $\Ket{H_{+A}}$ states for non-Clifford gates, or of preparing a collection of states on the Bloch sphere from the outset: the resource requirements will be similar, and the number of gate applications in the circuit will be reduced. In other words, there might be a tradeoff between the number of different state preparation devices and the number of gate elements in the computational circuit.

\begin{figure}
    \centering
    \includegraphics[width=\linewidth]{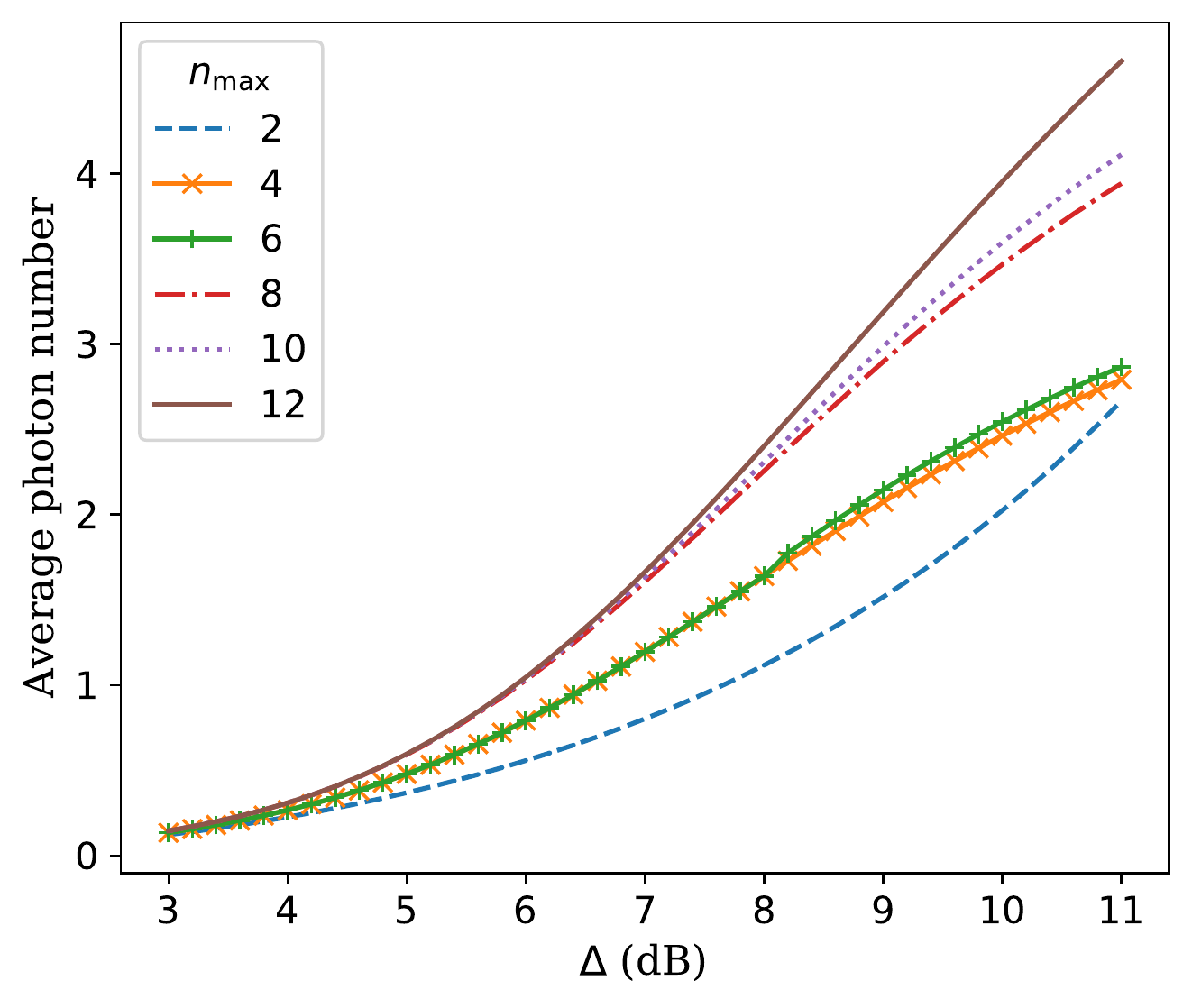}
    \caption{ Average photon number of the approximate states $\Ket{0_A}$ as a function of $\Delta$ (dB), the parameter that characterizes the target state $\Ket{0_{\Delta}}$ that $\Ket{0_A}$ is designed to approximate. Line colours/styles reflect different values of $n_{\max}$ for the core state used to construct $\Ket{0_A}$. As expected, higher quality states -- those with larger $\Delta$ in dB -- require more energy.}
    \label{fig:navg vs n,D}
\end{figure}

\subsubsection{Orthogonality}
For the ideal GKP states, the logical 0 and 1 are orthogonal; however, for physically realizable GKP states, $|\psi_G\rangle$, they have nonzero overlap due to the tails of the wavefunction existing outside the bins. We can check the orthogonality of the approximate 0 and 1 GKP states to see how close it is to the orthogonality between the normalizable 0 and 1 states. This is valuable even when the approximate states do not have high fidelity to the normalizable states; if their overlap is small, they can still be used to encode a qubit, although they may not preserve the error-correcting properties of the GKP states nor the simplicity of the canonical gate implementation. A small overlap is also the most basic necessary condition for error correction, as we will require a low probability of mistaking a 0 for a 1. 

\begin{figure}
    \centering
    \includegraphics[width=\linewidth]{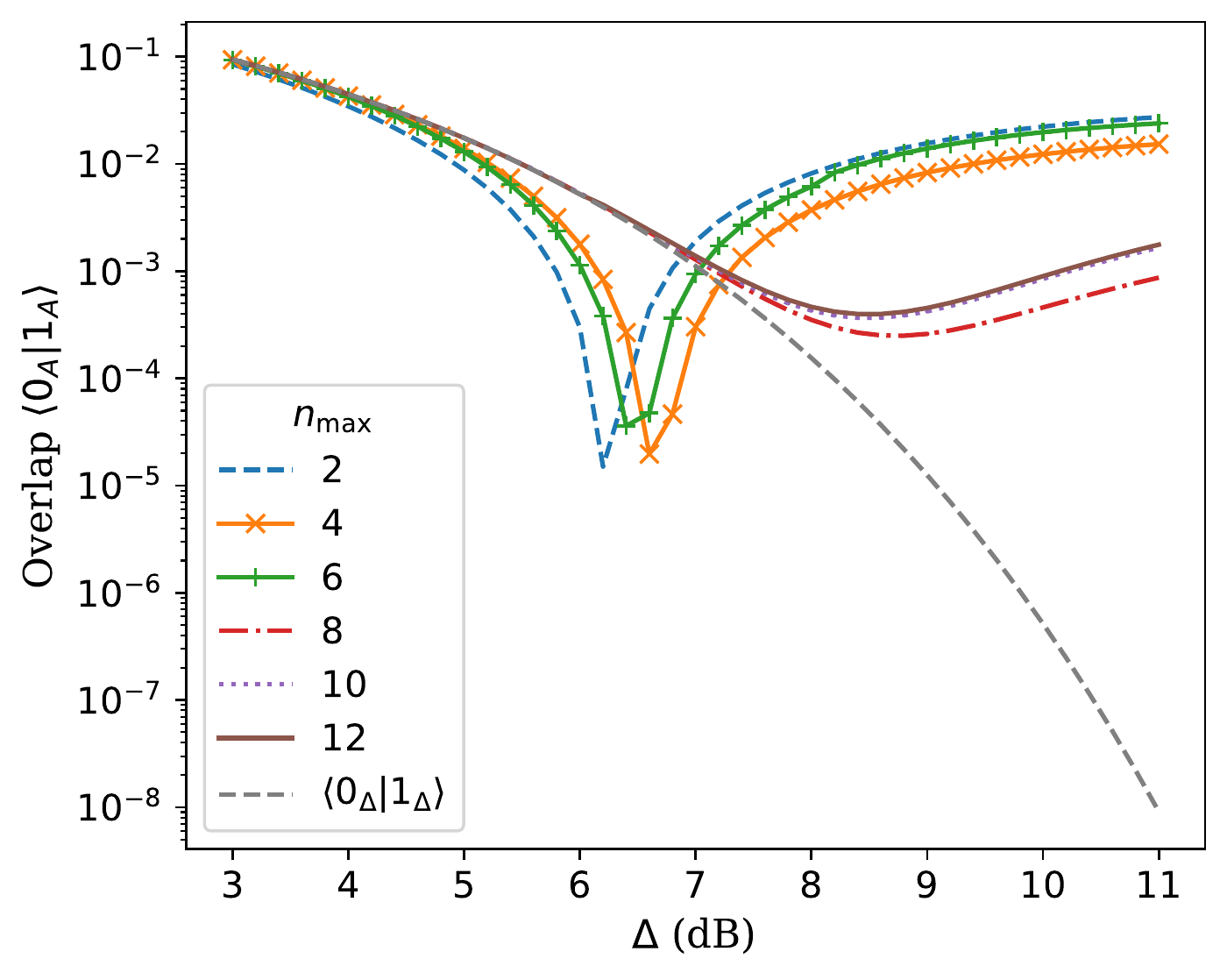}
    \caption{ Overlap $\Braket{1_A|0_A}$ between the approximate logical states as a function of $\Delta$ (dB), the parameter that characterizes the GKP states that the approximate states are targeting. Line colours/styles reflect different values of $n_{\max}$ for the core state used to construct $\Ket{0_A}$. For comparison, we also plot the overlap $\Braket{1_\Delta|0_\Delta}$ of the target states (grey/dashed). A small overlap is a minimal condition for being able to distinguish logical states.
    }
    \label{fig:overlap}
\end{figure}

In Fig.~\ref{fig:overlap} we plot the overlap $\Braket{1_A|0_A}$ as a function of $\Delta$, for different $n_{\max}$. We see that all the states roughly follow the trend of the target overlap for low dB values of $\Delta$, but as the squeezing of the target states increases, the approximate states become less orthogonal.  This is because as $\Delta$ increases, the fidelity gets worse for both $\Ket{0_A}$ and $\Ket{1_A}$ relative to their target states; specifically, while for larger $\Delta$ the $\Ket{\mu_\Delta}$ states consist of smooth, narrow peaks, the truncated core of the $\Ket{\mu_A}$ states mean that their wavefunctions have additional oscillation between the peaks, where they should instead be close to zero (see Fig.~\ref{fig:gkwavefunction} for an example wavefunction). This now increases the overlap $\Braket{1_A|0_A}$ because they gain non-zero contributions at $q$ values between the peak locations.

\subsubsection{Glancy-Knill property}\label{subsubsec:glancy}

\begin{figure}
    \centering
    \includegraphics[width=\linewidth]{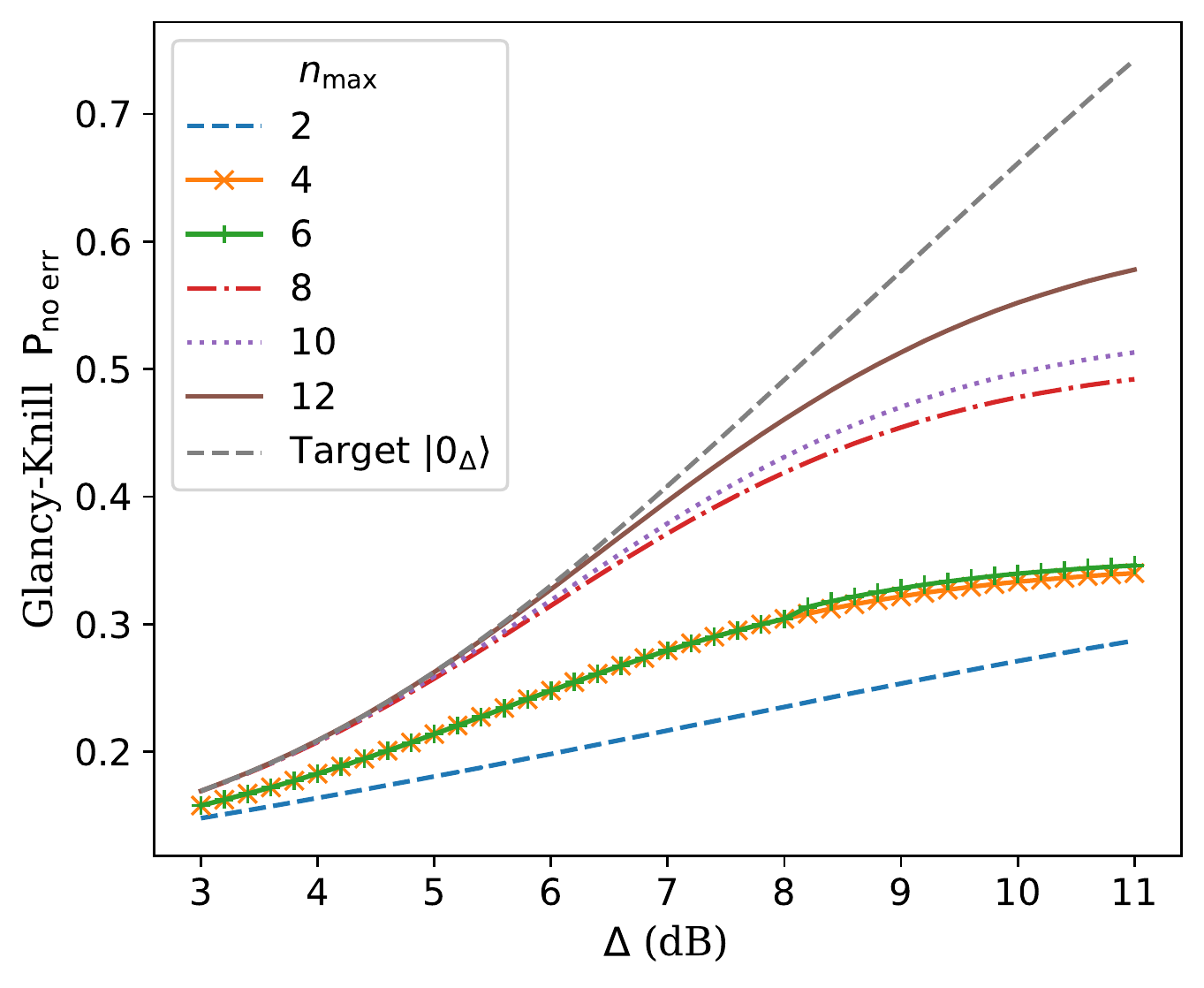}
    \caption{ The Glancy-Knill property, $P_\text{no err}$, which characterizes the probability a state would not yield a logical error if used as an ancilla for Steane error correction with GKP states (see Sec.~\ref{subsec:error_correction} for definition and details). The line colours/styles reflect $P_\text{no err}$ for the various $\Ket{0_A}$, each constructed with a different $n_{\max}$ in  the core state. We vary $\Ket{0_A}$ with $\Delta$, which parametrizes $\Ket{0_{\Delta}}$, the state $\Ket{0_A}$ approximates. Additionally, we provide $P_\text{no err}$ for $\Ket{0_{\Delta}}$ (grey/dashed). For $\Delta = 11$ dB and $n_{\max}=12$, we see that, even though $\Ket{0_A}$ can have 93\% fidelity to $\Ket{0_{\Delta}}$ (see Fig.~\ref{fig:fid vs n,D}), $P_\text{no err}$ drops by 16\%.}
    \label{fig:glancy}
\end{figure}

Using the Glancy-Knill condition in Eq.~\eqref{eq:glancy}, we compute $P_\text{no err}$ for $\Ket{0_A}$ with results depicted in Fig.~\ref{fig:glancy}. Even though, for example, $\Ket{0_A}$ with $n_{\max}=12$ has higher than 93\% fidelity for all $\Delta$ considered, this can still translate to a drop in $P_\text{no err}$ from a target value of 74\% to 58\%. One must therefore be cautious in how one constructs approximate GKP states, as their error correcting properties may differ significantly. While physical fidelity depends on the specific choice of targeted finite-energy GKP state, there may be some global properties, such as the Glancy-Knill condition or logical fidelity, that are more valuable than the choice of representation.

\begin{figure}
    \centering
    \includegraphics[width=\linewidth]{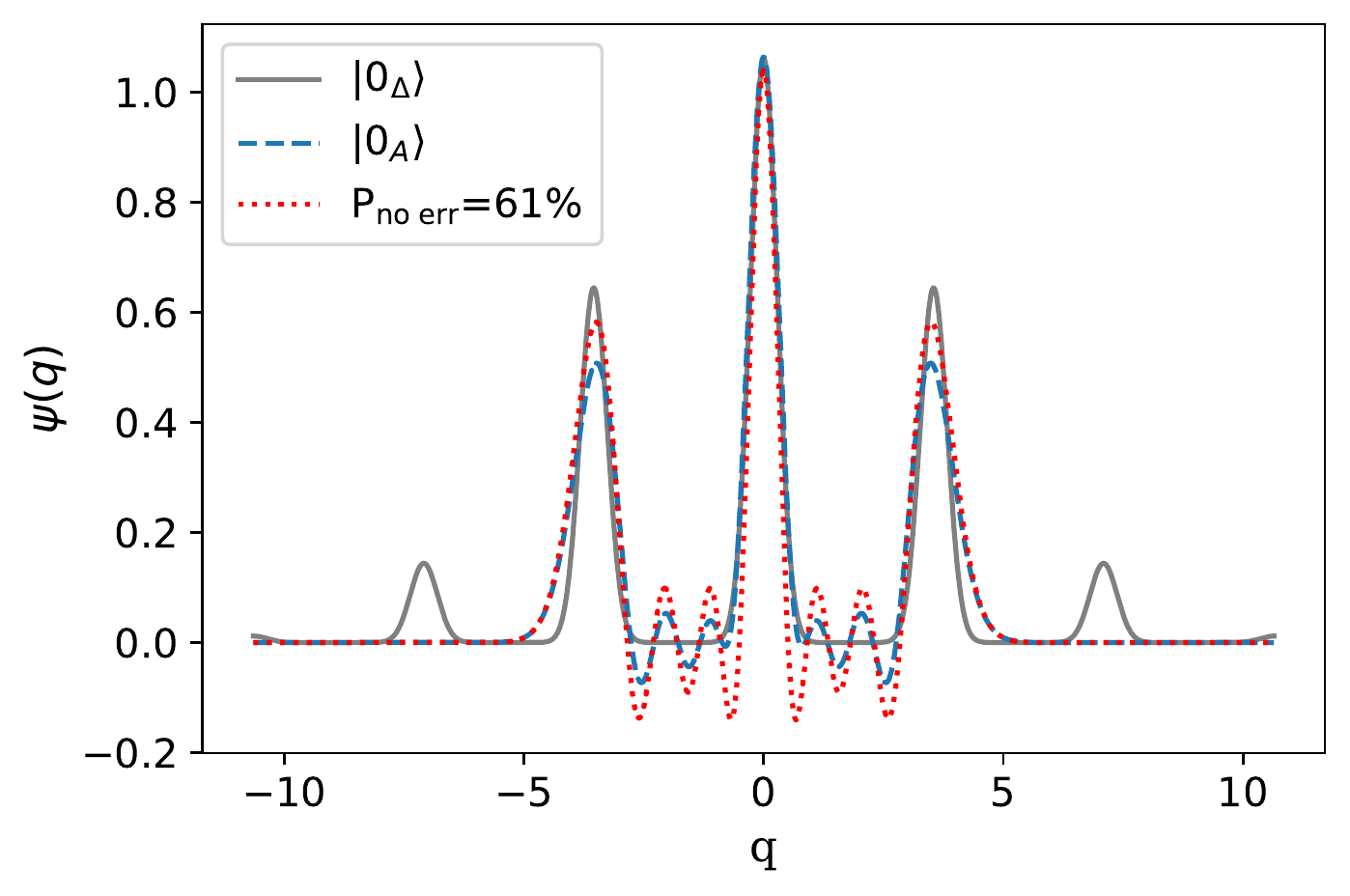}
    \caption{ Examples of different normalizable and approximate GKP wavefunctions and their performance under the Glancy-Knill property \eqref{eq:glancy}: $\Ket{0_{\Delta}}$ with $\Delta = 11$ dB, which yields $P_\text{no err} = 74\%$ (grey/solid); $\Ket{0_A}$ with $n_{\max} = 12$ that achieves $P_\text{no err} = 57\%$ by maximizing fidelity to $\Ket{0_{\Delta}}$ with $\Delta = 11$ dB (blue/dashed); and the state obtained from optimizing directly with respect to $P_\text{no err}$ with core states of $n_{\max}=12$, giving $P_\text{no err} = 61\%$ (red/dotted). Although the red/dotted wavefunction has higher $P_\text{no err}$, the blue/dashed wavefunction is better confined to the bin structure of the GKP states, meaning it has better logical encoding when using the modular subsystem decomposition which only depends on the bin structure of the wavefunction.}
    \label{fig:gkwavefunction}
\end{figure}

This motivates examining an additional question: given an $n_{\max}$, what is the best $P_\text{no err}$ one can achieve? This question does not require defining a target state, since the cost function in Algorithm \ref{algone} is simply replaced with $P_\text{no err}$. We found that, using a core state with $n_{\max} = 12$ and a squeezing of $r\approx 1.87$ dB, we could modestly increase $P_\text{no err}$ from 57\% to 61\%. In Fig.~\ref{fig:gkwavefunction}, we plot three wavefunctions corresponding to three different values of $P_\text{no err}$: $\Ket{0_{\Delta}}$ with $\Delta = 11$ dB, which yields $P_\text{no err} = 74\%$; $\Ket{0_A}$ with $n_{\max} = 12$ that achieved $P_\text{no err} = 57\%$ by maximizing fidelity to $\Ket{0_{\Delta}}$ with $\Delta = 11$ dB; and the state obtained from optimizing directly with respect to $P_\text{no err}$ and employing core states with $n_{\max}=12$, giving $P_\text{no err} = 61\%$. One interesting feature of the last wavefunction is that it has more support outside of the logical bins than the $\Ket{0_A}$ state we plotted. This means that, although $\Ket{0_A}$ is worse for error correction, it is better at encoding information in the logical subsystem of the modular decomposition. An additional issue with only using $P_\text{no err}$ as the cost function is that $P_\text{no err}$ can be perfectly satisfied by ideal GKP states, so the optimization procedure may be overly demanding and push towards the ideal states which we know to be non-normalizable, while we know that there exist finite energy GKP states that are suitable for computation \cite{clusterFT}.

It should also be noted that the Glancy-Knill condition is specifically a benchmark for using GKP states in error correction in quantum computation. It does not, for example, address the utility of GKP in quantum communication, where the values of $\Delta$ required for useful states are much lower. In~\cite{albert2018performance} the authors showed that GKP states worked best for correcting noise resulting from a loss channel. For states with an average photon number of less than two photons, square lattice GKP states with $\Delta = 6.4$ dB were shown to be better than codes designed to correct errors due to loss. For such a $\Delta$, with $n_{\max} = 12$, we found the fidelity of $\Ket{0_A}$ to $\Ket{0_{\Delta}}$ to be greater than 99.9\%.

\subsubsection{Logical subsystem Bloch sphere}

While the Glancy-Knill condition quantifies the error-correcting capability of the approximate GKP states, examining the logical subsystem in the modular decomposition provides a benchmark for the encoding properties of the states. These are two distinct characteristics: for example, if one takes an ideal GKP state and blurs it such that the delta peaks now become distributions over position, as long as those distributions are confined to the original modular bins of the delta functions of width $\sqrt{\pi}$, then the logical information has not been disturbed, since the results of a binned homodyne measurement will be the same. However, the states might be inadequate for error correction, since the blurring could easily extend beyond the $\sqrt{\pi}/6$ threshold.

\begin{figure}
    \centering
    \includegraphics[width=\linewidth]{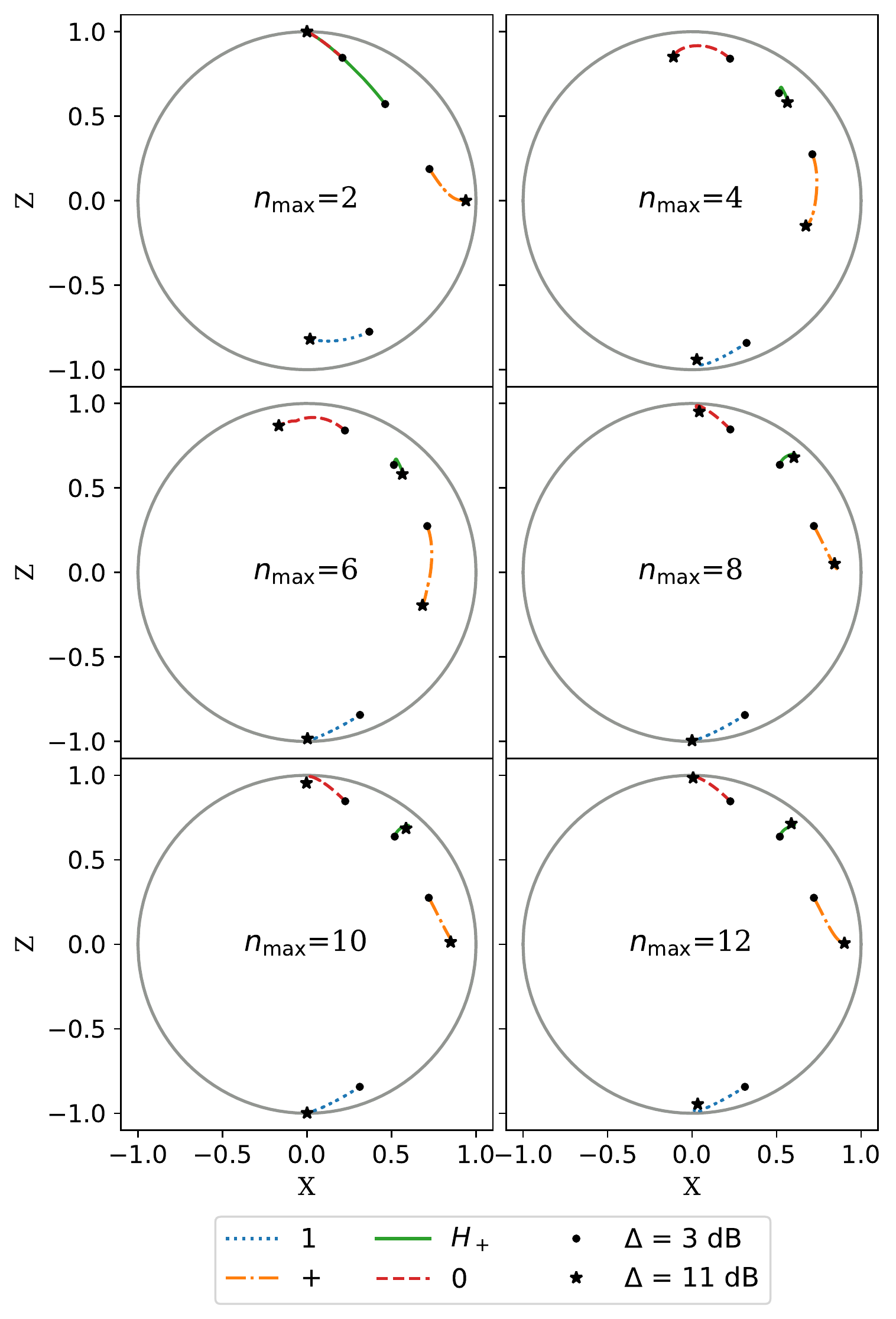}
    \caption{ Location of the logical subsystem of the approximate states $\Ket{\mu_A}$, where $\mu = 0,1,+,H_+$, on the X-Z slice of the Bloch sphere. The trajectories are formed as the $\Ket{\mu_A}$ vary with $\Delta$, which parametrizes the states $\Ket{\mu_\Delta}$ that the $\Ket{\mu_A}$ approximate. We denote where the trajectories begin and end using a point for $\Delta = 3$ dB and a star for $\Delta = 11$ dB. We repeat the plot for different choices of $n_{\max}$ for the core states of $\Ket{\mu_A}$. While in certain cases $\Ket{0_A}$ and $\Ket{+_A}$ can be mapped correctly on the Bloch sphere even with $n_{\max}=2$, if we want all $\mu$ to be well-mapped, we need core states with higher $n_{\max}$.}
    \label{fig:bloch approx}
\end{figure}

We obtain the logical subsystem state from Eq.~\eqref{eq:logical state}, except instead of $E(\epsilon)$ the error operators correspond to the envelopes associated with each approximate state. In Fig.~\ref{fig:bloch approx}, we plot the trajectories of the logical subsystems of the $\Ket{\mu_A}$ states on the Bloch sphere as the target $\Delta$ is varied for different $n_{\max}$. We see that, in some cases, the logical information can be relatively close to the target position on the Bloch sphere even when the fidelity is quite low. For instance, with $n_{\max} = 2$, for $\Ket{0_A}$ and $\Ket{+_A}$, the approximate states are basically squeezed states in $q$ and $p$. While this may not compromise the logical information in those states, we know that applications of Gaussian operations will keep the states almost Gaussian, and so we would not expect them to be suitable for universal computation. Combining the Bloch sphere picture with the Glancy-Knill results, we find that the better approximate states are provided, unsurprisingly, by increasing $n_{\max}$ and the target $\Delta$.

\subsection{Circuits for GKP state preparation}\label{subsec:circuits}
\subsubsection{Algorithm for optimal circuits}

Given the representation and characterization of approximate GKP states as single-mode Gaussian operations applied to truncated core states, we can now design GBS devices for producing the approximate states. As described in Sec.~\ref{subsec:nongauss}, our state preparation framework is to apply an interferometer, $U(\bar{\boldsymbol{\Theta}})$, to $N$ modes of displaced, squeezed vacuum states, $D(\boldsymbol{\alpha})S(\boldsymbol{z}\,)|\mathbf{0}\rangle$, perform a PNR measurement on $N-1$ modes, and postselect on a specific photodetection pattern, $\bar{\textbf{n}}$, to obtain the target state in the $N^{\text{th}}$ mode. 

As we summarize in Algorithm \ref{alg2} (see complete details in App.~\ref{subsec:alg2det}), for a given truncated core state, we train $\boldsymbol{\alpha},\boldsymbol{z},\bar{\boldsymbol{\Theta}}$ and $\bar{\mathbf{n}}$ for fixed numbers of modes using machine learning algorithms, the \textsf{strawberryfields}~\cite{Killoran2019} and \textsf{the walrus} simulators~\cite{Bjorklund2018, Gupt2019}. Even once we find the optimal circuit for producing the truncated core state, we still have to include the squeezing, $S(r)$; to avoid the need for inline squeezing, we can take the Gaussian unitary $\tilde{U} = S(r)U(\bar{\boldsymbol{\Theta}})D(\boldsymbol{\alpha})S(\boldsymbol{z})$ applied to vacuum states on all the modes and express it according to the Euler or Bloch-Messiah \cite{euler} decomposition as an equivalent Gaussian unitary of the form $U(\bar{\boldsymbol{\Theta}}')D(\boldsymbol{\alpha'})S(\boldsymbol{z}')$, where now all the squeezing is relegated to the source. This framework for training the circuits is analogous to the process described in~\cite{saba1}.

Since the truncated core states only have even Fock coefficients, we do not need to implement the initial displacement on each of the modes. We use the rectangular decomposition from \textsf{strawberryfields} to decompose $U(\bar{\boldsymbol{\Theta}})$ as an $N$-mode interferometer with $N(N-1)$ independent beamsplitter and phase-shifter parameters. While a completely general interferometer will also include rotations on each of the modes after the application of all the beamsplitters, since we are performing PNR measurements on $N-1$ of the modes, we neglect these rotations.

\begin{algorithm}[H]
\caption{Optimal Circuits for Approximate States}\label{alg2}
\begin{algorithmic}

\Function{cost}{$\boldsymbol{c}$, \textbf{x}, $\bar{\textbf{n}}$, handle} 
    
    $|\psi_\text{target}\rangle \gets \sum_n c_n |n\rangle$
    
    \# \textbf{x} = ($\boldsymbol{z},\bar{\boldsymbol{\Theta}}$)
    
    $|\psi_\text{out}\rangle \gets \langle \bar{\textbf{n}}| U(\bar{\boldsymbol{\Theta}})S(\boldsymbol{z}) |\textbf{0}\rangle$
    
    $\text{prob} \gets \langle\psi_\text{out}|\psi_\text{out}\rangle$
    
    $\text{fid} \gets |\langle \psi_\text{out}|\psi_\text{target}\rangle|^2/\text{prob}$
    
    \If{handle = `global'} 
    
        \Return $\text{fid} + 0.1\times \text{prob}$
        
    \ElsIf{handle = `local'} 
    
        \Return $\text{fid} + \text{prob}$
        
    \EndIf

\EndFunction

\Procedure{optimization}{$\boldsymbol{c}$, $\bar{\textbf{n}}$, N}

    \# for an $N$-mode circuit

    \# random squeezing and BS angles within ranges

    \textbf{initialize} $\textbf{x}_0$ = ($\boldsymbol{\zeta},\bar{\boldsymbol{\Theta}}$) 

    \# \textsf{basinhopping} is a global search algorithm

    $\textbf{x}_1 \gets \textsf{basinhopping}[\textbf{x}_0,\texttt{cost},args = (\bar{\textbf{n}}, \text{`global')}]$

    \# \texttt{local\_search} is a local maximization algorithm

    $\textbf{x}_2 \gets \texttt{local\_search}[\textbf{x}_1,\texttt{cost},args = (\bar{\textbf{n}}, \text{`local'})]$
 
\EndProcedure

\Procedure{redecompose\_circuit}{r,$\textbf{x}_2$}

    \# $|\psi_\text{approx}\rangle = S(r)\sum_n c_n |n\rangle$
    
    \# squeezing only applied to mode with $|\psi_\text{out}\rangle$
    
    $\boldsymbol{r} \gets (r, 0, ..., 0)$
    
    $\tilde{U} \gets S(\boldsymbol{r})U(\bar{\boldsymbol{\Theta}})S(\boldsymbol{z})$
    
    \# given a Gaussian unitary applied to vacuum, \texttt{euler} returns circuit parameters in the form $U(\bar{\Theta'})S(\boldsymbol{\zeta'})$
    
   $(\boldsymbol{z}',\bar{\boldsymbol{\Theta}}') \gets \texttt{euler}(\tilde{U})$

\EndProcedure

\end{algorithmic}
\end{algorithm}

\subsubsection{Circuit optimization results}

We now employ Algorithm~\ref{alg2} to find the circuits to produce $\Ket{0_A}$ for $\Delta =$ 10 dB, for $n_{\max} =$ 4, 8 and 12 photons. The minimum number of circuit modes we examined was 2, while for $n_{\max} =$ 4, 8, and 12, we examined circuits with up to 3, 4, and 5 modes, respectively. For each circuit, we checked all PNR measurement patterns $\bar{\mathbf{n}}$, such that the number of photodetections summed to $n_{\max}$. We restricted our squeezing parameter search to, at most, 12 dB of squeezing, which is within the state of the art~\cite{squeezing}. In nearly all cases, we found that the optimal fidelities were achieved by saturating the squeezing in at least one of the modes to $z \sim \pm$ 12 dB. We can increase the search over larger squeezing values in a straight-forward manner.

In Table~\ref{tab:fidprob1}, we provide our results for the best fidelities found using Algorithm~\ref{alg2} for different circuit sizes and values of $n_{\max}$. Additionally, we provide some other numerical results that still returned fidelities above 99\%, but with modestly higher probabilities. Even though increasing $N$ yields more independent parameters to tune to the target state, we see from our results that the increase in fidelity gained beyond three modes is marginal, with the corresponding probability of success vanishing. 

\begin{figure*}
    \centering
    \includegraphics[width=\textwidth]{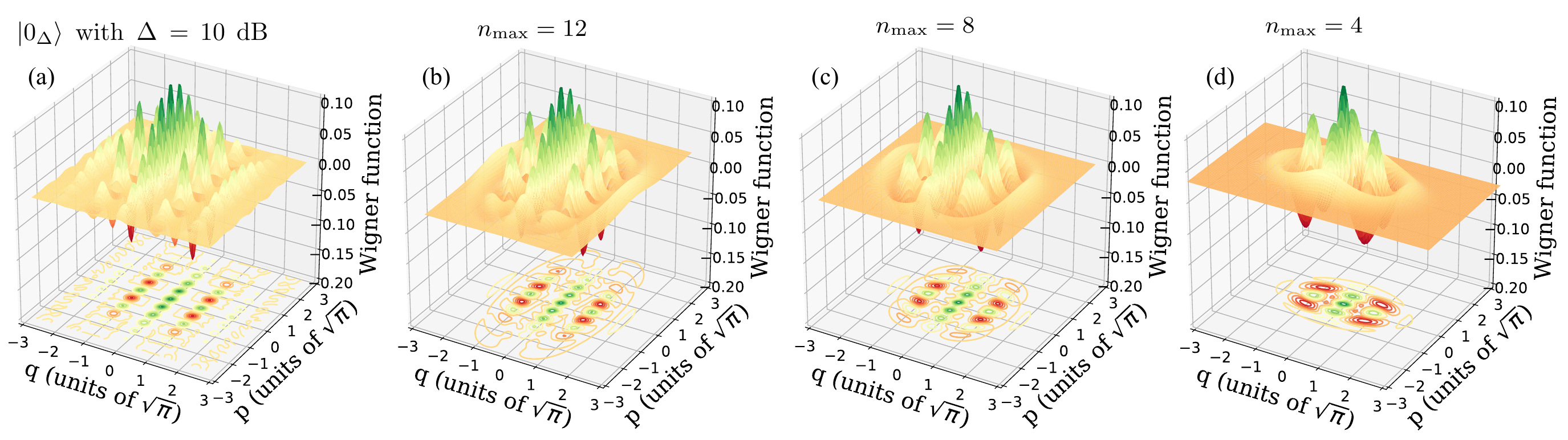}
    \caption{Wigner functions for (a) $\Ket{0_{\Delta}}$ with $\Delta =$ 10 dB, as well as for the optimal states output by the three-mode GBS devices designed to produce $\Ket{0_A}$ with $n_{\max} =$ (b) 12, (c) 8, and (d) 4 photons. These correspond to the starred results for $N=3$ in Table \ref{tab:fidprob1}. We see the peak structure gets better with increasing $n_{\max}$, but differences to $\Ket{0_{\Delta}}$ are still apparent further from the origin in phase space.}
    \label{fig:wigner}
\end{figure*}

In Fig.~\ref{fig:wigner}, we plot the Wigner function of $\Ket{0_{\Delta}}$ with $\Delta =$ 10 dB, as well as the Wigner functions of the optimal states (highest fidelity) output by the three-mode circuits designed to produce $\Ket{0_A}$ with $n_{\max} =$ 4, 8 and 12 photons. These correspond to the starred results for $N=3$ in Table~\ref{tab:fidprob1}. We see that, with increasing $n_{\max}$ -- that is, as the core state resource improves -- the number and sharpness of peaks approaches to that of $\Ket{0_{\Delta}}$. The difference is smallest near the origin in phase space.

\begin{table}
    \centering
    (a) $n_{\max}=4$
    
    \begin{tabular}[t]{ c | c | c | c }
        \hline
        $N$ & $1-$(Fidelity to $\Ket{0_A}$)  & Probability & $\bar{\mathbf{n}}$ \\
        \hline
        \hline
        2 & 0.33* & 6.8\% & (4)\\
        3 & $1\times10^{-5}$* & 2.1\% & (1,3)\\
          & $3\times10^{-4}$ & 2.2\% & (2,2)\\
        \hline
    \end{tabular}
    
    \vspace{0.3cm}

    (b) $n_{\max}=8$
    
    \begin{tabular}[t]{ c | c | c | c }
        \hline
        $N$ & $1-$(Fidelity to $\Ket{0_A}$) & Probability & \textbf{$\bar{n}$} \\ 
        \hline
        \hline
        2 & 0.34* & 4.7\% & (8)\\
        3 & $1\times10^{-3}$* & 0.41\% & (4,4)\\
        4 & $2\times10^{-6}$* & 0.14\% & (2,2,4)\\
          & $5\times10^{-6}$ & 0.19\% & (1,3,4)\\
        \hline
    \end{tabular}
    
    \vspace{0.3cm}

    (c) $n_{\max}=12$
    
    \begin{tabular}[t]{ c | c | c | c }
        \hline
        $N$ & $1-$(Fidelity to $\Ket{0_A}$)  & Probability & \textbf{$\bar{n}$} \\ 
        \hline
        \hline
        2 & 0.35* & 2.3\% & (12)\\
        3 & $3\times10^{-3}$* & 0.11\% & (5,7)\\
        4 & $4\times10^{-8}$* & $5.5\times10^{-5}$ & (3,3,6)\\
          & $2\times10^{-5}$ & $2.3\times10^{-4}$ & (2,4,6)\\
        5 & $7\times10^{-9}$* & $6.5\times10^{-5}$ & (1,1,3,7)\\
          & $7\times10^{-8}$ & $7.2\times10^{-5}$ & (1,2,3,6)\\
        \hline
    \end{tabular}
\caption{ Results for the GBS circuits optimized to produce an approximate GKP state $\Ket{0_A}$ constructed to approach $\Ket{0_{\Delta}}$ with $\Delta = 10$ dB. We present, as a function of number of circuit modes $N$, the best fidelities (starred) along with other points of comparably high fidelity and probability found using Algorithm \ref{alg2}, with corresponding probabilities and PNR measurement patterns $\bar{\mathbf{n}}$. We examine $\Ket{0_A}$ with core states of $n_{\max} =$ (a) 4, (b) 8, and (c) 12 photons.}\label{tab:fidprob1}
\end{table}

\subsection{Experimental imperfections}\label{subsec:exp_imperf}
\subsubsection{Stability analysis of optical elements}
\begin{figure}
    \centering
    \includegraphics[width=\linewidth]{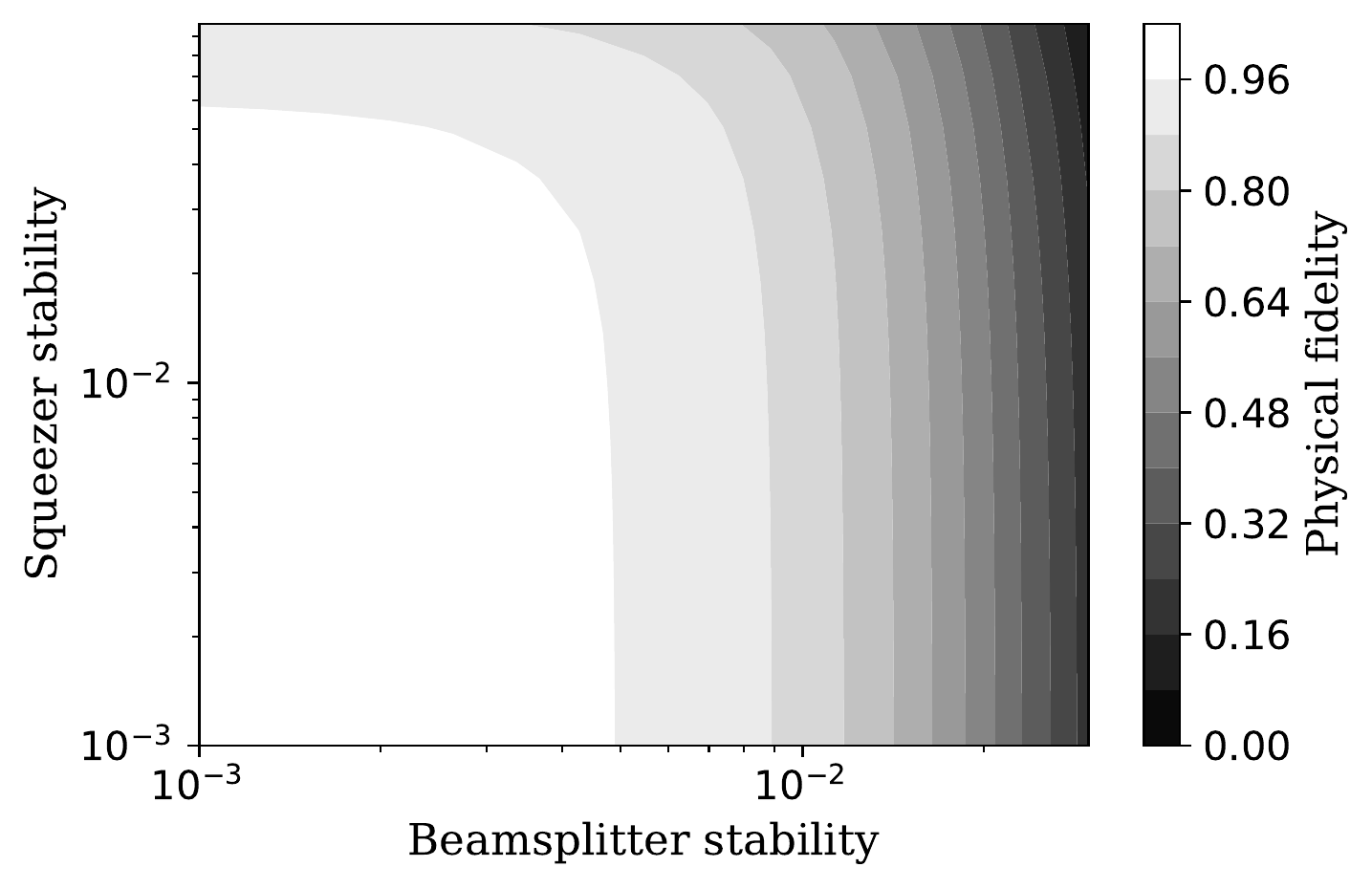}
    \caption{ Stability analysis for the optimal 3-mode circuit designed to produce  $\Ket{0_A}$ with a core state of $n_{\max} = 12$ targeting $\Ket{0_{\Delta}}$ with $\Delta = 10$ dB. We find the worst case fidelity between the circuit output and $\Ket{0_A}$ as a function of the circuit beamsplitter and squeezer stability parameters, which are defined as the relative error to the ideal parameters.}
    \label{fig:stability}
\end{figure}

The stability of the numerically computed optimal circuit parameters is important for experiment since, for example, one might have a given uncertainty in tuning the initial squeezing parameters and beamsplitter angles. As a benchmark of solution stability, we can find the worst-case fidelity within a small region in parameter space about the optimal solution. To illustrate this, let us take from Table~\ref{tab:fidprob1} as an initial guess the optimal solution for a three-mode circuit designed to produce the approximate state $\Ket{0_A}$ with a core state of $n_{\max} = 12$. We can then modify Algorithm~\ref{alg2} to minimize the fidelity and to only search within a region set by stability parameters. That is, given the optimal squeezing and beamsplitter parameters, $r_\text{opt},\theta_\text{opt},\phi_\text{opt}$, we search for the worst fidelity in a region $r_\text{opt}\pm\delta r/r_\text{opt},\theta_\text{opt}\pm\delta \theta/\theta_\text{opt},\phi_\text{opt}\pm\delta \phi/\phi_\text{opt}$. In Fig.~\ref{fig:stability} we show how much fidelity could change as a function of squeezing stability, $\delta r/r_\text{opt}$, and beamsplitter stability, $\delta \theta/\theta_\text{opt} = \delta \phi/\phi_\text{opt}$. We see that a much larger instability in the squeezing parameters can be tolerated compared to the beamsplitter parameters.

\subsubsection{Effects of photon loss}
\begin{figure}
    \centering
    \includegraphics[width=\linewidth]{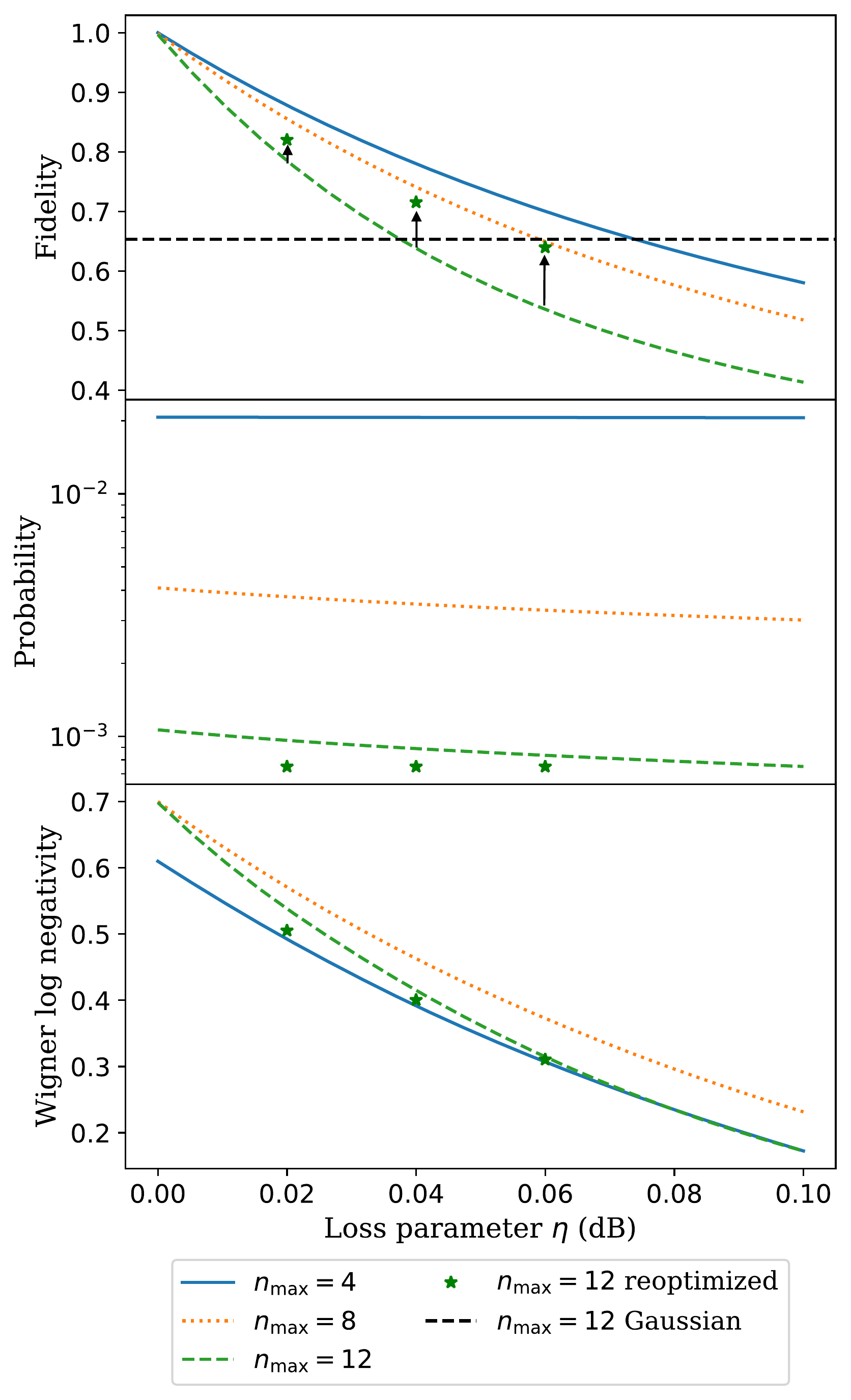}
    \caption{The effect of lossy optical components on the optimal GBS devices for GKP state preparation. We examine how the optimal circuits for $N=3$ modes from Table \ref{tab:fidprob1} for core states of $n_{\max}=4,8,12$ perform as a function of a single loss parameter, $\eta$. Our loss model consists of applying a loss channel parametrized by $\eta$ to each optical component, as well as to the mode outcoupling and PNR detectors. We plot how the fidelity, probability and Wigner log negativity change as loss increases. Additionally, for a few values of loss, we plot reoptimized results for lossy circuits designed to produce $n_{\max}=12$ (green stars with arrows indicating which values were reoptimized). We see an increase in fidelity as a result. We stop reoptimization once fidelity is on the same order as that of the Gaussian state with highest fidelity to the target, a threshold we plot with a dashed black line.}
    \label{fig:loss}
\end{figure}

In an experimental implementation of a GBS circuit device for state preparation, there will inevitably be loss in the optical components. Here we examine how such loss affects the results of the optimal circuit GKP states. We employ a simple loss model for our circuit: after each squeezer and each beamsplitter (with a complex transmissivity) of the interferometer in the rectangular decomposition, we apply a loss channel with loss parameter $\eta$. As the squeezers act on vacuum, they represent the only source of input light, so we capture the effect of lossy sources. Additionally, at the end of each mode, we apply a circuit out-coupling loss, $\eta$; this is followed by loss, also of magnitude $\eta$, before each PNR detector to account for detector inefficiency. The loss channel is modelled by coupling a beamsplitter of transmissivity $\sqrt{\eta}$ to an ancillary mode and then tracing out the mode. In Fig.~\ref{fig:loss}, we plot how the fidelity, probability, and Wigner log negativity of the optimal $N=3$ mode solutions from Table~\ref{tab:fidprob1} change as a function of the single loss parameter $\eta$. Notably, we see that with increasing $n_{\max}$, loss becomes increasingly detrimental to fidelity, as it affects higher-photon number components. The probability remains relatively stable, while the Wigner log negativity also decreases with higher loss.

As a proof-of-concept, we also reoptimize some of the circuits in the presence of loss. For the three-mode circuit designed to produce the $n_{\max}=12$ core state for $\Delta = 10$~dB, we re-ran a modified version of Algorithm~\ref{alg2} to find optimal circuits in the presence of loss. One change we made was to include Wigner log negativity in the cost function to ensure the states had non-Gaussian properties. Additionally, we skip the step of redecomposing the circuit, which we discuss in the next paragraph. In Fig.~\ref{fig:loss}, we plot the reoptimized results for $n_{\max}=12$ for three values of $\eta$. We stop at $\eta = 0.06$ dB because by that point we have dropped below the fidelity that can be achieved with only a Gaussian state, a threshold which we indicate with a black dashed line. Although the state still has nonzero Wigner log negativity, being in the regime where Gaussian states are approximating the state just as well as non-Gaussian ones significantly hampered our search for optimal states. 

There are several complications to our strategy for constructing GKP states with GBS devices that arise in the presence of loss. First, as already alluded to, there is the question of how to include the squeezing, $r$, applied to the core state, since the Euler decomposition is only valid for circuits without loss, so it cannot be applied perfectly in this situation. One option is to employ inline squeezing after the core state is produced by the lossy circuit, but this can be difficult experimentally \cite{inline_sq}. Another (suboptimal) strategy is to find the optimal circuit for preparing the core state in the presence of loss, remove the loss channels, add the squeezing $r$, redecompose the circuit with the Euler decomposition, and then finally reinsert the loss channels. Since the $r$ values applied to the core states of $|0_A\rangle$ with $n_{\max}=4,8,$ and 12 are relatively small (less than 3 dB), we do not expect a large drop in fidelity using this strategy. For example, we used this strategy for the reoptimized result in Fig.~\ref{fig:loss} with $\eta = 0.02$ dB, and found that the fidelity only decreased from 82\% to 81\%. 

A second complication to our strategy is the need to revisit the restriction that $n_{\max}$ photodetections ought to be detected to prepare a Fock superposition up to $n_{\max}$ with high fidelity. In the presence of loss, we might expect fewer photodetections are required as some photons could have been lost in the circuit. This leads to many more detection patterns to check, which we leave open to future study.

\section{Conclusions}
In this paper we have investigated, in considerable detail, the methodology for tracking abstract qubit computation encoded into grid states, which are among the earliest and most prominent bosonic codes. A bird's-eye view of our analysis reveals an attempt to understand how realistic and practical considerations affect the quality of the encoded qubit and thereby the information that it carries. Any general qubit computation can be broken down into state preparation, gate implementation, and finally a measurement readout. In the ideal case of a GKP or grid code, the state preparation is the only non-Gaussian resource required. The operations on these ideal states are specific Gaussian gates that implement Clifford unitaries on the encoded qubits, and the Pauli measurements can be realized through homodyne measurements. Our focus is solely on state preparation errors; we assume that our gates and measurements are error-free. We are able to track how an imperfect state preparation manifests itself in the computation and provide ideas around compilation and error mitigation for the physical implementation of the circuits.

To design good error correction protocols, we first need to adequately characterize and monitor the errors. To this end we deploy the modular subsystem decomposition, trace out the gauge mode, and follow a logical qubit encoded in a grid state through a computation. The qubit's journey begins with the ideal, infinite-energy states, which become physical following the application of an error map; we present several ways this can be done. The qubits then continue their travels through single and two-mode gates. We note that there are many physical operations that lead to the same logical operation on the encoded qubit, meaning a modified real-time mapping or a circuit compilation is necessary to find the best physical gates for the implementation of an abtract qubit circuit. What is meant by ``best'' is facilitated by the various figures of merit that we introduce in both the physical and logical space, the latter giving us a direct indicator of the information content. Finally, we offer a closer look at the standard Steane-type error correction on GKP qubits, including a discussion of recalibrating the gauge mode alongside an error-correction step.

In the second half of the paper, we focus on obtaining the explicit resources required to produce a class of approximate grid states using a photonic platform. More specifically, we find the fidelity-probability trade-off of producing these single-mode approximate GKP qubits for different circuit sizes, the circuits being measured in all but one of the modes of  multimode Gaussian states using PNR detectors. We provide detailed optimization tools for producing the target states and various ways the approximations manifest, such as in the non-orthogonality of the qubits. The stability analysis of these numerical optimization schemes suggest that the states output by the circuit are more robust to uncertainty in squeezing than beamsplitter angles. We furthermore investigate how loss decreases the quality of our approximate GKP states, and show how circuits can be reoptimized in the presence of loss. In addition to looking at the normalizable states through the figures of merit introduced above, we give a few more metrics that could further inform the suitability of these noisy qubits for computation.

In the lead-up to the implementation of useful algorithms on near-term noisy devices, we believe that the analytical and numerical toolboxes that we have developed are valuable in the context of bosonic codes. Since bosonic codes are viewed as promising candidates, especially in the context of lossy circuits, our analysis informs the construction of practical error correction protocols, which, when concatenated with more elaborate codes, would lead ultimately to fault-tolerant computation.

\begin{acknowledgements}
We are grateful to Hoi-Kwong Lo, Saikat Guha, Arne L. Grimsmo, Victor V. Albert, and colleagues at Xanadu for helpful discussions. This work is supported by Mitacs through the Mitacs Accelerate program grant. I.~T.\ is supported by an Ontario Graduate Scholarship. J.~E.~B.\ is supported by an Ontario Graduate Scholarship and the Lachlan Gilchrist Fellowship. N.~C.~M.\ is supported by the Australian Research Council Centre of Excellence for Quantum Computation and Communication Technology (Project No.\ CE170100012). All the numerical codes and data are available in Ref.~\cite{code}. For a distilled discussion of the topics in this paper, see the blog post~\cite{medium_blog}.
\end{acknowledgements}

\appendix

\section{Notation, nomenclature, convention, and units}
\label{sec:convention}
\begin{table}\
\begin{tabular}{|c|c|m{4cm}|c|}
\hline
Category & Symbol & \multicolumn{1}{|c|}{Description} & Eq.~\\
\hline 
\hline 
Ideal & $\Ket{\psi_{I}}$ & Infinite energy square-lattice GKP states & (\ref{eq:0I}) \\
\hline 
Normalized & $\Ket{\psi_{G}}$ & $G\Ket{\psi_{I}}$; ideal GKP states normalized with operator $G$ & \eqref{eq:0NG} \\
\cline{2-4} \cline{3-4} \cline{4-4} 
 & $\Ket{\psi_{\epsilon}}$ & $\Ket{\psi_{G}}$ with $G=E\left(\epsilon\right)\equiv e^{-\epsilon\hat{n}}$ & \eqref{eq:0Nen}\\
\cline{2-4} \cline{3-4} \cline{4-4} 
 & $\Ket{\psi_{\Delta,\kappa}}$ & Gaussians of width $\Delta$ enveloped
by a Gaussian of width $\kappa ^{-1}$ &  \eqref{eq:0Ndeltakappa} \\
\cline{2-4} \cline{3-4} \cline{4-4} 
 & $\Ket{\psi_{\Delta}}$ & $\Ket{\psi_{\Delta,\kappa}}$ with $\Delta=\kappa$; equals $\Ket{\psi_{\epsilon}}$
for small $\Delta$ & \eqref{eq:delta=kappa} \\
\hline 
Approximate & $\Ket{\psi_{A}}$ & Approximations to a given choice of normalizable states & (\ref{eq:core})\tabularnewline
\hline 
\end{tabular}
\caption{Notation for various kinds of GKP states referred to in the paper.}
\label{tab:not}
\end{table}

\begin{table}\
    \centering
    \begin{tabular}{c|c|c}
    $\Delta_{\text{dB}}$ & $\epsilon$ & $\frac{\Delta^3}{24} $\\
    \hline
    \hline
    0 & 1 & $2\ \times 10^{-2}$ \\
    2 & 0.631 & $2\ \times 10^{-2}$ \\
    4 & 0.398 & $1\ \times 10^{-2}$  \\
    6 & 0.251 & $5\ \times 10^{-3}$  \\
    8 & 0.158 & $3\ \times 10^{-3}$  \\
    10 & 0.100 & $1\ \times 10^{-3}$ \\
    12 & 0.0630 & $7\ \times 10^{-4}$ \\
    15 & 0.0316 & $2\ \times 10^{-4}$  \\
    20 & 0.0100 & $4\ \times 10^{-5}$
    \end{tabular}
    \caption{Conversion between selected squeezing values expressed in decibels, corresponding to the width of the peaks in normalizable GKP states $\Ket{\psi_\Delta}$ and the epsilon parameter in $\Ket{\psi_\epsilon}$ from Eq.~ $\eqref{eq:0Nen}$. The final column contains the error in the approximation between $\Delta$ and $\epsilon$, expressed as the magnitude of the third-order term in the expansion of $\tanh{\frac{\Delta}{2}}$ (see Eq.~\eqref{eq:tanh}). }
    \label{tab:e_db}
\end{table}

We use this section to disambiguate some of the notation and conventions we use in the paper.

\paragraph{Grid states.} See Table \ref{tab:not} for the notation we use for the different GKP states we refer to in the paper. 

\paragraph{Logical and physical gates.} Gates with a bar, as in $\bar{U}$, always refer to logical gates acting on qubits, and gates without a bar, as in $U$, are always physical gates acting on the oscillator space.

\paragraph{Squeezing.}
The convention for the squeezing parameter $r$ we use is such that the effect of the squeezing gate on the quadrature operators is
\begin{equation}
    S^{\dagger}(r)\hat{q}S(r) = e^{-r} \hat{q}, \qquad
    S^{\dagger}(r)\hat{p}S(r) = e^r \hat{p},
\end{equation}
meaning that, in the position representation,
\begin{equation}\label{eq:sq_conv}
    S(r) \psi(q) = e^{r/2} \psi(e^r q).
\end{equation}

Squeezing values are often expressed in decibels in this paper. The conversion we use for the parameter in the squeezing gate is
\begin{equation}
    r = \frac{\ln{10}}{20} r_\text{dB},
\end{equation}
and for the width $\Delta$ of the normalizable GKP state peaks, it is
\begin{equation}
    \Delta = 10^{-\frac{\Delta_{\text{dB}}}{20}}.
\end{equation}
Although we often work in the regime Delta = $\kappa$, it is $\kappa^{-1}$ that is the width of the overall envelope in the $\Ket{\psi_{\Delta,\kappa}}$ states. Therefore we are interested in low values of both $\Delta$ and $\kappa$ (large positive in dB). 

Lastly, we often use the envelope operator $E(\epsilon)$ from Eq.~\eqref{eq:0Nen} to express our normalizable states. In the regimes we consider, it is true that
\begin{equation}
    \epsilon \approx \Delta^{2},
\end{equation}
meaning
\begin{equation}
    \epsilon  \approx 10^{-\frac{\Delta_{\text{dB}}}{10}}.
\end{equation}
In Eq.~\ref{tab:e_db} we list the conversion between a few squeezing values expressed in $\Delta_\text{dB}$ and $\epsilon$, as well as the error in the approximation.

\section{Derivations}
\subsection{Dilation for the envelope operator \texorpdfstring{$E(\epsilon)$}{TEXT}}\label{subsec:envelope}
As stated in Sec.~\ref{subsec:norm_gkp}, a convenient method for constructing normalizable GKP states is by applying the operator $e^{-\epsilon\hat{n}}$ to the ideal GKP states. According to~\cite{noh2019}, this can be constructed as follows: first pass an ideal GKP state through a beamsplitter of transmissivity $t$ with a vacuum state:
\begin{equation}
\begin{split}
    |\psi_I\rangle|vac\rangle &= \int \frac{d^2\beta}{\pi} \langle \beta|\psi_I\rangle|\beta\rangle|vac\rangle\\
    &\overset{\text{BS}}{\rightarrow} \int \frac{d^2\beta}{\pi} \langle \beta|\psi_I\rangle|t\beta\rangle|r\beta\rangle\
\end{split}
\end{equation}
Then, post-select on measuring the second mode in the vacuum state to obtain

\begin{equation}
\begin{split}
&\int \frac{d^2\beta}{\pi} e^{-|r \beta|^2/2} \langle \beta|\psi_I\rangle|t \beta\rangle\\ =& \int \frac{d^2\beta}{\pi}  \langle \beta|\psi_I\rangle \sum_{n} e^{-|\beta|^2/2}\frac{\beta^n t^n}{n!}|n\rangle\\
 =& t^{\hat{n}} |\psi_I\rangle,
\end{split}
\end{equation}
Finally, make the association $e^{-\epsilon} \to t$.

\subsection{Readout from the logical subsystem}\label{subsec:readout}
A measurement in the computational basis of a qubit can be effected with a binned homodyne measurement of the $q$ quadrature of the corresponding GKP state. Here we show that this translates easily to the subsystem picture, meaning binned homodyne measurements allow direct access to the logical subsystem.

Just as with a measurement in the computational basis, a binned homodyne measurement must yield a binary output. Naturally, for the square lattice, the union of bins of width $\sqrt{\pi}$ centered at $2n\sqrt{\pi}$ ($(2n+1)\sqrt{\pi}$) corresponds to binary output 0 (1). We note that the POVMs corresponding to these bins can naturally be written in terms of displacements of the GKP states:
\begin{equation}\label{eq:log_POVMS}
    M_{\mu} = \int_{-\sqrt{\pi}/2}^{\sqrt{\pi}/2} d\beta X(\beta)|\mu_I\rangle\langle\mu_I| X^\dagger(\beta);\; \mu = 0,1; \beta \in \mathbb{R}.
\end{equation}
Recall the decomposition from Eq.~\eqref{eq:log_decomp}, and note that
\begin{equation}\label{eq:disp_log}
    X(\beta)|\mu_I\rangle = \begin{cases}
    \Ket{\mu}_{\mathcal{L}}\otimes X(\beta)\Ket{+_{I}}_{\mathcal{G}}, & \beta \in \text{0 bins},\\
            \bar{X}\Ket{\mu}_{\mathcal{L}}\otimes X(\beta)\Ket{+_{I}}_{\mathcal{G}}, & \beta \in \text{1 bins},
		 \end{cases}
\end{equation}
taking care to realize that this is not a decomposition of $X(\beta)$ but only a statement about its action on ideal GKP states. Here, the 0 (1) bins refer to the $q$ quadrature region of $[(2n-\frac{1}{2})\sqrt{\pi},(2n+\frac{1}{2})\sqrt{\pi}]$ ($[(2n+\frac{1}{2})\sqrt{\pi},(2n+\frac{3}{2})\sqrt{\pi}]$). 

Thus, Eq.~\eqref{eq:log_POVMS} can be rewritten as:
\begin{equation}
\begin{split}
    M_{\mu} &= |\mu\rangle\langle\mu|_{\mathcal{L}}\otimes\int_{-\sqrt{\pi}/2}^{\sqrt{\pi}/2} d\beta X(\beta)|+_I\rangle\langle +_I|_{\mathcal{G}} X^\dagger(\beta)\\
    &= |\mu\rangle\langle\mu|_{\mathcal{L}}\otimes\id_{\mathcal{G}},
\end{split}
\end{equation}
which is exactly a measurement on the logical subsystem in the computational basis.

\subsection{Conjugation and commutation with \texorpdfstring{$E(\epsilon)$}{TEXT}}
\label{subsec:comm_rel_E}
\begin{table}
\begin{centering}
\begin{tabular}{c|c}
\hline 
$U$ & $\tilde{E}\left(\epsilon\right) = UE\left(\epsilon\right)U^{\dagger}$ \\
\hline 
\hline 
$X\left(\alpha\right)$ & $e^{-\frac{\epsilon}{2}\left[\left(\hat{q}-\alpha\right)^{2}+\hat{p}^{2}\right]}$ \\
\hline 
$Z\left(\alpha\right)$ & $e^{-\frac{\epsilon}{2}\left[\hat{q}^{2}+\left(\hat{p}-\alpha\right)^{2}\right]}$ \\
\hline 
$P\left(s\right)$ & $e^{-\frac{\epsilon}{2}\left[\hat{q}^{2}+\left(\hat{p}-s\hat{q}\right)^{2}\right]}$ \\
\hline 
$R\left(\phi\right)$ & $E\left(\epsilon\right)$\\
\hline 
SUM$\left(g\right)$ & $e^{-\frac{\epsilon}{2}\left[\hat{q}_{1}^{2}+\left(\hat{p}_{1}-g\hat{p}_{2}\right)^{2}\right]}e^{-\frac{\epsilon}{2}\left[\left(g\hat{q}_{1}+\hat{q}_{2}\right)^{2}+\hat{p}_{2}^{2}\right]}$\\
\hline 
$S\left(r\right)$ & $e^{-\frac{\epsilon}{2}\left[e^{2r}\hat{q}^2 +e^{-2r}\hat{p}^2\right]}$\\
\hline 
$B\left(\theta,\phi\right)$ & $e^{-\epsilon\hat{n}_{1}}e^{-\epsilon\hat{n}_{2}}$\\
\hline 
\end{tabular}
\par\end{centering}
\caption{Physical operators $U$ and conjugated envelope operator $\tilde{E}\left(\epsilon\right)=UE\left(\epsilon\right)U^{\dagger}$ introduced in Eq.~\ref{eq:0Nen}.}
\label{tab:e_conj}
\end{table}

In Table \ref{tab:e_conj}, we display how the envelope operator $E(\epsilon)$ from \eqref{eq:0Nen} transforms under conjugation
with Gaussian operations through $E\left(\epsilon\right)\to UE\left(\epsilon\right)U^{\dagger}$. In addition to this
we might wish to see how the operations themselves change: $U\to E\left(-\epsilon\right)UE\left(\epsilon\right)$. Note that there are some mathematical difficulties in working with $E(\epsilon)$ that one should be wary of, since it is an exponential of an unbounded operator. For a discussion and a rigorous treatment of some the issues that arise, see, for example,~\cite{Biagi2018}.

To derive the commutation relations, we first show that, for any
$k\in\mathbb{N},$
\begin{align}
\left[\hat{n},\hat{a}^{k}\right] & =-k\hat{a}^{k}\label{eq:n-a-k}\\
\left[\hat{n},\hat{a}^{\dagger k}\right] & =k\hat{a}^{\dagger k}.
\end{align}
For this, proceed by induction. First, using the identity $\left[AB,C\right]=A\left[B,C\right]+\left[A,C\right]B$,
we can see that
\begin{equation}
\left[\hat{n},\hat{a}\right]=\left[\hat{a}^{\dagger}\hat{a},\hat{a}\right]=\hat{a}^{\dagger}\cancelto{0}{\left[\hat{a},\hat{a}\right]}+\cancelto{-1}{\left[\hat{a}^{\dagger},\hat{a}\right]}\hat{a}=-\hat{a}.
\end{equation}

Now choose $j\in\mathbb{N}$, and assume $\left[\hat{n},\hat{a}^{j-1}\right]=-a^{j-1}$.
In light of the related identity $\left[A,BC\right]=\left[A,B\right]C+B\left[A,C\right]$,
we have
\begin{align}
\left[\hat{n},\hat{a}^{j}\right] & =\left[\hat{n},\hat{a}^{j-1}\hat{a}\right]\\
 & =\left[\hat{n},\hat{a}^{j-1}\right]\hat{a}+\hat{a}^{j-1}\left[\hat{n},\hat{a}\right]\\
 & =-\left(j-1\right)\hat{a}^{j-1}\hat{a}+\hat{a}^{j-1}\left(-\hat{a}\right)\\
 & =-j\hat{a}^{j}.
\end{align}
To show $\left[\hat{n},\hat{a}^{\dagger k}\right]=\hat{a}^{\dagger k}$,
simply take the Hermitian conjugate of both sides of (\ref{eq:n-a-k}).

Let us now rewrite these commutation relations in a more helpful form:
\begin{align}
\left[r\hat{n},s\hat{a}^{k}\right] & =\left(-kr\right)\left(s\hat{a}^{k}\right)\\
\left[r\hat{n},t\hat{a}^{\dagger k}\right] & =\left(kr\right)\left(t\hat{a}^{\dagger k}\right).
\end{align}
For relations that look like this, that is $\left[X,Y\right]=bY$,
there is a braiding identity
\begin{equation}
e^{X}e^{Y}=e^{e^{b}Y}e^{X}.
\end{equation}
Thus we may write
\begin{align}
 & e^{r\hat{n}}e^{s\hat{a}^{k}}=e^{e^{-kr}s\hat{a}^{k}}e^{r\hat{n}}\\
\implies & e^{r\hat{n}}e^{s\hat{a}^{k}}e^{-r\hat{n}}=e^{s\left(e^{-r}\hat{a}\right)^{k}}\\
 & e^{r\hat{n}}e^{t\hat{a}^{\dagger k}}=e^{e^{kr}t\hat{a}^{\dagger k}}e^{r\hat{n}}\\
\implies & e^{r\hat{n}}e^{t\hat{a}^{\dagger k}}e^{-r\hat{n}}=e^{t\left(e^{r}\hat{a}^{\dagger}\right)^{k}}.
\end{align}

Now consider an arbitrary single-mode operator of the form
\[
U=e^{p_{k}\left(\hat{a}^{\dagger},\hat{a}\right)},
\]
where $p_{k}$ is a $k$-th degree polynomial. We can insert $E(-\epsilon)E(\epsilon)=\id$ between any two operators in this polynomial, implying that
\begin{align}
    &E(-\epsilon)p_{k}\left(\hat{a}^\dagger, \hat{a}\right)E(\epsilon) 
    \\
    =& p_{k}\left[E(-\epsilon)\hat{a}^\dagger E(\epsilon), E(-\epsilon) \hat{a} E(\epsilon) \right]\\
    =& p_{k}\left(e^\epsilon \hat{a}, e^{-\epsilon} \hat{a}^\dagger \right).
\end{align}
Again using the fact $E(-\epsilon)E(\epsilon)=\id$, we see that
\begin{align}
E\left(-\epsilon\right)e^{p_{k}\left(\hat{a}^{\dagger},\hat{a}\right)}E\left(\epsilon\right) & =e^{E\left(-\epsilon\right)p_{k}\left(\hat{a}^{\dagger},\hat{a}\right)E\left(\epsilon\right)}\\
 & =e^{p_{k}\left(e^{\epsilon}\hat{a}^{\dagger},e^{-\epsilon}\hat{a}\right)}.\label{eq:poly_a}
\end{align}

Therefore
we conclude that the envelope conjugates across any operator in the form
(\ref{eq:poly_a}) -- an exponential of an arbitrary polynomial of
the creation and anihilation operators -- at the expense of changing
the inputs via
\begin{align}
\hat{a} & \to e^{-\epsilon}\hat{a}\\
\hat{a}^{\dagger} & \to e^{\epsilon}\hat{a}^{\dagger}.
\end{align}

\subsection{Measures of non-Gaussianity} \label{subsec:non-class}

One way to characterize genuine non-Gaussianity  of a quantum state is through the negativity of the Wigner function~\cite{Kenfack2004, Albarelli2018}. The Wigner function is a phase space quasiprobability distribution defined through
\begin{equation}
W_{\rho}\left(q,p\right)=\frac{1}{\pi}\int_{-\infty}^{\infty}\Bra{q+x}\rho\Ket{q-x}e^{-2ipx}dx,
\end{equation}
and the Wigner negativity is the area of the negative part of the
Wigner function:
\begin{equation}
W\left(\rho\right)=\int dqdp\left|W_{\rho}\left(q,p\right)\right|-1.
\label{eq:wig_log_neg}
\end{equation}

The Wigner logarithmic negativity is then
\begin{equation}
W_N\left(\rho\right)=\log\left[W\left(\rho\right)+1\right].
\end{equation}

Although pure states with a vanishing Wigner negativity must be Gaussian, there exist non-Gaussian mixed states with a positive Wigner function \cite{brocker1995mixed}.

\subsection{Glancy-Knill probability of no error}\label{subsec:GKderiv}
In~\cite{glancy}, the authors find that GKP error correction can be performed perfectly with ideal GKP states displaced by less than $\sqrt{\pi}/6$. Thus, to determine the probability of successful error correction with normalizable GKP states, they first expand an arbitrary oscillator state in a basis of displaced ideal GKP states:
\begin{equation}
    \Ket{\psi} = \int_{-\sqrt{\pi}}^{\sqrt{\pi}}du\int_{-\sqrt{\pi}/2}^{\sqrt{\pi}/2}dv \langle u,v\Ket{\psi} |u,v\rangle
\end{equation}
where $\Ket{u,v} = \pi^{-1/4}e^{-iu\hat{p}}e^{-iv\hat{x}}|0_I\rangle$ is reminiscent of the modular basis later explored in \cite{ketterer2016quantum} and $\pi^{-1/4}$ is a normalization factor that ensures completeness:
\begin{equation}
    \int_{-\sqrt{\pi}}^{\sqrt{\pi}}du\int_{-\sqrt{\pi}/2}^{\sqrt{\pi}/2}dv \Ket{u,v}\Bra{u,v} = \id.
\end{equation}
Then, the probability of finding $\Ket{\psi}$ within a displaced region of less than $\sqrt{\pi}/6$, so that the errors introduced by the ancilla are small enough for the error correction to go through, is given by
\begin{equation}
    P_{\text{no error}}=\int_{-\sqrt{\pi}/6}^{\sqrt{\pi}/6}du\int_{-\sqrt{\pi}/6}^{\sqrt{\pi}/6}dv |\langle u,v\Ket{\psi}|^2.
\end{equation}
Expanding, we find
\begin{equation} 
\label{eq:gk_inter}
\begin{split}
    P_{\text{no error}} =
    \pi^{-1/2}
    \sum_{s,t=-\infty}^\infty 
    \int_{-\sqrt{\pi}/6}^{\sqrt{\pi}/6}dv \, e^{2iv(s-t)\sqrt{\pi}} \\ 
    \times \int_{-\sqrt{\pi}/6}^{\sqrt{\pi}/6}du \, \psi^*(2t\sqrt{\pi}+u)\psi(2s\sqrt{\pi}+u).
\end{split}
\end{equation}
Performing the integral over $v$ and taking $t\rightarrow t+s$ and $u\rightarrow u-2s\sqrt{\pi}$, we recover Eq.~\eqref{eq:glancy}.  Eq.~\eqref{eq:gk_inter} can be interpreted as follows: For $s=t$, one gets the probability that $\psi(x)$ lies within $\frac{\sqrt{\pi}}{6}$ of all integer multiples of $2\sqrt{\pi}$ in position space, that is, where $\Ket{0_I}$ has support. For $s \neq t$, one gets the probability that $\psi(x)$ lies within $\frac{\sqrt{\pi}}{6}$ of integer multiples of $2\sqrt{\pi}$ in momentum, without overcounting the correctable position regions.

It is straightforward to extend the formula to error correction with arbitrary mixed states:
\begin{align}
P_{\text{no error}}= & \frac{\pi}{3}\sum_{s,t}\text{sinc}\left(\frac{\pi t}{3}\right)\times\nonumber \\
 & \int_{\sqrt{\pi}(2s-\frac{1}{6})}^{\sqrt{\pi}(2s+\frac{1}{6})}du\rho(u, 2t\sqrt{\pi}+u).
\end{align}
where $\rho(x,x')$ is the density matrix in the position basis.

\section{Numerical techniques}\label{sec:numdet}

 In addition to the \textsf{strawberryfields} and \textsf{the walrus} libraries, we employed packages from \textsf{scipy} libraries for special functions, numerical integration, and optimization algorithm implementations. Here we present details of our implementation of the algorithms. 

\subsection{Algorithm \ref{algone} details}\label{subsec:alg1det}
 
 To initialize the normalizable state $\Ket{\mu_\Delta}$, we build a numerical wavefunction on a discretized position space as follows:
\begin{enumerate}
    \item To be safe, we take 7 standard deviations of $1/\Delta$ as the range in $q$ space so that the heights of the peaks at the edge of the range will be negligible.
    \item We solve for the integer number of peaks separated by $\alpha = \sqrt{\pi}$ that fit into that range, and choose the number of points in the discretization of the range of $q$ space to be 100 times the number of peaks, so that each peak is well-resolved.
    \item With the array of $q$ values in hand, we build the $\Ket{\mu_\Delta}$ wavefunction by summing together Gaussians of width $\Delta$ centred at each of the $n\sqrt{\pi}$ within the range of $q$, and weighted according to the value of $\mu \in \{0,1,+,H_+\}$.
    \item We numerically integrate the function using the \textsf{numpy trapz} function to determine the normalization factor.
\end{enumerate}  

We define the \texttt{cost} function to be the fidelity between the target state $\Ket{\mu_\Delta}$ and the approximate state of the form of a squeezed finite superposition of Fock states.
We construct a numerical wavefunction for $\Ket{\psi} = S(r) \sum_n c_n |n\rangle$. Using the same $q$ array as the $\Ket{\mu_\Delta}$ wavefunction, we sum the weighted $q$ space wavefunctions for the Fock states (built using the \textsf{scipy eval\_hermite} function). Then, we apply the squeezing using the convention from Eq.~\eqref{eq:sq_conv}. The fidelity, $|\Braket{\mu_\Delta|\psi}|^2$, is evaluated as a numerical integration with the \textsf{numpy trapz} function.

In the optimization algorithm, we employ the \textsf{basinhopping} algorithm available from the \textsf{scipy optimize} library. An overview of this algorithm is available in \cite{saba1}, with a more detailed description available in the library documentation and references therein \cite{scipy,basinhopping}. Broadly, \textsf{basinhopping} consists of two phases: 
\begin{enumerate}
    \item A local minimization (we minimize the negative of the \texttt{cost} function) over the optimization parameters $(r,\boldsymbol{c})$ given an initial guess, performed using the sequential least-squares programming method. We impose the constraint that $\boldsymbol{c}$ needs to be normalized and only has even components up to $n_{\max}$. This step produces a candidate for the global optimum (not a true global optimum).
    \item A stochastic "hop" is performed in parameter space once a local minimum is found. After the hop is made, the local minimization phase is repeated with the new point in parameter space as the initial guess. The candidate global optimum is updated based on an acceptance test. We found 40 hops to be a suitable number.
\end{enumerate}
When the algorithm terminates we are provided with an optimal squeezing parameter, $r_{\text{opt}}$, and a vector of coefficients, $\boldsymbol{c}_{\text{opt}}$, consistent with our choice of $n_{\max}$.

Importantly, we find that \textsf{basinhopping} is helpful: in some cases the optimal squeezing parameter and the truncated coefficients of the core state do not vary smoothly as we change $\Delta$, even though the fidelity appears to be changing smoothly. That is, two or more different regions of parameter space can yield comparable fidelities, so if one region becomes better than another there is an apparent jump in parameter space. The global properties of the approximate states, such as the fidelity and the average photon number, still vary smoothly because the interplay between the squeezing and the core state coefficients combine to yield similar final states even if the optimal squeezing parameters and coefficients are undergoing discontinuous jumps. Put differently, even though the stellar representation ~\cite{Chabaud2019} of a state is unique, when we apply a truncation to the core state, there can be many choices for states that meet a fidelity threshold to the target state. Thus, when one chooses the state with the highest fidelity among a collection of states that meet a fidelity threshold, the states can come from very different regions of parameter space even if $\Delta$ is only changing slightly. 

To speed up the search for optimal parameters, for a given $\mu$ and $n_{\max}$, we pass the optimal results from the previous value of $\Delta$ as the initial guess for the next value. Since the \textsf{basinhopping} global search algorithm employs random jumps in parameter space, the initial guess does not exclude finding better local optima in other regions of parameter space.

\subsection{Algorithm \ref{alg2} details}\label{subsec:alg2det}

To run Algorithm \ref{alg2} for finding the optimal circuit-produced state relative to an approximate state $\Ket{\mu_A}$, we supply the squeezing parameter, $r$; the core state Fock coefficients, $\boldsymbol{c}$, of $\Ket{\mu_A}$; the number of modes for the GBS device; and a post-selection pattern, $\bar{\textbf{n}}$, consistent with the need for the number of photodetections in $\bar{\textbf{n}}$ to total the $n_{\max}$ of the core state. We loop Algorithm \ref{alg2} over all choices of $\bar{\textbf{n}}$ up to permutations of modes, as permutations can be absorbed into the interferometer. 

There are three essential parts of Algorithm \ref{alg2}. First, we define a \texttt{cost} function which is called in the \texttt{optimization} procedure. Second, the \texttt{optimization} procedure returns the optimal circuit parameters (relative to the \texttt{cost} function) for producing the core state of $\Ket{\mu_A}$, $\boldsymbol{c}$. Third, the \texttt{redecompose\_circuit} procedure uses the optimal circuit parameters from \texttt{optimization} as well as the squeezing parameter, $r$, associated with $\Ket{\mu_A}$ to find new GBS device parameters, where $r$ is offloaded to the beginning of the circuit instead of having to be applied in-line after the interferometer to the core state. The algorithm outputs are the parameters for a GBS device that produces a state given a post-selection pattern, $\bar{\textbf{n}}$, along with the fidelity of that state to $\Ket{\mu_A}$ and the success probability of obtaining that photon detection pattern.

The \texttt{optimization} procedure runs in two steps:
\begin{enumerate}
    \item After initializing a random set of circuit parameters, $\textbf{x}_0$, consistent with the number of modes (constraining the initial squeezed light source to within 12 dB to maintain realistically achievable values), we employ the \textsf{basinhopping} algorithm already described in App.~\ref{subsec:alg1det} to find a global candidate solution that maximizes the \texttt{cost} function labelled by the "global" handle. This global \texttt{cost} function is the fidelity between the core state of $\Ket{\mu_A}$, $\boldsymbol{c}$, and the state output by the circuit given the post-selection pattern, $\bar{\textbf{n}}$. Additionally, we add to the cost function the probability of the postselection pattern weighted by a factor of 0.1; if we solely optimized fidelity, we sometimes found solutions with probabilities smaller than $10^{-6}$. We found the \textsf{L-BFGS-B} algorithm within \textsf{basinhopping} to be the most efficient local minimization algorithm in our case. We employed 50 hops to search the parameter space. 
    
    \item The optimal circuit parameters, $\textbf{x}_{1}$, from \textsf{basinhopping} are passed to \texttt{local\_search}, which employs the \textsf{scipy minimize} function to find the optimal solution to the \texttt{cost} function with the "local" handle. This local \texttt{cost} function is almost the same as the global \texttt{cost} function, but now the weight for the probability of the state is increased to 1 so that a solution in the neighbourhood of $\textbf{x}_{1}$ with comparable fidelity but higher probability is returned. The \textsf{scipy minimize} function stops running when the first local minimum is found instead of the iterative process of random hops implemented in \textsf{basinhopping}. This change ensures only the neighbourhood of $\textbf{x}_{1}$ is searched instead of the entire parameter space.
\end{enumerate}

We now move on to how we calculate the \texttt{cost} function, including how to calculate the output of the GBS circuit. The \texttt{cost} function is given the $\Ket{\mu_A}$ core state coefficients, $\boldsymbol{c}$, initialized as a vector in the Fock basis that will be used to compute fidelity; the squeezing and interferometer parameters; and the postselection pattern, $\bar{\textbf{n}}$, for the GBS device. The algorithm proceeds as follows:
\begin{enumerate}
    \item Using a \textsf{strawberryfields} engine we apply the squeezing to each mode followed by the interferometer in the rectangular decomposition. As the state of the modes at this point is still Gaussian, we use the \textsf{gaussian} backend for the engine, which returns the mean and covariance matrix for the resulting state.
    \item The mean and covariance matrix, along with $\bar{\textbf{n}}$, is passed to the \textsf{state\_vector} function from \textsf{the walrus} library to get $|\psi_\text{out}\rangle$, the coefficients of the output state in the Fock basis (up to the dimension of $\boldsymbol{c}$), and before they have been normalized by the square root of the probability of obtaining $\bar{\textbf{n}}$.
    \item The probability is computed in the following way:
    \begin{enumerate}
        \item The mean and covariance matrix of all the modes before PNR measurements, as well as the indices of the modes we intend to measure, are passed to the \textsf{strawberryfields reduced\_gaussian} function, which returns the mean and covariance matrix of $\rho_{n-1}$, the reduced $(n-1)$-mode state of all but the output mode. $\rho_{n-1}$ is still Gaussian, as it is the partial trace of a Gaussian state.
        \item Using the mean and covariance matrix of $\rho_{n-1}$, we can use the \textsf{density\_matrix\_element} function of \textsf{the walrus} and find the diagonal element $\langle\bar{\textbf{n}}|\rho_{n-1}|\bar{\textbf{n}}\rangle$ which is exactly the probability of finding the post-selection pattern $\bar{\textbf{n}}$.
    \end{enumerate}
    \item We normalize $|\psi_\text{out}\rangle$ with the probability obtained in the previous step.
    \item We take the inner product between $|\psi_\text{out}\rangle$ and $\boldsymbol{c}$ in the Fock basis to compute the fidelity, and use the already calculated probability to compute the relevant \texttt{cost} function being called.
\end{enumerate}

Finally, to implement \texttt{redecompose\_circuit}, we find the Gaussian unitary associated with $S(r)U(\bar{\boldsymbol{\Theta}})D(\boldsymbol{\alpha})S(\boldsymbol{z})$ and use the \textsf{strawberryfields GaussianTransform} function to find $U(\bar{\boldsymbol{\Theta}}')D(\boldsymbol{\alpha'})S(\boldsymbol{z'})$. 

\section{Approximate GKP \texorpdfstring{$Z$}{TEXT}, \texorpdfstring{$X$}{TEXT}, and \texorpdfstring{$H$}{TEXT} eigenstates}

\subsection{Further comments on \texorpdfstring{$\Ket{0_A}$}{TEXT}}\label{subsec:further0}

Our results for constructing the $\Ket{0_A}$ state using $n_{\max} = 4$ and 6 were identical until $\Delta \approx 8$ dB, meaning the $n=6$ component was not required; after that point, there was a jump in parameter space and the $n=6$ state was used to attain better fidelity than $n_{\max} = 4$, albeit only slightly. Here we explain why it makes sense that the results for $n_{\max} = 4$ and 6 are so close, and why we see a small jump in parameter space for $n_{\max} = 6$.

\begin{figure}
    \centering
    \includegraphics[width=\linewidth]{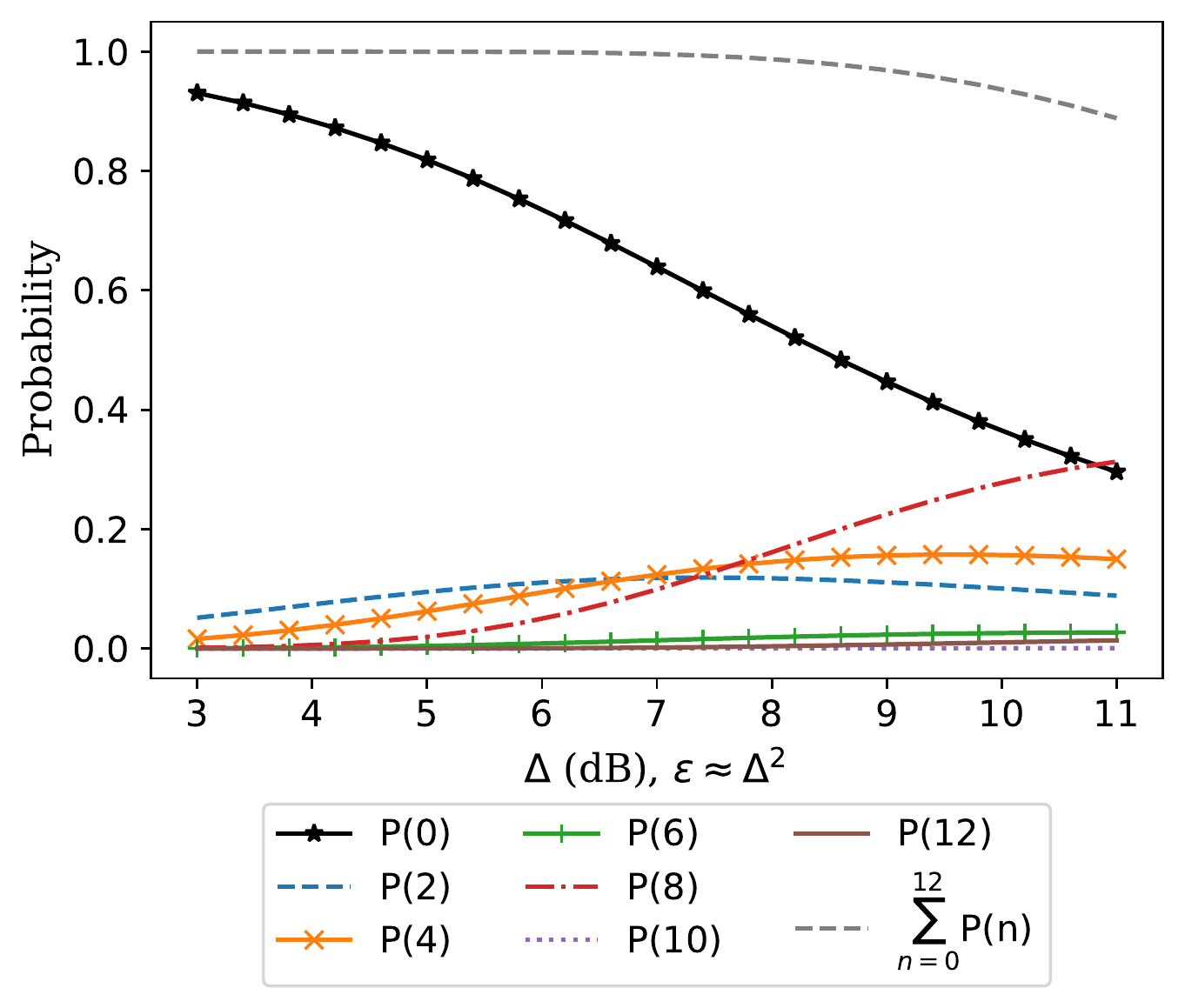}
    \caption{ Probabilities of different even Fock states in $|0_\epsilon\rangle$ up to $n=12$ as a function of $\epsilon \approx \Delta^2$. We see that the probability of measuring 8 photons is higher than 6; thus, when constructing the GKP states from squeezed core states, increasing $n_{\max}$ of the core state from 4 to 6 provides little to no advantage compared to increasing $n_{\max}$ to 8.}
    \label{fig:fock_probs}
\end{figure}

First, we can use the $|0_\epsilon\rangle$ state to understand the distribution of the Fock coefficients for $\Ket{0_{\Delta}}$, since the states are almost indistinguishable for $\epsilon = \Delta^2$. In Fig.~\ref{fig:fock_probs}, we plot the probabilities of detecting the even Fock states in $\Ket{0_\epsilon}$ up to $n=12$ as a function of $\Delta$, for $\epsilon = \Delta^2$. We see that, up to $\Delta \approx 8$ dB, the $n=2$ and 4 states are the most probable states after the vacuum, and beyond $\Delta \approx 8$ dB, the $n=8$ state becomes more probable. The probabilities for $n=6,10$, and 12 remain significantly smaller over the whole range. Thus, if one wants to increase the $n=8$ component using only squeezing applied to a core state of $n_{\max}=6$, one will also end up increasing the $n=6$ component in the distribution. If one wants to keep the $n=6$ component small, one option is to truncate the core state at $n=4$; as it turns out, this is the optimal option found numerically up to $\Delta \approx 8$ dB, leading to similar results between $n_{\max} = 4$ and 6 in that regime.

\begin{figure}
    \centering
    \includegraphics[width=\linewidth]{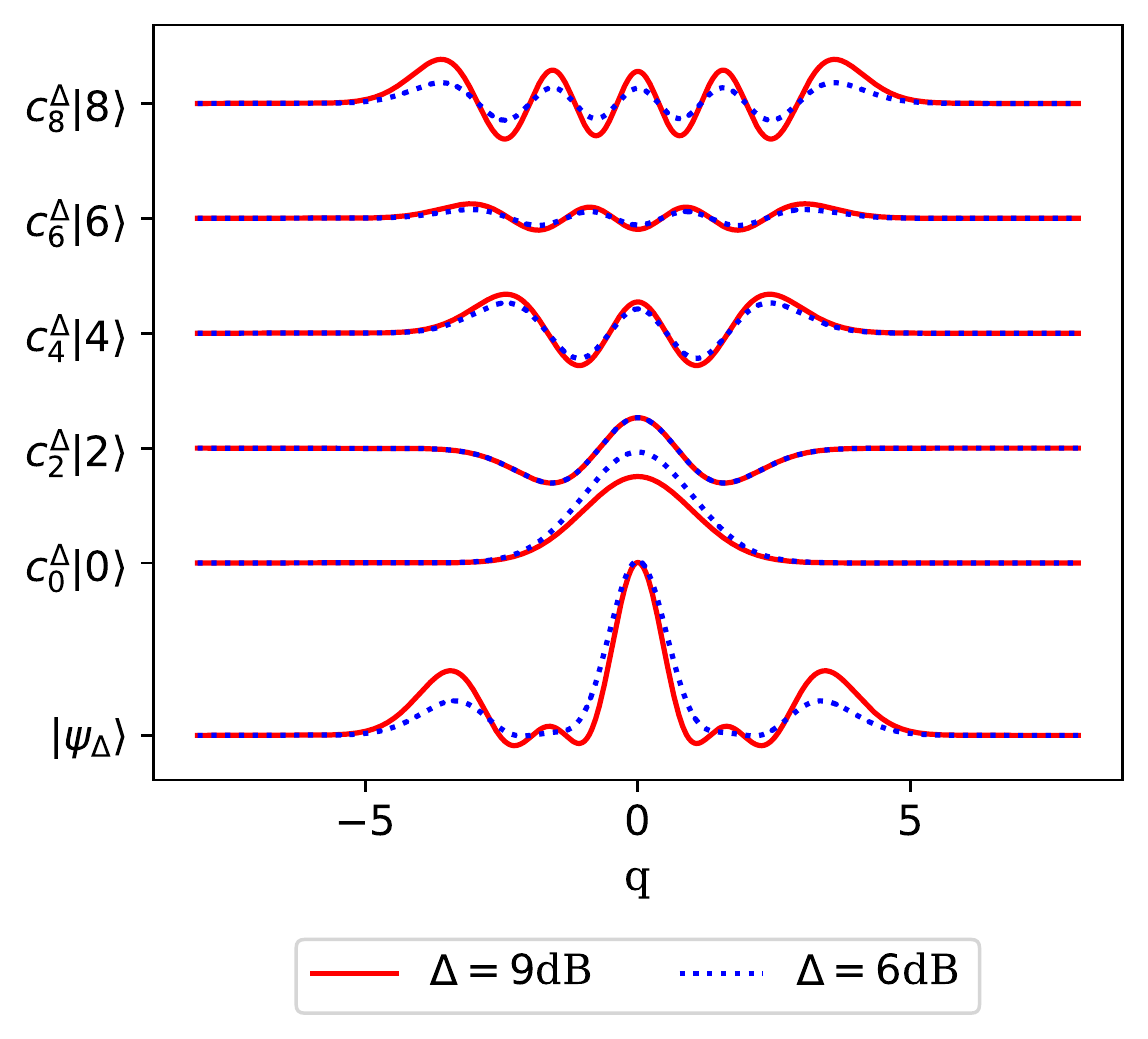}
    \caption{ Wavefunctions produced from summing the first eight Fock states in the distribution for $|0_\epsilon\rangle$ (using $\epsilon \approx \Delta^2$), as well as the wavefunctions for the weighted Fock states in the superposition. We see the important contribution of the 8 photon state to the outer set of peaks for the normalizable GKP state. Thus, without the 8-photon state in the core of $\Ket{0_A}$, the $n=4$ or 6 states are needed to build the outer peaks, leading to worse fidelities.}
    \label{fig:fock_decomp}
\end{figure}

We can also use the wavefunctions of the Fock states to understand why the $n=2,4$, and 8 components are weighted higher than $n=6$. In Fig.~\ref{fig:fock_decomp}, we plot the wavefunction produced by the sum of the Fock state components in $\Ket{0_{\Delta}}$ up to $n=8$, as well as the wavefunctions for the weighted Fock states in the superposition. We start to see the peak structure of the GKP wavefunctions -- a central peak and two outer peaks at the correct locations -- in addition to some wiggles in between due to the truncation. Importantly, we see that the outer set of peaks is created almost exclusively by the $n=8$ Fock state, with the $n=2$ and 4 states working to narrow the central peak. Thus, were one to construct $\Ket{0_{\Delta}}$ by only using a core state up to $n=6$ and then stretching the Fock wavefunctions through squeezing, building the outer peaks of the $\Ket{0_{\Delta}}$ state would require stretching the $n=6$ or the $n=4$ state to align with the GKP grid. But using the $n=6$ state and weighting it heavily would shrink the central peak, since the $n=6$ state has a dip at $q=0$. Thus, the better option for matching the symmetry of the $\Ket{0_{\Delta}}$ state is to stretch the $n=4$ state, which has outer peaks and a central peak. This is further evidence for why we see no difference between the results for $\Ket{0_A}$ using $n_{\max} = 4$ and 6 for $\Delta < 8$ dB.

For $\Delta > 8$ dB, we see a small jump in the $(r,\boldsymbol{c})$  parameter space for the $n_{\max} = 6$ state, although the resulting wavefunction does not change much, leading to only slight jumps in the fidelity and average photon numbers. In that regime, all coefficients in the core state are now nonzero, and the squeezing applied to the core state as compared to $\Delta < 8$ dB is slightly greater. We understand this jump to come from the fact that the achievable fidelity to the target state is already relatively low ($\approx 90\%$) for this range of $\Delta$; in other words, since the core state resource available from $n_{\max}$ is not too high, we are simply trying to find the best fidelity among several poorly contending regions of parameter space. There are two basins in parameter space that provide very similar fidelities when $\Delta=8$ dB; as $\Delta$ increases, one basin overtakes the other to provide marginally higher fidelity, leading to a jump from the first basin to the second.

\subsection{Results for \texorpdfstring{$\Ket{1_A},\Ket{+_A}$, and $\Ket{H_{+A}}$}{TEXT}}\label{subsec:1pm}

In Fig.~\ref{fig:fid vs n,D1pm}, we provide the numerical results for the construction of the $\Ket{1_A}$, $\Ket{+_A}$, and $\Ket{H_{+A}}$ states using Algorithm~\ref{algone}. All the trends for fidelity to the target states are comparable to those of the $\Ket{0_A}$ states. The $\Ket{+_A}$ state is effectively as resource-intensive to create as the $\Ket{0_A}$ state given their relation by a Fourier transform. Finally, the $\Ket{H_{+A}}$ state has squeezing and average photon numbers on the same order as $\Ket{0_A}$ and $\Ket{+_A}$, so it should also be comparatively demanding to prepare. 

 Notice that the results for the $\Ket{0_A}$ in the main text and $\Ket{+_A}$ states are nearly indistinguishable. Since the states are closely relatable by a Fourier transform \footnote{They would be exactly relatable if we chose the $E(\epsilon)$ envelope since $E(\epsilon)$ commutes with phase space rotations.}, $r$ for the $\Ket{0_A}$ state is approximately $-r$ for the $\Ket{+_A}$ state: one is squeezed in $q$ and the other in $p$. Additionally, since the state $\Ket{n}$ is an eigenfunction of the Fourier transform with eigenvalue $(-i)^n$, we expect the coefficients of the Fock states in the core state superposition to be related by phases of the form $(-i)^n$. In fact, we do see in our results for $\Ket{0_A}$ and $\Ket{+_A}$ that the core Fock states with $n \bmod 4 = 0$, for which $(-i)^n = 1$, have almost the same coefficients, while all the states wth $n \bmod 4 = 2$, for which $(-i)^n = -1$, have coefficients that differ effectively by a phase of $-1$. As a result of the states having such similar properties, their global properties end up being very close.  

\begin{figure*}
    \centering
    \includegraphics[width=\textwidth]{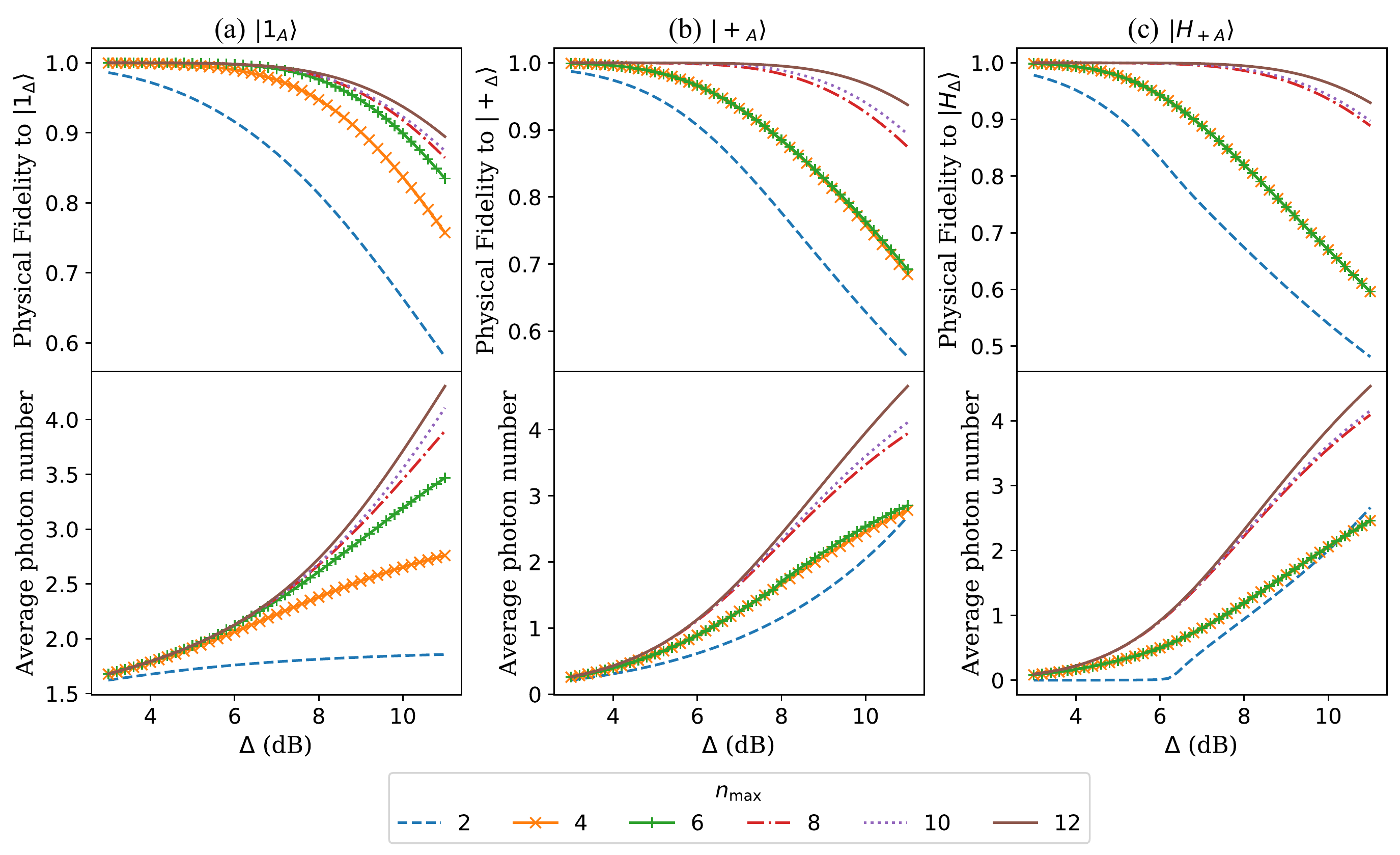}
    \caption{ Fidelity of approximate states $\Ket{\mu_A}$ to normalizable states $\Ket{\mu_\Delta}$; and photon number vs. $\Delta$ for (by column) $\mu$ = (a) 1, (b) +, and (c) $H_+$. Line colours/styles in each plot represent the results for different $n_{\max}$ in the core states of $\Ket{\mu_A}$. The trend in all cases is that a greater $n_{\max}$ leads to better fidelity to the target state and higher energy; if the quality of the target state is improved by increasing the dB value of $\Delta$, for a fixed $n_{\max}$, the fidelity gets worse, and the required energy gets higher. We expect comparable circuit resources (see Sec.~\ref{subsec:circuits}) are required to construct approximate states for any point on the Bloch sphere.}
    \label{fig:fid vs n,D1pm}
\end{figure*}

\section{Additional characterization of approximate GKP states}\label{sec:add_feat}

\subsection{Projectors}
\begin{figure*}
    \centering
    \includegraphics[width=\textwidth]{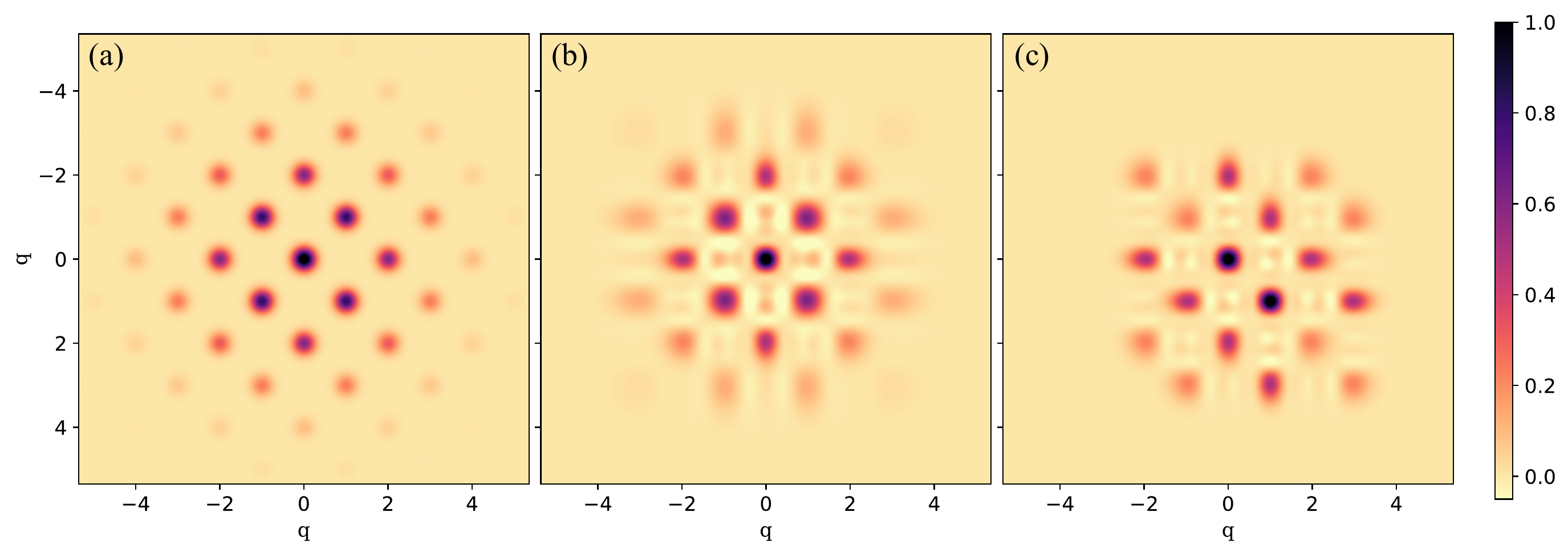}
    \caption{ Single-qubit codespace projectors $P(q,q')$ in $q$ basis (in units of $\sqrt{\pi}$) for three cases: (a) The normalizable GKP states $\Ket{0_\Delta}\Bra{0_\Delta}+\Ket{1_\Delta}\Bra{1_\Delta}$, with $\Delta = 10$ dB. (b) Our approximate GKP states $\Ket{0_A}\Bra{0_A}+\Ket{1_A}\Bra{1_A}$, designed to target the normalizable states with $\Delta = 10$ dB, constructed using squeezing applied to a core state with $n_{\max} = 12$. Here we assume we prepare both $\Ket{0_A}$ and $\Ket{1_A}$. (c) Our approximate GKP states $\Ket{0_A}\Bra{0_A}+ X(\sqrt{\pi})\Ket{0_A}\Bra{0_A}X^\dagger(\sqrt{\pi})$ with the same parameters as (b), except here we assume we prepare only $\Ket{0_A}$ and create the logical 1 state by displacing $\Ket{0_A}$. We see that the symmetries of the target projector are best preserved when preparing both $\Ket{0_A}$ and $\Ket{1_A}$, although that symmetry may not be required for encoding.}
    \label{fig:proj_eg}
\end{figure*}

Projectors provide a valuable understanding as well as some visual intuition for how the code subspace behaves. Here we consider projectors defined by
\begin{enumerate}
    \item The normalizable states $\Ket{\psi_\Delta}$: $\Ket{0_\Delta}\Bra{0_\Delta}+\Ket{1_\Delta}\Bra{1_\Delta}$;
    \item The approximate states $\Ket{\psi_A}$: $\Ket{0_A}\Bra{0_A}+\Ket{1_A}\Bra{1_A}$;
    \item $\Ket{0_A}$ and displaced $\Ket{0_A}$:
    $$\Ket{0_A}\Bra{0_A}+X(\sqrt{\pi})\Ket{0_A}\Bra{0_A}X^\dagger(\sqrt{\pi}).$$
\end{enumerate}
We consider the third kind of projector since one may want to prepare just one GKP $Z$ eigenstate and generate the other through gate application. In Fig.~\ref{fig:proj_eg}, we provide some examples of these projectors plotted in the $q$ basis. We take $\Delta = 10$ dB, and choose the approximate states that are meant to target the $\Delta = 10$~dB states with core states of $n_{\max} = 12$. Note that the projector $\Ket{0_A}\Bra{0_A}+\Ket{1_A}\Bra{1_A}$ maintains the symmetry of the target state projector, while the displacement in $\Ket{0_A}\Bra{0_A}+ X(\sqrt{\pi})\Ket{0_A}\Bra{0_A}X^\dagger(\sqrt{\pi})$ breaks it. As we know from the modular subsystem decomposition, a loss of such symmetry does not affect the logical encoding. Of course, because the approximate states only have $\sim 95\%$ fidelity to the target states, the sharp peaks get blurrier further away from the origin in $q$ space.

\subsection{Quantum error correction matrix}
One object that encompasses the error-correcting properties of general codes is the quantum error correction (QEC) matrix~\cite{NielsenChuang,albert2018performance}. For an error map with Kraus operators $\{E_k\}$ and code states $|0_G\rangle$ and $|1_G\rangle$, the QEC matrix elements are defined by
\begin{equation}
\label{eq:qec}
    \gamma_{kk'} \equiv P_G E^\dagger_k E_{k'}P_G,
\end{equation}
where $P_G$ is the projector constructed from the code states, as defined in App.~\ref{sec:add_feat}.

The $2 \times 2$ matrix $\gamma$ can then be expanded in a basis of Pauli matrices:
\begin{equation}
    \gamma_{kk'} = \varepsilon_0 \id + \varepsilon_x\sigma_x + \varepsilon_y\sigma_y + \varepsilon_z\sigma_z.
\end{equation}
If $\gamma$ is proportional to the identity, then there exists a recovery which can perfectly correct the error~\cite{NielsenChuang}. When this is not the case, errors can only be approximately corrected. However, the magnitude of the coefficients $\varepsilon_i$ offer a convenient interpretation of the error's effects; namely, $\varepsilon_x$, $\varepsilon_y$, and $\varepsilon_z$ are the probability of bit, bit-phase, and phase flip errors~\cite{albert2018performance}.

\begin{figure*}
    \centering
    \includegraphics[width=\textwidth]{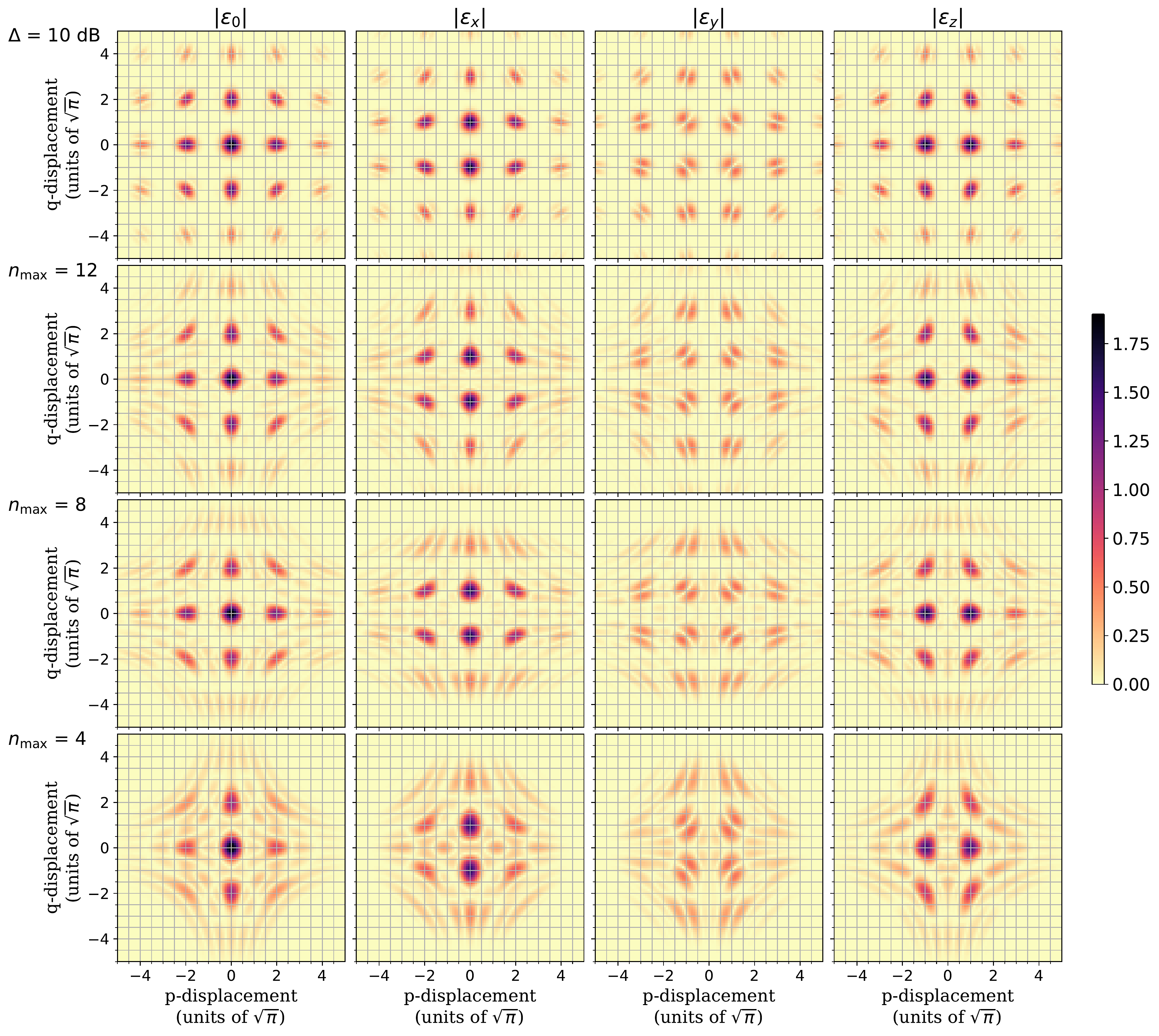}
    \caption{ Visualization of the quantum error correction (QEC) matrices (Eq.~\eqref{eq:qec}) of normalizable and approximate GKP states, $\Ket{\psi_\Delta}$ and $\Ket{\psi_A}$, subject to displacement errors. Each row of subfigures corresponds to a different choice of code states. In each subfigure, the y (x) axis corresponds to a value of displacement along $q$ ($p$) in phase space, so that a single point corresponds to a net displacement error. The QEC matrix associated with this displacement error can then be decomposed in terms of the identity and Pauli operators. Each column corresponds to the magnitude of the coefficients in the decomposition, which can be interpretted as no error, bit flip, bit-phase flip, and phase flip probabilities. The first row corresponds to $\Ket{\psi_\Delta}$ with $\Delta = 10$ dB. The second through fourth rows correspond to $\Ket{\psi_A}$ meant to approximate $\Ket{\psi_\Delta}$ with $\Delta = 10$ dB with core states of $n_{\max}$ = 12, 8, and 4 photons.}
    \label{fig:QEC}
\end{figure*}

Ideal square-lattice GKP states were designed to correct displacement errors in phase space of up to $\sqrt{\pi}/2$. Thus, we examine the QEC matrix for displacement errors applied to the normalizable and approximate states. In Fig.~\ref{fig:QEC}, we calculate $P_G D(\beta)P_G$ as a function of displacement, $\beta$, expand it in terms of the Pauli matrices, and then plot the magnitudes of $\varepsilon_i$ as functions of $\beta$. We do this for four choices of states (that is, four different $P_G$): the normalizable GKP states with $\Delta = 10$~dB and the approximate GKP states for the same $\Delta$ but with $n_{\max} = 4,8$, and 12.

The first row of plots in Fig.~\ref{fig:QEC} corresponds to the normalizable states $\Ket{0_\Delta}$ with $\Delta = 10$~dB. We see that, for small displacements, the largest magnitude is $\varepsilon_0$, meaning the matrix is basically proportional to the identity, meaning these displacement are correctable. A displacement by $\sqrt{\pi}$ in $q$ ($p$) leads to the dominant terms becoming $\varepsilon_x$ ($\varepsilon_z$), corresponding to a bit (phase) flip. Further away from the origin, the blobs become more smeared, indicating that a displacement of that magnitude is more likely to lead to an uncorrectable error.

As we progress down the column, we see how the approximate states perform. The radii of the blobs become bigger as $n_{\max}$ decreases, meaning the probability of error increases. Moreover, the blobs further from the origin become quite smeared in phase space, meaning only the first multiples of $\sqrt{\pi}/2$ will be correctable.

\clearpage

\bibliography{refs}
\end{document}